\documentclass[onecolumn]{aa} 
\usepackage{graphicx} 
\usepackage{subcaption} 
\usepackage{txfonts} 
\usepackage{amsmath} 
\usepackage{amssymb} 
\usepackage{multirow} 
\usepackage{hyperref} 
\usepackage{xcolor}

\hypersetup{
    colorlinks=true,
    linkcolor=blue,
    citecolor=blue,
    filecolor=blue,      
    urlcolor=blue,
}

\begin{document}

   \title{The dynamical evolution of close-in binary systems formed by a super-Earth and its host star}


   \subtitle{Case of the Kepler-21 system}

   \author{S. H. Luna
          \inst{1,2}\thanks{Corresponding author (Santiago Luna): \texttt{santiagohluna@gmail.com}}
          \and
           H. D. Navone
           \inst{1,3}
          \and
           M. D. Melita
          \inst{2,4}
          }

   \institute{Facultad de Ciencias Exactas, Ingeniería y Agrimensura, Universidad Nacional de Rosario, Av. Pellegrini 250, Rosario, Argentina.
         \and
             Instituto de Astronomía y Física del Espacio (IAFE). CONICET-Universidad de Buenos Aires. Buenos Aires, Argentina.
         \and
             Instituto de Física de Rosario (IFIR). CONICET-Universidad Nacional de Rosario. Rosario, Argentina.
         \and Facultad de Ciencias Astronómicas y Geofísicas, Universidad Nacional de La Plata. Paseo del Bosque s/n. La Plata. Argentina.
             }

   \date{Received ...; accepted ...}

 
  \abstract
  {}
  {The aim of this work is to develop a formalism for the study of the secular evolution of a binary system which includes interaction due to the tides that each body imparts on the other. We also consider the influence of the $J_2$-related secular terms on the orbital evolution and the torque, caused by the triaxiality, on the rotational evolution, both of which are associated only to one of the bodies. We apply these set of equations to the study of the orbital and rotational evolution of a binary system composed of a rocky planet and its host star in order to characterize the dynamical evolution at work, particularly near spin-orbit resonances.}
  {We used the equations of motion that give the time evolution of the orbital elements and the spin rates of each body to study the time evolution of the Kepler-21 system as an example of how the formalism that we have developed can be applied.}
  {We obtained a set of equations of motion without singularities for vanishing eccentricities and inclinations. This set gives, on one hand, the time evolution of the orbital elements due to the tidal potentials generated by both members of the system as well as the triaxiality of one of them. On the other hand, it gives the time evolution of the stellar spin rate due to the corresponding tidal torque and of the planet's rotation angle due to both the tidal and triaxiality-induced torques. We found that for the parameters and the initial conditions explored here, the tidally and triaxiality-induced modifications of the tidal modes can be more significative than expected and that the time of tidal synchronization strongly depends on the values of the rheological parameters.}
   {}


   \maketitle 

\section{\label{sec:intro}Introduction}

Recent developments and improvements in the field of exoplanet detection methods have allowed for the discovery of many Earth-sized planets, so-called super-Earths, which have masses between $1$ and $10 \, M_{\oplus}$. This has given rise to an interest in other, related topics, such as the dynamical and thermal evolution \citep{stamenkovicetal2012,shojiykurita2014,makarovetal_2018,renaudhenning2018}, the presence of atmospheres \citep{seager2010}, the internal structure and composition \citep{valenciaetal2006,valenciaetal2007b,valenciaetal2007c}, and the possibility that these types of celestial bodies offer for harboring life.

In this work, we are interested in the first of these problems, that is, the dynamical evolution of a binary system formed by a super-Earth and its host star, modeled as bodies that are capable of experiencing internal stresses and energy dissipation. Our aim is to develop a set of equations of motion to study the orbital and rotational evolution of such a system due to the tides that each celestial body induces on the other and the triaxiality of  the planet itself. We are particularly interested in the dynamical evolution near spin orbit-resonances. For this reason, we have derived the equations of motion that give, on one hand, the time evolution of the orbital elements due to the secular terms of both the tidal and the $J_2$-related disturbing potentials and, on the other hand, the time evolution of the stellar rotation rate and that of the rotation angle of the planet, which are both due to the secular terms of the corresponding tidal torque and to the triaxiality-induced torque on the latter.

Certainly, we expect that close-in planets ends their rotational history in the 1:1 spin-orbit resonance, that is, when $\dot{\theta} = n,$ where $\dot{\theta}$ is the spin rate of the considered body and $n$ is the the mean orbital frequency, which, in the unperturbed two point-masses problem, is defined by the mathematical expression of the Kepler's third law: 
\begin{equation}\label{ec:defn} 
G \left( m_1 + m_2 \right) = n^2 a^3 
,\end{equation} where $G$ is the universal gravitation constant. Yet, as we shall see later, a rotating body may get trapped in higher-than-synchronous rotation states \citep{efroimsky2012} and that the eccentricity plays a key role in the possibility of capture in spin-orbit resonances, as shown by \citet{gold_peale_1966,efroimsky2012,makarov_2012,makarovetal_2012,noyellesetal2014}.

Resonant capture in spin-orbit resonances is a phenomenon that occurs due to the interplay between the tidal and the triaxiality-induced torques. The former torque is due to the tides that each component of the binary system induces on the other, while the latter is characterized by the principal moments of inertia of at least one of them. Naturally, here, we consider super-Earths as triaxial bodies because they probably have a rocky composition, in contraposition to the stars, which are expected to be fluid.

This work is organized as follows: In Sect.~\ref{sec:generalities}, we present the equations of motion in its general form as were derived by \citet{boue_efroimsky2019}. We also explicitly give  the tidal and triaxial potentials and the rheological models for each component. In Sect.~\ref{sec:eom}, we derive the equations that have to be solved numerically in order to obtain the time evolution of the particular system under study, which have the particularity of being regular for very small and null eccentricities and inclinations. In Sect.~\ref{sec:prev}, we discuss the choice of the values of the physical and rheological parameters involved and the initial conditions for the numerical integration. In Sect.~\ref{sec:charact_dynamical_ev}, we study the dynamical evolution of the Kepler-21 system as an example of application of the developed methodology. Also, we present and discuss the results obtained from the numerical integration of the equations of motion mentioned above and evaluate the possibility of detecting tidal effects. Finally, in Sect.~\ref{sec:conclusions}, we offer a final discussion and draw our conclusions.

\section{\label{sec:generalities}Generalities}

To describe the dynamical evolution, including the orbital and rotational motions, of a binary system constituted of a rocky planet and its host star, we make use of the equations of motion derived by \citet{boue_efroimsky2019}, which are valid for the case when the tidal dissipation within both bodies is considered, including the disturbing potential due to the permanent-triaxial shape of the planet.

We consider a system of two bodies, the host star and a rocky planet. Each body is thought of as a nearly spherical mass distribution characterized by its principal moments of inertia, total mass, and mean radius. The former is assumed to have total mass, $m_1$, and radius, $R_1$, and the latter has the total mass, $m_2$, and radius, $R_2$. The corresponding principal moments of inertia of each body is represented by $A_k$, $B_k$, and $C_k$ (with $k=1,2$ and $A<B<C$). The index, $k$, identifies each body: $k=1$ corresponds to the star-related parameters and variables, while $k=2$ corresponds to the planet-related ones.

The mutual orbit is parameterized by the six classical orbital elements: the major semiaxis ($a$), the eccentricity ($e$), the mean anomaly ($M$), the argument of pericenter ($\omega$), the inclination ($i$), and the longitude of the ascending node ($\Omega$). However, as we allow for the tidal dissipation within both bodies, then the two reference systems corresponding to each one must be considered in order to evaluate the forces each body exerts on the other. In other words, when the force of the planet on the star is to be evaluated, we have to consider the orbit of the former as seen from the latter, and vice versa when evaluating the force of the star on the planet. As both orbits are defined with respect to the corresponding equatorial planes of each body, then the aforementioned reference systems are related by a rotation from one system to an inertial reference system and then to the system attached to the other body. However, this rotations only affect the angles that give the orientation of the orbits in the space, that is, $\omega$, $i$ and $\Omega$, while $a$, $e,$ and $M$ are the same in both systems \citep{boue_efroimsky2019}.

We assume that each body rotates about its maximal inertia axis. The rotational state of the star and the planet are completely determined by the corresponding Euler angles, that is, the precession, $\psi_k$, nutation, $\varepsilon_k$, and proper rotation angles, $\theta_k$ -- the latter being reckoned from a fixed direction in an inertial reference system -- as well as their time rates $\dot{\psi}_k$, $\dot{\varepsilon}_k$ and $\dot{\theta}_k$, respectively. As we see later in this paper, under the gyroscopic approximation, it is enough to follow the time evolution of $\varepsilon_k$ and $\theta_k$ \citep{boue_efroimsky2019}.

\subsection{\label{ss:evorb}Orbital evolution}

The orbital dynamics is described by a set of Lagrange-type planetary equations which give, on one hand, the time evolution of the major semiaxis, the eccentricity, and the mean anomaly along with, on the other hand, the time evolution of the inclination, the argument of the pericenter and the longitude of the ascending node of the common orbit as seen from each body. The respective equations were derived by \citet{boue_efroimsky2019}, which correspond to Eqs. (116)-(121) of the cited work and we reproduce them here:
\begin{subequations}
\begin{equation}\label{ec:dadt_gral} 
\frac{\mathrm{d}a}{\mathrm{d}t} = \frac{2}{n \, a} \frac{\partial R}{\partial M}, 
\end{equation} 
\begin{equation}\label{ec:dedt_gral} 
\frac{\mathrm{d}e}{\mathrm{d}t} = \frac{1-e^2}{n \, a^2 \, e} \frac{\partial R}{\partial M} - \frac{\sqrt{1- e^2}}{n \, a^2 \, e} \left(\frac{\partial R}{\partial \omega_1} + \frac{\partial R}{\partial \omega_2} \right), 
\end{equation} 
\begin{equation}\label{ec:dMdt_gral} \frac{\mathrm{d} M}{\mathrm{d}t} = n - \frac{2}{n \, a} \frac{\partial R}{\partial a} - \frac{1-e^2}{n \, a^2 \, e} \frac{\partial R}{\partial e} 
,\end{equation} 
\begin{multline}\label{ec:di1dt_gral}
\frac{\mathrm{d}i_1}{\mathrm{d}t} = \frac{\beta}{C_1 \, \dot{\theta}_1 \sin i_1} \left(\frac{\partial R}{\partial \omega_1} - \cos i_1 \frac{\partial R}{\partial \Omega_1} \right) - \frac{1}{n \, a^2 \sqrt{1-e^2} \sin i_1} \left(\frac{\partial R}{\partial \Omega_1} - \cos i_1 \frac{\partial R}{\partial \omega_1} \right) \\ - \frac{\sin (\omega_1 - \omega_2)}{n a^2 \sqrt{1-e^2}} \frac{\partial R}{\partial i_2} - \frac{\cos (\omega_1 - \omega_2)}{n a^2 \sqrt{1-e^2} \sin i_2} \left(\frac{\partial R}{\partial \Omega_2} - \cos i_2 \frac{\partial R}{\partial \omega_2} \right), 
\end{multline} 
\begin{multline}\label{ec:di2dt_gral} 
\frac{\mathrm{d}i_2}{\mathrm{d}t} = \frac{\beta}{C_2 \, \dot{\theta}_2 \sin i_2} \left(\frac{\partial R}{\partial \omega_2} - \cos i_2 \frac{\partial R}{\partial \Omega_2} \right) - \frac{1}{n \, a^2 \sqrt{1-e^2} \sin i_2} \left(\frac{\partial R}{\partial \Omega_2} - \cos i_2 \frac{\partial R}{\partial \omega_2} \right) \\ - \frac{\sin (\omega_2 - \omega_1)}{n a^2 \sqrt{1-e^2}} \frac{\partial R}{\partial i_1} - \frac{\cos (\omega_2 - \omega_1)}{n a^2 \sqrt{1-e^2} \sin i_1} \left(\frac{\partial R}{\partial \Omega_1} - \cos i_1 \frac{\partial R}{\partial \omega_1} \right), 
\end{multline} 
\begin{multline}\label{ec:dw1dt_gral} 
\frac{\mathrm{d} \omega_1}{\mathrm{d}t} = - \frac{\beta}{C_1 \, \dot{\theta}_1 \sin i_1} \frac{\partial R}{\partial i_1} + \frac{\sqrt{1-e^2}}{n \, a^2 \, e} \frac{\partial R}{\partial e} - \frac{\cos i_1}{n \, a^2 \, \sqrt{1-e^2} \sin i_1} \frac{\partial R}{\partial i_1} \\ - \frac{\cos i_1 \cos \left(\omega_1 - \omega_2 \right)}{n \, a^2 \, \sqrt{1-e^2} \sin i_1} \frac{\partial R}{\partial i_2} - \frac{\cos i_1 \sin \left(\omega_1 - \omega_2 \right)}{n \, a^2 \, \sqrt{1-e^2} \sin i_1 \sin i_2} \left(\cos i_2 \frac{\partial R}{\partial \omega_2} - \frac{\partial R}{\partial \Omega_2}\right), 
\end{multline} 
\begin{multline}\label{ec:dw2dt_gral} 
\frac{\mathrm{d} \omega_2}{\mathrm{d}t} = - \frac{\beta}{C_2 \, \dot{\theta}_2 \sin i_2} \frac{\partial R}{\partial i_2} + \frac{\sqrt{1-e^2}}{n \, a^2 \, e} \frac{\partial R}{\partial e} - \frac{\cos i_2}{n \, a^2 \, \sqrt{1-e^2} \sin i_2} \frac{\partial R}{\partial i_2} \\ - \frac{\cos i_2 \cos \left(\omega_2 - \omega_1 \right)}{n \, a^2 \, \sqrt{1-e^2} \sin i_2} \frac{\partial R}{\partial i_1} - \frac{\cos i_2 \sin \left(\omega_2 - \omega_1 \right)}{n \, a^2 \, \sqrt{1-e^2} \sin i_2 \sin i_1} \left(\cos i_1 \frac{\partial R}{\partial \omega_1} - \frac{\partial R}{\partial \Omega_1}\right), 
\end{multline} 
\begin{multline}\label{ec:dOm1dt_gral} 
\frac{\mathrm{d} \Omega_1}{\mathrm{d}t} = \frac{\beta \cos i_1}{C_1 \, \dot{\theta}_1 \sin i_1} \frac{\partial R}{\partial i_1} + \frac{\beta \, \cos \varepsilon_1}{C_1 \, \dot{\theta}_1\sin \varepsilon_1} \left(\sin \Omega_1 \cot i_1 \frac{\partial R}{\partial \Omega_1} - \cos \Omega_1 \frac{\partial R}{\partial i_1} - \frac{\sin \Omega_1}{\sin i_1} \frac{\partial R}{\partial \omega_1}\right) + \frac{1}{n \, a^2 \, \sqrt{1-e^2} \sin i_1} \frac{\partial R}{\partial i_1} \\ + \frac{\cos \left(\omega_2 - \omega_1 \right)}{n \, a^2 \, \sqrt{1-e^2} \sin i_1} \frac{\partial R}{\partial i_2} + \frac{\sin \left(\omega_1 - \omega_2\right)}{n \, a^2 \, \sqrt{1-e^2} \sin i_1 \sin i_2} \left(\cos i_2 \frac{\partial R}{\partial \omega_2} - \frac{\partial R}{\partial \Omega_2}\right), 
\end{multline} 
\begin{multline}\label{ec:dOm2dt_gral} 
\frac{\mathrm{d} \Omega_2}{\mathrm{d}t} = \frac{\beta \cos i_2}{C_2 \, \dot{\theta}_2 \sin i_2} \frac{\partial R}{\partial i_2} + \frac{\beta \, \cos \varepsilon_2}{C_2 \, \dot{\theta}_2\sin \varepsilon_2} \left(\sin \Omega_2 \cot i_2 \frac{\partial R}{\partial \Omega_2} - \cos \Omega_2 \frac{\partial R}{\partial i_2} - \frac{\sin \Omega_2}{\sin i_2} \frac{\partial R}{\partial \omega_2}\right) + \frac{1}{n \, a^2 \, \sqrt{1-e^2} \sin i_2} \frac{\partial R}{\partial i_2} \\ + \frac{\cos \left(\omega_1 - \omega_2 \right)}{n \, a^2 \, \sqrt{1-e^2} \sin i_2} \frac{\partial R}{\partial i_1} + \frac{\sin \left(\omega_2 - \omega_1\right)}{n \, a^2 \, \sqrt{1-e^2} \sin i_2\sin i_1} \left(\cos i_1 \frac{\partial R}{\partial \omega_1} - \frac{\partial R}{\partial \Omega_1}\right), 
\end{multline}
\end{subequations} where $\beta = m_1 \, m_2 / \left(m_1 + m_2\right)$ is the reduced mass, $R$ is the disturbing function, $\omega_1$, $i_1$, and $\Omega_1$ are, respectively, the argument of pericenter, the inclination, and the longitude of the ascending node of the mutual orbit with respect to the star's equatorial plane, and $\omega_2$, $i_2$ and $\Omega_2$ are the corresponding elements defined with respect to the planet's equator.

As we are considering a system of two bodies, there is only one orbital plane and the inclinations of the equatorial plane of each body with respect to the orbital plane are, indeed, the obliquities. But, as pointed out by \citet{boue_efroimsky2019}, these quantities are called inclinations, in terms of the continuity of Kaula's notation. On the other hand, the names of obliquity and inclination of the orbital plane depends on the problem  that is being considered. In our case, what would be called inclination of the planet's orbit as seen from the star ($i_1$) is simply the orbital inclination and, conversely, the inclination of the orbit of the star as seen from the planet ($i_2$) can simply be referred to as the obliquity of the planet.

It worthwhile to indicate that the orbital elements parameterizing the positions of each body with respect to the other are not osculating. As it is widely known, in the unperturbed two-body problem, the position, $\vec{r,}$ and the velocity, $\dot{\vec{r}}$, vectors are functions of the orbital elements, that is: $\vec{r} = \vec{f} (a,e,M,i,\omega,\Omega)$ and $\dot{\vec{r}} = \vec{g} (a,e,M,i,\omega,\Omega)$ being, by definition,
\begin{equation}\label{ec:rel_gdfdt} 
\vec{g} = \left(\frac{\partial \vec{f}}{\partial t}\right)_{c_k = \mathrm{const.}}, \end{equation} where $c_k = a,e,i,M,\omega,\Omega$ for $k=1,\ldots,6$. Under perturbation, the position vector remains equal to $\vec{f}$ but the velocity vector reads: 
\begin{equation}\label{ec:pert_vel} \dot{\vec{r}} = \frac{\mathrm{d} \vec{r}}{\mathrm{d}t} = \left(\frac{\partial \vec{f}}{\partial t}\right)_{c_k = \mathrm{const.}} + \sum_{k=1}^6 \frac{\partial \vec{f}}{\partial c_k} \frac{\mathrm{d} c_k}{\mathrm{d}t} = \vec{g} + \vec{\Phi},
\end{equation} where $\vec{\Phi}$ is known as the gauge function  and it is completely arbitrary \citep{efroimsky2005a,efroimsky2005b,kopeikinetal2011}. In other words, we can fix a particular functional form of the convective term of Eq.~\eqref{ec:pert_vel}: 
\begin{equation}\label{ec:gauge} 
\sum_{k=1}^6 \frac{\partial \vec{f}}{\partial C_k} \frac{\mathrm{d} C_k}{\mathrm{d}t} = \vec{\Phi}. 
\end{equation} The usual practice is to set $\vec{\Phi} = \vec{0}$, which is also called the Lagrange's constraint. In such a case, we have $\dot{\vec{r}} = \vec{g}$, meaning that the perturbed velocity is tangential to the conic parameterising the instantaneous position of each celestial body with respect to the other. Of course, the corresponding orbital elements are the osculating ones.

However, within the formalism developed by \citet{boue_efroimsky2019}, the orbital elements are defined in the equatorial planes of each body, that is, they choose to describe the relative motions of the bodies in the coprecessing reference frames whose origins coincide with the center of mass of each body. In such situations, it is more convenient to choose $\vec{\Phi} = \vec{\mu} \times \vec{r}$, where $\vec{\mu}$ is the precession velocity vector \citep{efroimsky_goldreich_2004,efroimsky2005a,efroimsky2005b}. Thus, the velocity vector of each body is given by $\dot{\vec{r}} = \vec{g} + \vec{\mu} \times \vec{f}$. Later on, we can perform the transformation from the set $(\vec{r},\dot{\vec{r}})$ to the set $(a,e,M,i,\omega,\Omega)$ to get the osculating elements, which have a more direct interpretation. The orbital elements appearing in Eqs.~\eqref{ec:dadt_gral} -- \eqref{ec:dOm2dt_gral} are called contact elements.

\subsection{\label{ss:evrot}Rotational evolution}

In order to close the system of equations of motion, at least, in principle, it is necessary to add the following equations describing the rotational evolution, which correspond to Eqs. (122) and (123) of the work by \citet{boue_efroimsky2019}: 
\begin{equation}\label{ec:depsdt_gral} 
\frac{\mathrm{d} \varepsilon_k}{\mathrm{d}t} = - \frac{\beta}{C_k \, \dot{\theta}_k} \left(\cos \Omega_k \cot i_k \frac{\partial R}{\partial \Omega_k} + \sin \Omega_k \frac{\partial R}{\partial i_k} - \frac{\cos \Omega_k}{\sin i_k} \frac{\partial R}{\partial \omega_k}\right) 
,\end{equation} 
\begin{equation}\label{ec:d2thdt2_gral} 
\frac{\mathrm{d}^2 \theta_k}{\mathrm{d}t^2} = - \frac{\beta}{C_k} \frac{\partial R}{\partial \Omega_k}. 
\end{equation} where $k=1,2$, $\varepsilon_k$ is the angle between each body's spin axis and the $z$ direction of the inertial reference frame, and, again, $\theta_k$ are the sidereal angles. It should be remembered that the last equations and Eqs.~\eqref{ec:dadt_gral} -- \eqref{ec:dOm2dt_gral} are obtained under the so-called gyroscopic approximation. This means that $\left \vert \dot{\theta} \right \vert \gg \left \vert \dot{\psi} \right \vert \, \mathrm{,} \, \left \vert \dot{\varepsilon} \right \vert$ is assumed. The latter assertion is equivalent to stating that the rotational angular momentum vector coincides with the principal axis of inertia the body rotates around.

Another important and necessary assumption is that we also neglect the time variation of the moments of inertia of the considered bodies. This is justified because even though the tidal theory considers the deformation of celestial bodies -- as we see later -- these deformations (assumed to be linear) are always very small, that is, $\epsilon < 10^{-6}$, where $\epsilon$ is a typical value of the strain \citep{frouard_efroimsky_2018}.

\subsection{\label{ss:distpots}Disturbing potentials}

The disturbing function, $R$, present in the equations of motion is given by \citep{boue_efroimsky2019}: 
\begin{equation}\label{ec:disturbingfunc} 
R = - \frac{1}{\beta} \left[ m_1 \left[U_2 \left(\vec{r}_{2 \rightarrow 1}, \vec{r}_{2 \rightarrow 1}^{*} \right) + V_2 \left(\vec{r}_{2 \rightarrow 1}\right) \right] + m_2 U_1 \left(-\vec{r}_{1 \rightarrow 2},-\vec{r}_{1 \rightarrow 2}^{\star}\right) \right] 
,\end{equation} where $U_2 \left(\vec{r}_{2 \rightarrow 1}, \vec{r}_{2 \rightarrow 1}^{*} \right)$ is the tidal disturbing potential generated by the (deformed) planet and $U_1 \left(-\vec{r}_{1 \rightarrow 2},-\vec{r}_{1 \rightarrow 2}^{\star}\right)$ is the corresponding tidal potential generated by the star. As we are interested in the dynamical evolution near the spin-orbit resonances, we add the permanent-triaxiality-induced potential of the planet $V_2 \left(\vec{r}_{2 \rightarrow 1}\right)$, as well as in the influence of the dynamical form factor on the secular evolution of the system. The first one is given by \citep{efroimsky2012}: 
\begin{equation}\label{ec:pottidalplanet} U_2 \left(\vec{r}_1, \vec{r}_1^{*} \right) = - \, \frac{G m_1}{a^{*}} \sum_{l=2}^{\infty} \sum_{m=0}^l \sum_{p=0}^l \sum_{q=-\infty}^{\infty} \sum_{h=0}^{l} \sum_{s=-\infty}^{\infty} U^{(2)}_{lmpqhs},
\end{equation} where 
\begin{multline}\label{ec:pottidalplanetlmpqhs} U^{(2)}_{lmpqhs} = \left(\frac{R_2}{a}\right)^{l+1} \left(\frac{R_2}{a^{*}}\right)^l \frac{(l-m)!}{(l+m)!} (2-\delta_{0m}) \, F_{lmp} (i_2^{*}) \, G_{lpq}
(e^{*}) \, F_{lmh} (i_2) \, G_{lhs} (e) \left[\cos \left[ \left(v_{lmpq}^{(2),*} - m \, \theta_{2}^{*}\right) - \left(v_{lmhs}^{(2)} - m \, \theta_{2} \right) \right] K_{\mathrm{R}}^{(2)} \left(l,\omega^{(2)}_{lmpq}\right) \right. \\ \left. + \sin \left[ \left(v_{lmpq}^{(2),*} - m \, \theta_{2}^{*}\right) - \left(v_{lmhs}^{(2)} - m \, \theta_{2} \right) \right] K_\mathrm{I}^{(2)} \left(l,\omega^{(2)}_{lmpq}\right) \right].
\end{multline} The tidal potential generated by the star is given by an expression analogous to Ec.~\eqref{ec:pottidalplanet} and \eqref{ec:pottidalplanetlmpqhs}: 
\begin{equation}\label{ec:pottidalstar} 
U_1 \left(-\vec{r}_{1 \rightarrow 2},-\vec{r}_{1 \rightarrow 2}^{\star}\right) = - \, \frac{G m_2}{a^{\star}} \sum_{l=2}^{\infty} \sum_{m=0}^l \sum_{p=0}^l \sum_{q=-\infty}^{\infty} \sum_{h=0}^{l} \sum_{s=-\infty}^{\infty} U^{(1)}_{lmpqh},
\end{equation} with
\begin{multline}\label{ec:pottidalstarlmpqhs} 
U^{(1)}_{lmpqhs} = \left(\frac{R_1}{a}\right)^{l+1} \left(\frac{R_1}{a^{\star}}\right)^l \frac{(l-m)!}{(l+m)!} (2-\delta_{0m}) \, F_{lmp} (i_1^{\star}) \, G_{lpq} (e^{\star}) \, F_{lmh} (i_1) \, G_{lhs} (e) \left[\cos \left[ \left(v_{lmpq}^{(1),\star} - m \, \theta_1^{\star}\right) - \left(v^{(1)}_{lmhs} - m \, \theta_1 \right) \right] K_{\mathrm{R}}^{(1)} \left(l,\omega^{(1)}_{lmpq}\right) \right. \\ \left. + \sin \left[ \left(v_{lmpq}^{(1),\star} - m \, \theta_1^{\star}\right) - \left(v^{(1)}_{lmhs} - m \, \theta_1 \right) \right] K_\mathrm{I}^{(1)} \left(l,\omega^{(1)}_{lmpq}\right) \right], 
\end{multline} where $\delta_{0m}$ is the Kronecker's delta, $F_{lmp} (i_k)$ and $G_{lpq} (e)$ are the inclination functions \citep{goodwagner2008} and the eccentricity functions \citep{giacaglia1976b}, respectively. The $v_{lmpq}^{(k)}$ represent linear combinations of $\omega_k$, $\Omega_k$ and $M$, expressed via: 
\begin{equation}\label{ec:vlmpq} 
v_{lmhs}^{(k)} = (l-2h) \, \omega_k + (l-2h+s) \, M + m \, \Omega_k 
\end{equation} and analog expressions for $v_{lmpq}^{(k),*}$ and $v_{lmpq}^{(k),\star}$.

Within the framework of the Darwin-Kaula formalism of bodily tides, we may consider the presence of three bodies in order to take the tidal dissipation into account: the primary -- whose shape and deformation is considered -- and two ``secondaries,'' the tide-rising and the tidally disturbed ones, considered to be point masses. Let $\vec{r}_{2 \rightarrow 1}$ be the position vector of the body 1 (the star) with respect to the body 2 (the planet), and $\vec{r}_{1 \rightarrow 2}$ the corresponding position vector of the planet with respect to the star, both expressed in the corotating frame of one of the bodies. In the case of the planet being the primary, the tide-rising secondary is located at $\vec{r}_{2 \rightarrow 1}^{*}$, while the tide disturbed one is at $\vec{r}_{2 \rightarrow 1}$. Conversely, when the star is considered to be the primary, then the tide-rising secondary is at $-\vec{r}_{1 \rightarrow 2}^{\star}$ while the tide-disturbed one is at $-\vec{r}_{1 \rightarrow 2}$. It worthwhile indicating that as we are considering a two-body system, in both cases the two secondaries coincide with one other, that is, in the first case, both secondaries coincide with the star and, in the second case, both coincide with the planet. By virtue of the \citet{kaula1961} transformation, the set of coordinates for each secondary is expressed through the corresponding set of orbital elements that parameterizes their orbits \citep{efrowill2009}.

In the above expressions -- Eqs.~\eqref{ec:pottidalplanetlmpqhs} and \eqref{ec:pottidalstarlmpqhs} -- $K_{\mathrm{R}}^{k} (l,\chi^{k}_{lmpq})$ and $K_{\mathrm{I}}^{k} (l,\chi^{k}_{lmpq})$ are known as quality functions \citep{makarov_2012} and given by: 
\begin{align} 
K_{\mathrm{R}} (l,\omega) &= \frac{3}{2} \frac{1}{l-1} \frac{\left(\Re \left[ \bar{J} (\chi) \right] + B_l \right) \Re \left[ \bar{J} (\chi) \right]+ \left(\Im \left[ \bar{J} (\chi) \right]\right)^2}{\left(\Re \left[ \bar{J} (\chi) \right] + B_l \right)^2 + \left(\Im \left[ \bar{J} (\chi) \right]\right)^2} \label{ec:reklJ}, \\ 
K_\mathrm{I} (l,\omega) &= - \frac{3}{2} \frac{1}{l-1} \frac{B_l \, \Im \left[ \bar{J} (\chi) \right]}{\left(\Re \left[ \bar{J} (\chi) \right] + B_l \right)^2 + \left(\Im \left[ \bar{J} (\chi) \right]\right)^2} \, \mathrm{sign} \left(\omega \right), \label{ec:imklJ} 
\end{align} where $\Re (\bar{z})$ and $\Im (\bar{z})$ denotes the real and imaginary parts of the complex number $\bar{z}$ and $\mathrm{sign} (x)$ is the signum function. We omitted the indices for the sake of simplicity. Then, $\chi^{(k)}_{lmpq}$ are the tidal forcing frequencies, that is, they are the physical frequencies at which the deformations are excited. They are defined as the absolute value of the tidal modes, $\omega^{(k),*}_{lmpq}$, \citep{efroimsky2012,efromak2013}, mathematically: 
\begin{equation}\label{ec:chis} 
\chi^{(k)}_{lmpq} = \left\vert \, \omega^{(k),*}_{lmpq} \, \right\vert = \left\vert \, (l-2p) \, \dot{\omega}^{*}_{k} + (l-2p+q) \dot{M}^{*} + m \left(\dot{\Omega}^{*}_{k} - \dot{\theta}^{\, *}_{k} \right) \, \right\vert 
,\end{equation} where $\dot{\theta}^{\, *}_{k}$ is the spin rate of the primary that is being considered. If both the apsidal and the nodal precessions are neglected, the tidal modes can be expressed as: 
\begin{equation}\label{ec:tidalmodesapprox} 
\omega^{(k),*}_{lmpq} \approx (l-2p+q) \, n^{*} - m \, \dot{\theta}^{*}_{k} 
,\end{equation} where $n^{*}_{k}$ is the mean orbital frequency of the tide-rising secondary. Consequently, the $\omega^{(k),*}_{lmpq}$'s can be either positive or negative depending on whether the primary's spin rate is slower or faster than the mean rate at which the tide-rising secondary describes its orbit, respectively, while the tidal frequencies are always positive. $B_l$ is given by \citep{efroimsky2015}: 
\begin{equation}\label{ec:defAl}
 B_l = \frac{2 l^2 + 4 l + 3}{l \, g_{k} \, \rho_{k} \, R_{k}} 
,\end{equation} where $\rho_{k}$ is the mean density, $R_{k}$ is the mean radius, and $g_{k} = Gm_{k}/R_{k}^2$ is the surface gravity of the body, $k$. Finally, $\bar{J} (\chi)$ is the complex compliance and is defined by: 
\begin{equation}\label{ec:defJcomplex} 
\bar{J} (\chi) = \int_0^\infty \dot{J} (t-t') \exp{\left[- \mathrm{i} \chi (t-t')\right]} \, 
\mathrm{d}t' 
,\end{equation} where the over-dot means differentiation with respect to $t'$. The functional form of the kernel $J (t-t')$ depends on the rheology of the material that composes the primary or, in other words, on the rheological model that describe the creep response of the body as a whole. Realistic materials are thought to have a mixture of elastic, viscous (or viscoeslastic), and hereditary behaviors \citep{renaudhenning2018}, that is, the deformation is directly proportional to the stress and simultaneously depends on the time derivative of the latter and on its past values, respectively. Thus, $J (t-t')$ is given generally by: 
\begin{equation}\label{ec:kernJgral} 
J (t-t') = J(0) \, \Theta (t-t') + \mbox{viscous and hereditary terms} 
,\end{equation} with $J(0)$ as the value of the unrelaxed compliance, corresponding to the elastic response and inverse of the elastic rigidity, $\mu (0)$, and $\Theta (t-t')$ is the Heaviside step function \citep{efroimsky2012}.

It worthwhile indicating that the complex compliance enters in the constitutive equation that relates the stress and strain tensors together. In a linear regime, and under the assumption that the material the primary is made of is an homogeneous incompressible and isotropic medium, the relationship between the components of the tensors, written in the frequency domain, is: 
\begin{equation}\label{ec:ecconstfrec} 
2 \, \bar{u}_{\gamma \nu} (\chi) = \bar{J} (\chi) \, \bar{\sigma}_{\gamma \nu} (\chi), 
\end{equation} where $\bar{u}_{\gamma \nu}$ and $\bar{\sigma}_{\gamma \nu}$ are the complex counterparts of the components of the strain and stress tensors, respectively. We refer the reader to the work by \citet{efroimsky2012} for a comprehensive explanation of this subject and detailed derivations of the equations given above.

The permanent-triaxiality-induced potential is a special case of the more general expression of the potential generated by the so-called coefficients of the gravitational field (CGF), $C_{lm}$, and $S_{lm}$, which is given by \citep{frouard_efroimsky_2017}: 
\begin{equation}\label{ec:pottriaxplanet}
V_{2} (\vec{r}_{1}) = - \frac{G \,m_{1}}{a} \sum_{l=2}^{\infty} \sum_{m=0}^l \sum_{p=0}^l \sum_{q=-\infty}^{\infty} V^{(2)}_{lmpq},
\end{equation} where 
\begin{equation}\label{ec:pottriaxplanetlmpq} 
V^{(2)}_{lmpq} = \left(\frac{R_{2}}{a}\right)^l F_{lmp} (i_{2}) \, G_{lpq} (e) \left[C_{lm}^{(2)} \cos \left(v^{(2)}_{lmpq} - m \, \theta_{2} + \phi_{lm} \right) + S_{lm}^{(2)} \sin \left(v^{(2)}_{lmpq} - m \, \theta_{2} + \phi_{lm} \right) \right],
\end{equation} and where $\phi_{lm} = \left[(-1)^{l-m} - 1 \right] \frac{\pi}{4}$. The leading terms of the expansion given in Eqs.~\eqref{ec:pottriaxplanet} and \eqref{ec:pottriaxplanetlmpq} are those related with the triaxiality of the planet, namely the terms associated with the $C_{lm}$ coefficients for which $l=2$ and $m=0,2$, specifically, $C_{20}$ and $C_{22}$. The latter are related with the aforementioned principal moments of inertia through:
\begin{align}\label{ec:rel_Clm_ABC}
C_{20} &= \frac{1}{2 \, m_2 \, R_2^2} \left[A + B - 2 C\right] = - J_2, \\
C_{22} &= \frac{B-A}{4 \, m_2 \, R_2^2},
\end{align} where $J_2$ is known as the dynamic form factor, which usually is three orders of magnitude larger than the rest of the CGF, at least for the case of the Earth \citep{torge2012}. In this work, we consider the influence of the $J_2$-related secular terms on the time evolution of the orbital elements. The $C_{22}$ coefficient is kept because is the dominating factor of the triaxiality-induced torque \citep{makarov_2012,makarovetal_2012} which, together with the tidal torque, drive the time evolution of the proper rotation angle of the planet. \citet{frouard_efroimsky_2017} point out that the latter torque is also responsible for the longitudinal libration, which causes extra dissipation. However this effect is out of the scope of this work and we will consider it in the future.

We only consider the triaxiality of the planet for two main reasons: 1) The tidal interaction is expected to affect more strongly its dynamical evolution than that of the star, in particular the rotational evolution where there are probabilities related to the planet getting trapped in a spin-orbit resonance; 2) We can reasonably neglect the triaxiality (i.e. the deviation from a rotational symmetry) of a fluid body, such as a star. In that case, we assume that $A_1 = B_1$ and, thus, the $C_{22}$-related terms of the gravity potential of the star are nullified, as can be confirmed by looking at Eq.~\eqref{ec:rel_Clm_ABC}. As only the triaxiality of the planet is to be considered, we can safely suppress the corresponding subscripts of the principal moments of inertia, $A$ and $B$.

Regarding the $J_2$-related terms of the corresponding CGF disturbing potential, even though the oblateness of the star could have a relevant impact on the dynamics of close-in systems, such as the one studied in this work \citep{correiaetal2011}, we have left the consideration of this effect for a future work.

\section{\label{sec:eom}Development of the analytical model}

The principal considerations that guides the development of our analytical model consist of the following:
\begin{itemize}
      \item As a first step, we consider the secular evolution of the orbital elements. In this sense,  our model takes into account the secular terms of the  derivatives of the additional tidal potential and the $J_2$-related secular terms of the CGF disturbing potential.
      \item Concerning the time evolution of the star's spin rate, we consider that its angular acceleration is caused by the secular terms of the tidal torque acting on it.
      \item We then consider that the time evolution of the planet's spin rate is dominated by both the secular terms of the corresponding tidal and permanent-triaxiality-induced torques.
      \item Finally, we  limit the expansions of the disturbing potentials to the quadrupole approximation, i.e. $l=2$. A justification of this assumption is given in Sect.~\ref{sec:prev}.
\end{itemize}

The equations describing the orbital and rotational dynamics given in Eqs.~\eqref{ec:dadt_gral}--\eqref{ec:dOm2dt_gral}, \eqref{ec:depsdt_gral}, and \eqref{ec:d2thdt2_gral} are somewhat general and, therefore, we find it necessary at this point to rewrite them in a form that is more convenient to our purposes. As we consider the secular evolution of the orbital elements, we  make use of the averaged equations of motion:
\begin{subequations}\label{ec:ec_mov_orb_gral_avg}
\begin{equation}\label{ec:dadt_gral_avg} 
\left\langle \frac{\mathrm{d}a}{\mathrm{d}t} \right\rangle = \frac{2}{n \, a} \left\langle\frac{\partial R}{\partial M}\right\rangle, 
\end{equation} 
\begin{equation}\label{ec:dedt_gral_avg} 
\left\langle \frac{\mathrm{d}e}{\mathrm{d}t} \right\rangle = \frac{\sqrt{1-e^2}}{n \, a^2 \, e} \left[ \sqrt{1-e^2} \left\langle\frac{\partial R}{\partial M}\right\rangle - \left\langle\frac{\partial R}{\partial \omega_1}\right\rangle - \left\langle\frac{\partial R}{\partial \omega_2}\right\rangle\right],
\end{equation} 
\begin{equation}\label{ec:dMdt_gral_avg} 
\left\langle \frac{\mathrm{d} M}{\mathrm{d}t} \right\rangle = n - \frac{2}{n \, a} \left\langle\frac{\partial R}{\partial a}\right\rangle - \frac{1-e^2}{n \, a^2 \, e} \left\langle\frac{\partial R}{\partial e}\right\rangle
,\end{equation} 
\begin{equation}\label{ec:di1dt_gral_avg}
\left\langle \frac{\mathrm{d}i_1}{\mathrm{d}t} \right\rangle = \frac{1}{\sin i_1} \left[\frac{\beta}{C_1 \, \dot{\theta}_1} \left(\left\langle\frac{\partial R}{\partial \omega_1}\right\rangle - \cos i_1 \left\langle\frac{\partial R}{\partial \Omega_1}\right\rangle \right) - \frac{1}{n \, a^2 \sqrt{1-e^2}} \left(\left\langle\frac{\partial R}{\partial \Omega_1}\right\rangle - \cos i_1 \left\langle\frac{\partial R}{\partial \omega_1}\right\rangle \right)\right], 
\end{equation} 
\begin{equation}\label{ec:di2dt_gral_avg} 
\left\langle \frac{\mathrm{d}i_2}{\mathrm{d}t} \right\rangle = \frac{1}{\sin i_2} \left[\frac{\beta}{C_2 \, \dot{\theta}_2} \left(\left\langle\frac{\partial R}{\partial \omega_2}\right\rangle - \cos i_2 \left\langle\frac{\partial R}{\partial \Omega_2}\right\rangle \right) - \frac{1}{n \, a^2 \sqrt{1-e^2}} \left(\left\langle\frac{\partial R}{\partial \Omega_2}\right\rangle - \cos i_2 \left\langle\frac{\partial R}{\partial \omega_2}\right\rangle \right)\right], 
\end{equation} 
\begin{equation}\label{ec:dw1dt_gral_avg} 
\left\langle \frac{\mathrm{d} \omega_1}{\mathrm{d}t} \right\rangle =  \frac{\sqrt{1-e^2}}{n \, a^2 \, e} \left\langle\frac{\partial R}{\partial e}\right\rangle - \left[\frac{\beta}{C_1 \, \dot{\theta}_1} + \frac{\cos i_1}{n \, a^2 \, \sqrt{1-e^2}}\right] \frac{1}{\sin i_1} \left\langle\frac{\partial R}{\partial i_1}\right\rangle,
\end{equation} 
\begin{equation}\label{ec:dw2dt_gral_avg} 
\left\langle \frac{\mathrm{d} \omega_2}{\mathrm{d}t} \right\rangle = \frac{\sqrt{1-e^2}}{n \, a^2 \, e} \left\langle\frac{\partial R}{\partial e}\right\rangle - \left[\frac{\beta}{C_2 \, \dot{\theta}_2} + \frac{\cos i_2}{n \, a^2 \, \sqrt{1-e^2}}\right] \frac{1}{\sin i_2} \left\langle\frac{\partial R}{\partial i_2}\right\rangle,
\end{equation} 
\begin{equation}\label{ec:dOm1dt_gral_avg} 
\left\langle \frac{\mathrm{d} \Omega_1}{\mathrm{d}t} \right\rangle = \left[\frac{\beta \cos i_1}{C_1 \, \dot{\theta}_1} + \frac{1}{n \, a^2 \, \sqrt{1-e^2}}\right] \frac{1}{\sin i_1} \left\langle\frac{\partial R}{\partial i_1}\right\rangle,
\end{equation} 
\begin{equation}\label{ec:dOm2dt_gral_avg} 
\left\langle \frac{\mathrm{d} \Omega_2}{\mathrm{d}t} \right\rangle = \left[\frac{\beta \cos i_2}{C_2 \, \dot{\theta}_2} + \frac{1}{n \, a^2 \, \sqrt{1-e^2}}\right] \frac{1}{\sin i_2} \left\langle\frac{\partial R}{\partial i_2}\right\rangle.
\end{equation}
\end{subequations} In this context, the angular brackets on the left-hand side of Eqs.~\eqref{ec:ec_mov_orb_gral_avg} represent the averaging over the mean anomaly, the arguments of pericenters, and the longitudes of the ascending nodes, while the angular brackets on the right-hand side of the aforementioned equations indicate that we are considering the secular terms of the disturbing function's derivatives.

Since the equations of motion depends on the gradient of the disturbing function, $R$, among other factors, the calculation of the derivatives of $R$ with respect to the orbital elements deserves some discussion. As we point out in the previous subsection, $\vec{r}_{2 \rightarrow 1}$ correspond to the position of the tide-disturbed secondary while the tide-rising one is located at $\vec{r}_{2 \rightarrow 1}^{*}$ when the planet is considered to be the primary body. Accordingly, $-\vec{r}_{1 \rightarrow 2}$ and $-\vec{r}_{1 \rightarrow 2}^{\star}$ are the radius-vectors of the tide-disturbed and tide-rising secondaries, respectively, when the roles of the planet and the star are reversed with respect to the former case. It should be stressed that we always consider the motion of the tide-disturbed secondary and, thus, the general rule that is to be followed in order to find the aforementioned derivatives is the following: first, we have to differentiate the disturbing function with respect to the orbital elements of the tide-disturbed secondary and then identify both secondaries with each other, that is, to set $\vec{r}_{2 \rightarrow 1} = \vec{r}_{2 \rightarrow 1}^{*}$ and $\vec{r}_{1 \rightarrow 2} = \vec{r}_{1 \rightarrow 2}^{\star}$ in order to properly evaluate  the force per unit mass each body exerts on the other. It may be noted that as the arguments of the sine and cosine functions remain the same, once the identification is performed, we obtain: 
\begin{equation}\label{ec:argsincos}
\left(v_{lmpq}^{k,*} - m \, \theta_{k}^{*}\right) - \left(v_{lmhs}^{k} - m \, \theta_{k} \right) = 2(h-p) \, \omega_{k} + (2 h - 2 p + q - s) \, M.
\end{equation} By virtue of Eq.~\eqref{ec:argsincos}, it can be seen that the derivatives of the tidal potentials may be split into two types of terms, namely, the secular and the oscillating ones. The former comprises the terms independent of $M$ and $\omega_k$, that is, those for which $h=p$ and $q=s$, with the remainder being part of the latter type of terms. As we can easily see, for the case of the secular terms, the arguments of the sine and cosine given by Eq.~\eqref{ec:argsincos} are identically null and, as a consequence, the derivatives of the disturbing function become independent of one of the quality functions.

It can be shown by Eq.~\eqref{ec:pottriaxplanetlmpq} that the $J_2$-related secular term of the CGF disturbing potential corresponds to the one with $(lmpq) = (2010)$ \citep{kaula2000,efroimsky2005b}, that is:
\begin{equation}\label{ec:pottriaxplanet2010}
\begin{split}
\left\langle V_{lmpq}^{(2)}\right\rangle &= V_{2010}^{(2)} \\
                                         &= - J_2 \left(\frac{R_2}{a}\right)^2 F_{201} (i_2) \, G_{210} (e).
\end{split}
\end{equation}

In general terms, analogously to Eqs.~\eqref{ec:pottidalplanet}, \eqref{ec:pottidalstar}, and \eqref{ec:pottriaxplanet}, the disturbing potentials can be expressed as the product of a dimensional factor multiplied by a sum of adimensional terms. Taking into account the dependence of each perturbing potential upon the corresponding orbital elements, which can be expressed through:
\begin{align*}
U_1 &= U_1 (a,e,M,i_1,\omega_1,\Omega_1), & U_2 &= U_2 (a,e,M,i_2,\omega_2,\Omega_2), \\
    &                                     & V_2 &= V_2 (a,e,M,i_2,\omega_2,\Omega_2),
\end{align*} the derivatives of the disturbing function can be calculated as follows. Since the three disturbing potentials considered in this work depend on $a$, $e,$ and $M$, we have, on one hand:
\begin{equation}\label{ec:dRdaeM_gral}
\frac{\partial R}{\partial (a,e,M)} = - \frac{1}{\beta} \left[m_1 \left(\frac{\partial U_2}{\partial (a,e,M)} + \frac{\partial V_2}{\partial (a,e,M)}\right) + m_2 \frac{\partial U_1}{\partial (a,e,M)}\right],
\end{equation} while, on the other hand, the corresponding expression of the derivatives with respect to the inclinations, the arguments of the pericenter and the longitudes of the ascending nodes (as seen from each coordinate system) are:
\begin{subequations}\label{ec:dRdiwO_gral}
\begin{align}
\frac{\partial R}{\partial (i_1,\omega_1,\Omega_1)} &= - \frac{m_2}{\beta} \frac{\partial U_1}{\partial (i_1,\omega_1,\Omega_1)} \label{ec:dRdiwO1_gral} \\
\frac{\partial R}{\partial (i_2,\omega_2,\Omega_2)} &= - \frac{m_1}{\beta} \left[\frac{\partial U_2}{\partial (i_2,\omega_2,\Omega_2)} + \frac{\partial V_2}{\partial (i_2,\omega_2,\Omega_2)} \right].\label{ec:dRdiwO2_gral}
\end{align}
\end{subequations} 

Specifically, the secular terms of the derivatives of the disturbing potentials with respect to the major semiaxis can be expressed as:
\begin{subequations}
\begin{align}
\left\langle \frac{\partial U_1}{\partial a} \right\rangle &= \frac{G \, m_2}{a^2} \sum_{l=2}^\infty \sum_{m=0}^l \sum_{p=0}^l \sum_{q=-\infty}^{\infty} (l+1) \left\langle U_{lmpq}^{(1)} \right\rangle \label{ec:dU1da} \\
\left\langle \frac{\partial U_2}{\partial a} \right\rangle &= \frac{G \, m_1}{a^2} \sum_{l=2}^\infty \sum_{m=0}^l \sum_{p=0}^l \sum_{q=-\infty}^{\infty} (l+1) \left\langle U_{lmpq}^{(2)} \right\rangle \label{ec:dU2da} \\
\left\langle \frac{\partial V_2}{\partial a} \right\rangle &= \frac{G \, m_2}{a^2} 3 \, V_{2010}^{(2)} \label{ec:dV2da}
\end{align}
\end{subequations} where $\left\langle U_{lmpq}^{(1)} \right\rangle$ and $\left\langle U_{lmpq}^{(2)} \right\rangle$ are the secular terms of the tidal potentials of both the star and the planet, respectively. These terms are obtained through Eqs~\eqref{ec:pottidalplanetlmpqhs} and \eqref{ec:pottidalstarlmpqhs} by setting $h=p$ ans $q=s$, as previously explained. Even though we have pointed out before that we aim to limit the expansions to $l=2$, we still consider all multipoles in the expressions of the derivatives of the disturbing potentials as a matter of convenience in order to ease the exposition. Later, when we give the final expression of the equations of motion, we provide explicit expressions that correspond to the quadrupolar approximation.

 The expressions of the derivatives of the disturbing potentials with respect to the eccentricity and the mean anomaly are:
\begin{subequations}
\begin{align}
\left\langle\frac{\partial U_1}{\partial (e,M)}\right\rangle &= - \frac{G \, m_2}{a} \sum_{l=2}^\infty \sum_{m=0}^l \sum_{p=0}^l \sum_{q=-\infty}^{\infty} \left\langle \frac{\partial U_{lmpq}^{(1)}}{\partial (e,M)}\right\rangle \label{ec:dU1deM} \\
\left\langle\frac{\partial U_2}{\partial (e,M)}\right\rangle &= - \frac{G \, m_1}{a} \sum_{l=2}^\infty \sum_{m=0}^l \sum_{p=0}^l \sum_{q=-\infty}^{\infty} \left\langle \frac{\partial U_{lmpq}^{(2)}}{\partial (e,M)}\right\rangle \label{ec:dU2deM}, \\
\left\langle\frac{\partial V_2}{\partial e}\right\rangle &= - \frac{G \, m_2}{a} \frac{\partial V_{2010}^{(2)}}{\partial e}. \label{ec:dV2deM}
\end{align}
\end{subequations} The rest of the derivatives, those with respect to $i_k$, $\omega_k$, and $\Omega_k$, can be expressed as follows:
\begin{subequations}
\begin{align}
\left\langle\frac{\partial U_1}{\partial (i_1,\omega_1,\Omega_1)}\right\rangle &= - \frac{G \, m_2}{a} \sum_{l=2}^\infty \sum_{m=0}^l \sum_{p=0}^l \sum_{q=-\infty}^{\infty} \left\langle\frac{\partial U_{lmpq}^{(1)}}{\partial (i_1,\omega_1,\Omega_1)}\right\rangle \label{ec:dU1diwO}, \\
\left\langle\frac{\partial U_2}{\partial (i_2,\omega_2,\Omega_2)}\right\rangle &= - \frac{G \, m_1}{a} \sum_{l=2}^\infty \sum_{m=0}^l \sum_{p=0}^l \sum_{q=-\infty}^{\infty} \left\langle\frac{\partial U_{lmpq}^{(2)}}{\partial (i_2,\omega_2,\Omega_2)}\right\rangle \label{ec:dU2diwO}, \\
\left\langle\frac{\partial V_2}{\partial i_2}\right\rangle &= - \frac{G \, m_2}{a} \frac{\partial V_{2010}^{(2)}}{\partial i_2}. \label{ec:dV2diwO}
\end{align}
\end{subequations} The full expressions of the derivatives of the adimensional terms are given in Appendix~\ref{sec:sum_terms_adim}.

The next step is to write down the expressions of the derivatives of the disturbing function with respect to the orbital elements. By inserting Eqs.~\eqref{ec:dU1da}, \eqref{ec:dU2da}, \eqref{ec:dV2da}, \eqref{ec:dU1deM}, \eqref{ec:dU2deM}, and \eqref{ec:dV2deM} in Eq.~\eqref{ec:dRdaeM_gral} we have:
\begin{subequations}\label{ec:dRdaeM_sec}
\begin{align}
\left\langle \frac{\partial R}{\partial a} \right\rangle &= - n^2 a \left(\sum_{l=2}^\infty \sum_{m=0}^l \sum_{p=0}^l \sum_{q=-\infty}^{\infty} (l+1) \left[ \frac{m_2}{m_1} \left\langle U_{lmpq}^{(1)} \right\rangle + \frac{m_1}{m_2} \left\langle U_{lmpq}^{(2)} \right\rangle\right] + 3 \, V_{2010}^{(2)} \right), \label{ec:dRda} \\
\left\langle \frac{\partial R}{\partial e} \right\rangle &= n^2 a^2 \left( \sum_{l=2}^\infty \sum_{m=0}^l \sum_{p=0}^l \sum_{q=-\infty}^{\infty} \left[ \frac{m_2}{m_1} \left\langle \frac{\partial U_{lmpq}^{(1)}}{\partial (e,M)}\right\rangle + \frac{m_1}{m_2} \left\langle \frac{\partial U_{lmpq}^{(2)}}{\partial (e,M)}\right\rangle\right] + \frac{\partial V_{2010}^{(2)}}{\partial e}\right), \label{ec:dRde} \\
\left\langle \frac{\partial R}{\partial M} \right\rangle &= n^2 a^2 \sum_{l=2}^\infty \sum_{m=0}^l \sum_{p=0}^l \sum_{q=-\infty}^{\infty} \left[ \frac{m_2}{m_1} \left\langle \frac{\partial U_{lmpq}^{(1)}}{\partial (e,M)}\right\rangle + \frac{m_1}{m_2} \left\langle \frac{\partial U_{lmpq}^{(2)}}{\partial (e,M)}\right\rangle \right]. \label{ec:dRdM}
\end{align}
\end{subequations} The partial derivatives of the disturbing function with respect to the inclinations, arguments of pericenters and longitudes of the corresponding ascending nodes can be expressed in a similar way. Inserting Eqs.~\eqref{ec:dU1diwO}, \eqref{ec:dU2diwO}, and \eqref{ec:dV2diwO} again in Eqs.~\eqref{ec:dRdiwO_gral}, we obtain:
\begin{subequations}\label{ec:dRdiwO_sec}
\begin{align}
\left\langle\frac{\partial R}{\partial (i_1,\omega_1,\Omega_1)}\right\rangle &= \kappa \frac{m_2}{m_1} \sum_{l=2}^\infty \sum_{m=0}^l \sum_{p=0}^l \sum_{q=-\infty}^{\infty} \left\langle\frac{\partial U_{lmpq}^{(1)}}{\partial (i_1,\omega_1,\Omega_1)}\right\rangle, \label{ec:dRdiwO1_sec} \\
\left\langle\frac{\partial R}{\partial i_2}\right\rangle &= \kappa \left[\frac{m_1}{m_2} \sum_{l=2}^\infty \sum_{m=0}^l \sum_{p=0}^l \sum_{q=-\infty}^{\infty} \left\langle\frac{\partial U_{lmpq}^{(2)}}{\partial i_2}\right\rangle + \frac{\partial V_{2010}^{(2)}}{\partial i_2} \label{ec:dRdi2_sec} \right], \\
\left\langle\frac{\partial R}{\partial (\omega_2,\Omega_2)}\right\rangle &= \kappa \frac{m_1}{m_2} \sum_{l=2}^\infty \sum_{m=0}^l \sum_{p=0}^l \sum_{q=-\infty}^{\infty} \left\langle\frac{\partial U_{lmpq}^{(2)}}{\partial (\omega_2,\Omega_2)}\right\rangle, \label{ec:dRdwO2_sec}
\end{align}
\end{subequations} where 
\begin{equation}\label{ec:def_kappa}
\begin{split}
\kappa &= \frac{G \, m_1 m_2}{\beta \, a} \\
       &=  n^2 a^2.
\end{split}
\end{equation} By inserting Eqs.~\eqref{ec:dRdaeM_sec} and \eqref{ec:dRdiwO_sec} in Eq.~\eqref{ec:ec_mov_orb_gral_avg} we obtain the equations describing the orbital motion. If we also take into account Eq.~\eqref{ec:pottriaxplanet2010} we obtain:
\begin{subequations}\label{ec:evorb_avg}
\begin{equation}\label{ec:dadt_avg} 
\left\langle \frac{\mathrm{d}a}{\mathrm{d}t} \right\rangle = 2 \, n \, a \sum_{l=2}^\infty \sum_{m=0}^l \sum_{p=0}^l \sum_{q=-\infty}^{\infty} \left[ \frac{m_2}{m_1} \left\langle \frac{\partial U_{lmpq}^{(1)}}{\partial M}\right\rangle + \frac{m_1}{m_2} \left\langle \frac{\partial U_{lmpq}^{(2)}}{\partial M}\right\rangle \right],
\end{equation} 
\begin{equation}\label{ec:dedt_avg} 
\left\langle \frac{\mathrm{d}e}{\mathrm{d}t} \right\rangle = n \frac{\sqrt{1-e^2}}{e} \sum_{l=2}^\infty \sum_{m=0}^l \sum_{p=0}^l \sum_{q=-\infty}^{\infty} \left[\sqrt{1-e^2} \left( \frac{m_2}{m_1} \left\langle \frac{\partial U_{lmpq}^{(1)}}{\partial M}\right\rangle + \frac{m_1}{m_2} \left\langle \frac{\partial U_{lmpq}^{(2)}}{\partial M}\right\rangle\right) - \frac{m_2}{m_1} \left\langle\frac{\partial U_1}{\partial \omega_1}\right\rangle - \frac{m_1}{m_2} \left\langle\frac{\partial U_2}{\partial \omega_2}\right\rangle\right],
\end{equation} 
\begin{multline}\label{ec:dMdt_avg} 
\left\langle \frac{\mathrm{d} M}{\mathrm{d}t} \right\rangle = n + 2 \, n \left[\sum_{l=2}^\infty \sum_{m=0}^l \sum_{p=0}^l \sum_{q=-\infty}^{\infty} (l+1) \left(\frac{m_2}{m_1} \left\langle U_{lmpq}^{(1)} \right\rangle + \frac{m_1}{m_2} \left\langle U_{lmpq}^{(2)} \right\rangle\right) + 3 \, V_{2010}^{(2)}\right] \\ - n \frac{1-e^2}{e} \left[\sum_{l=2}^\infty \sum_{m=0}^l \sum_{p=0}^l \sum_{q=-\infty}^{\infty} \frac{m_2}{m_1} \left\langle \frac{\partial U_{lmpq}^{(1)}}{\partial e}\right\rangle + \frac{m_1}{m_2} \left\langle \frac{\partial U_{lmpq}^{(2)}}{\partial e}\right\rangle + \frac{\partial V_{2010}^{(2)}}{\partial e}\right],
\end{multline} 
\begin{equation}\label{ec:di1dt_avg}
\left\langle \frac{\mathrm{d}i_1}{\mathrm{d}t} \right\rangle = \frac{1}{\sin i_1} \frac{m_2}{m_1} \sum_{l=2}^\infty \sum_{m=0}^l \sum_{p=0}^l \sum_{q=-\infty}^{\infty} \left[\frac{G \, m_1 \, m_2}{a \, C_1 \, \dot{\theta}_1} \left(\left\langle\frac{\partial U_{lmpq}^{(1)}}{\partial \omega_1}\right\rangle - \cos i_1 \left\langle\frac{\partial U_{lmpq}^{(1)}}{\partial \Omega_1}\right\rangle \right) - \frac{n}{\sqrt{1-e^2}} \left(\left\langle\frac{\partial U_{lmpq}^{(1)}}{\partial \Omega_1}\right\rangle - \cos i_1 \left\langle\frac{\partial U_{lmpq}^{(1)}}{\partial \omega_1}\right\rangle \right)\right], 
\end{equation} 
\begin{equation}\label{ec:di2dt_avg} 
\left\langle \frac{\mathrm{d}i_2}{\mathrm{d}t} \right\rangle = \frac{1}{\sin i_2} \frac{m_1}{m_2} \sum_{l=2}^\infty \sum_{m=0}^l \sum_{p=0}^l \sum_{q=-\infty}^{\infty} \left[\frac{G \, m_1 \, m_2}{a \, C_2 \, \dot{\theta}_2} \left(\left\langle\frac{\partial U_{lmpq}^{(2)}}{\partial \omega_2}\right\rangle - \cos i_2 \left\langle\frac{\partial U_{lmpq}^{(2)}}{\partial \Omega_2}\right\rangle \right) - \frac{n}{\sqrt{1-e^2}} \left(\left\langle\frac{\partial U_{lmpq}^{(2)}}{\partial \Omega_2}\right\rangle - \cos i_2 \left\langle\frac{\partial U_{lmpq}^{(2)}}{\partial \omega_2}\right\rangle \right)\right],
\end{equation} 
\begin{multline}\label{ec:dw1dt_avg} 
\left\langle \frac{\mathrm{d} \omega_1}{\mathrm{d}t} \right\rangle = n \frac{\sqrt{1-e^2}}{e} \left(\sum_{l=2}^\infty \sum_{m=0}^l \sum_{p=0}^l \sum_{q=-\infty}^{\infty} \frac{m_2}{m_1} \left\langle \frac{\partial U_{lmpq}^{(1)}}{\partial e}\right\rangle + \frac{m_1}{m_2} \left\langle \frac{\partial U_{lmpq}^{(2)}}{\partial e}\right\rangle + \frac{\partial V_{2010}^{(2)}}{\partial e} \right) \\ - \frac{m_2}{m_1} \left(\frac{G \, m_1 \, m_2}{a \, C_1 \, \dot{\theta}_1} + \frac{n \cos i_1}{\sqrt{1-e^2}}\right) \sum_{l=2}^\infty \sum_{m=0}^l \sum_{p=0}^l \sum_{q=-\infty}^{\infty} \frac{1}{\sin i_1} \left\langle\frac{\partial U_{lmpq}^{(1)}}{\partial i_1}\right\rangle,
\end{multline} 
\begin{multline}\label{ec:dw2dt_avg} 
\left\langle \frac{\mathrm{d} \omega_2}{\mathrm{d}t} \right\rangle = n \frac{\sqrt{1-e^2}}{e} \left(\sum_{l=2}^\infty \sum_{m=0}^l \sum_{p=0}^l \sum_{q=-\infty}^{\infty} \frac{m_2}{m_1} \left\langle \frac{\partial U_{lmpq}^{(1)}}{\partial e}\right\rangle + \frac{m_1}{m_2} \left\langle \frac{\partial U_{lmpq}^{(2)}}{\partial e}\right\rangle + \frac{\partial V_{2010}^{(2)}}{\partial e} \right) \\ - \left(\frac{G \, m_1 \, m_2}{a \, C_2 \, \dot{\theta}_2} + \frac{n \cos i_2}{\sqrt{1-e^2}}\right) \frac{1}{\sin i_2} \left[\frac{m_1}{m_2} \sum_{l=2}^\infty \sum_{m=0}^l \sum_{p=0}^l \sum_{q=-\infty}^{\infty} \left\langle\frac{\partial U_{lmpq}^{(1)}}{\partial i_2}\right\rangle + \frac{\partial V_{2010}^{(2)}}{\partial i_2}\right],
\end{multline} 
\begin{equation}\label{ec:dOm1dt_avg} 
\left\langle \frac{\mathrm{d} \Omega_1}{\mathrm{d}t} \right\rangle = \frac{m_2}{m_1} \left(\frac{G \, m_1 \, m_2 }{a \, C_1 \, \dot{\theta}_1} \cos i_1 + \frac{n}{\sqrt{1-e^2}}\right) \sum_{l=2}^\infty \sum_{m=0}^l \sum_{p=0}^l \sum_{q=-\infty}^{\infty} \frac{1}{\sin i_1} \left\langle\frac{\partial U_{lmpq}^{(1)}}{\partial i_1}\right\rangle,
\end{equation} 
\begin{equation}\label{ec:dOm2dt_avg} 
\left\langle \frac{\mathrm{d} \Omega_2}{\mathrm{d}t} \right\rangle = \left(\frac{G \, m_1 \, m_2 }{a \, C_2 \, \dot{\theta}_2} \cos i_2 + \frac{n}{\sqrt{1-e^2}}\right) \frac{1}{\sin i_2} \left[\frac{m_1}{m_2} \sum_{l=2}^\infty \sum_{m=0}^l \sum_{p=0}^l \sum_{q=-\infty}^{\infty} \left\langle\frac{\partial U_{lmpq}^{(2)}}{\partial i_2}\right\rangle + \frac{\partial V_{2010}^{(2)}}{\partial i_2}\right].
\end{equation}
\end{subequations}

With regard to the rotational evolution, since Eqs.~\eqref{ec:evorb_avg} depend neither on $\varepsilon_1$ nor $\varepsilon_2$, we only need Eqs.~\eqref{ec:d2thdt2_gral} to close the set of equations describing the dynamical evolution of binary systems such as the ones we consider in this work. However, if we insert the expression of the derivative of the disturbing function with respect to the corresponding longitudes of the ascending nodes, given by Eq.~\eqref{ec:dRdiwO_sec}, we obtain just the equations for the angular accelerations due to the secular terms of the tidal potentials, that is, those of the tidal torques. That equation is enough to describe rotational evolution for the star, but not for the planet. Since the latter could cross one or several spin-orbit resonances, we should take into account the permanent-triaxiality-induced torque. This torque arises from the $C_{22}$-related terms of the CGF disturbing potential, which are given by:
\begin{equation}\label{ec:pottriaxplanet22}
V_{2}^{(lm) = (22)} (\vec{r}_{1}) = - \frac{G \,m_{1}}{a} \sum_{p=0}^2 \sum_{q=-\infty}^{\infty} V^{(2)}_{22pq},
\end{equation} where
\begin{equation}\label{ec:pottriaxplanetlmpq22} 
V^{(2)}_{22pq} = C_{22} \left(\frac{R_{2}}{a}\right)^2 F_{22p} (i_{2}) \, G_{2pq} (e) \cos \left(v^{(2)}_{22pq} - 2 \, \theta_{2}\right).
\end{equation} Thus, the equations governing the time evolution of the spin rate of both the star and the planet, respectively, are obtained as follows: By inserting in Eq.~\eqref{ec:d2thdt2_gral} the expression of the corresponding derivative, given by Eq.~\eqref{ec:dRdiwO1_sec}, we arrive at:
\begin{subequations}\label{ec:evrot_sp}
\begin{equation}\label{ec:thpp_star}
\frac{\mathrm{d}^2 \theta_1}{\mathrm{d}t} = - \frac{G \, m_2^2}{a \, C_1} \sum_{l=2}^\infty \sum_{m=0}^l \sum_{p=0}^l \sum_{q=-\infty}^{\infty} \left\langle\frac{\partial U_{lmpq}^{(1)}}{\partial \Omega_1}\right\rangle.
\end{equation} Next, we consider the corresponding derivative of Ec.~\eqref{ec:dRdiwO2_gral}, in which the derivative of the tidal potential of the planet is given by Eq.~\eqref{ec:dU2diwO}, into which we insert Eq.~\eqref{ec:pottriaxplanet22} and differentiate. Then, by replacing the result in Eq.~\eqref{ec:d2thdt2_gral}, we obtain:
\begin{equation}\label{ec:thpp_planet}
\frac{\mathrm{d}^2 \theta_2}{\mathrm{d}t} = - \frac{G \, m_1^2}{a \, C_2} \sum_{l=2}^\infty \sum_{m=0}^l \sum_{p=0}^l \sum_{q=-\infty}^{\infty} \left\langle\frac{\partial U_{lmpq}^{(2)}}{\partial \Omega_2}\right\rangle + 2 \frac{G \, m_1 \, m_2}{a \, C_2} C_{22} \left(\frac{R_2}{a}\right)^2 \sum_{p=0}^2 \sum_{q=-\infty}^{\infty} F_{22p} (i_2) G_{2pq} (e) \sin \left(v^{(2)}_{22pq} - 2 \, \theta_{2}\right).
\end{equation}
\end{subequations}

\subsection{Simplifications available for low inclinations and eccentricities}

Up to this point, we have presented the analytical tools in a general form which allow for a description of the dynamical evolution of a binary system that considers the tidal deformation of both of its components and the triaxiality of one of them. Based on the observation that the disturbing potentials and their corresponding gradients -- which build up both the disturbing function and its derivatives, respectively -- are represented as an infinite series, it becomes necessary to examine the leading terms in order to perform the mandatory truncation that allows for the implementation of the formulae in a computer code for its practical application. To that end, we take into account the quadrupolar approximation mentioned at the beginning of Sect.~\ref{sec:eom}.

We now focus on the inclination functions. These are defined by \citep{goodwagner2008}: 
\begin{equation}\label{ec:defincfuncs} 
F_{lmp} (i) = \frac{(l+m)!}{2^l (l-p)! p!} \sum_{j} (-1)^j \binom{2l-2p}{j} \binom{2p}{l-m-j} \cos^{3l-m-2p-2j} \left(\frac{i}{2}\right) \sin^{m-l+2p+2j} \left(\frac{i}{2}\right),
\end{equation} where the limits of summation are $\max (0,l-m-2p)$ and $\min (l-m,2(l-p))$. In Table~\ref{tab:inc_fns}, where we show the inclination functions for $l=2$, it can be seen that for low inclinations $F_{201} (i)$ and $F_{220} (i)$ tend towards $1/2$ and $3$, respectively, as $i$ tends towards zero while all the others tend towards zero. In consequence, the terms corresponding to $(lmpq) = (201q)$ and $(lmpq) = (220q)$ of the disturbing potentials dominate the dynamical evolution in the case of low inclinations. For this reason, the latter are given in Appendix~\ref{sec:sum_terms_adim}.

\begin{table*}
 \caption{Inclination functions for $l=2$ and their first derivatives.}
 \label{tab:inc_fns}
 \centering
 \begin{tabular}{c c c c c} 
 \hline\hline
 $m$ & $p$ & $F_{2mp} (i)$ & $\frac{\mathrm{d}F_{2mp} (i)}{\mathrm{d}i}$ & $\frac{1}{\sin i} \left(F_{2mp} (i) 
\frac{\mathrm{d}F_{2mp} (i)}{\mathrm{d}i}\right)$ \\ 
 \hline
 0 & 0 & $ \frac{3}{8} \sin^2 i$                     & $\frac{3}{4} \sin i \cos i$ & $\frac{9}{32} \sin^2 i \cos i$ \\
 0 & 1 & $-\frac{3}{4} \sin^2 i + \frac{1}{2}$       & $-\frac{3}{2} \sin i \cos i$ & $-\frac{3}{2} \cos i \left(\frac{1}{2} - 
\frac{3}{4} \sin^2 i\right)$ \\
 0 & 2 & $ \frac{3}{8} \sin^2 i$                     & $\frac{3}{4} \sin i \cos i$ & $\frac{9}{32} \sin^2 i \cos i$ \\
 1 & 0 & $-\frac{3}{4} \left(1+\cos i\right) \sin i$ & $-\frac{3}{4} \left[\cos i + \cos (2 i)\right]$ & $\frac{9}{16} \left(1+\cos 
i\right) \left[\cos i + \cos (2 i)\right]$ \\
 1 & 1 & $ \frac{3}{2} \sin i \cos i$                & $\frac{3}{2} \cos (2 i)$ & $\frac{9}{4} \cos i \cos (2i)$ \\
 1 & 2 & $ \frac{3}{4} \left(1-\cos i\right) \sin i$ & $\frac{3}{4} \left[\cos i -  \cos (2 i)\right]$ & $\frac{9}{16}  \left(1-\cos 
i\right) \left[\cos i -  \cos (2 i)\right]$ \\
 2 & 0 & $ \frac{3}{4} \left(1+\cos i\right)^2$      & $-\frac{3}{2} (1+\cos i) \sin i$ & $-\frac{9}{8}  \left(1+\cos i\right)^3$ \\
 2 & 1 & $ \frac{3}{2} \sin^2 i$                     & $3 \sin i \cos i$ & $\frac{9}{2} \sin^2 i \cos i$\\
 2 & 2 & $ \frac{3}{4} \left(1-\cos i\right)^2$      & $\frac{3}{2} (1-\cos i) \sin i$ & $\frac{9}{8} \left(1-\cos i\right)^3$ \\
\hline 
\end{tabular} 
\end{table*}

At this point, it is necessary to briefly discuss on the truncation of the sums over $q$ in order to make practical computation possible. In general terms, the eccentricity functions are defined as:
\begin{equation}\label{ec:def_ecc_funs_gral}
\left(\frac{r}{a}\right)^{-(l+1)} \exp \left[\mathrm{i} \, (l-2p) \, \nu \right] = \sum_{q=-\infty}^{\infty} G_{lpq} (e) \exp \left[\mathrm{i} \, (l-2p+q) \, M\right],
\end{equation} where
\begin{equation}\label{ec:def_rsa}
\frac{r}{a} = \frac{1-e^2}{1+e \cos \nu}.
\end{equation} where $\nu$ is the true anomaly. Thus, the eccentricity functions are defined as the coefficients of the expansion in Fourier series of the left-hand side of Eq.~\eqref{ec:def_ecc_funs_gral}. Of course, the derivatives of the eccentricity functions are also needed. These can be defined in an analogous way as the coefficients of the Fourier series expansion of the derivative of the left-hand side of Eq.~\eqref{ec:def_ecc_funs_gral} with respect to the eccentricity \citep{giacaglia1976b}:
\begin{equation}\label{ec:def_decc_funs_de_gral}
\frac{\partial}{\partial e} \left[ \left(\frac{r}{a}\right)^{-(l+1)} \exp \left[\mathrm{i} \, (l-2p) \, \nu \right] \right] = \sum_{q=-\infty}^{\infty} \frac{\mathrm{d}G_{lpq} (e)}{\mathrm{d}e} \exp \left[\mathrm{i} \, (l-2p+q) \, M\right],
\end{equation} where
\begin{equation}\label{ec:def_drsade}
\frac{\partial}{\partial e} \left[ \left(\frac{r}{a}\right)^{-(l+1)} \exp \left[\mathrm{i} \, (l-2p) \, \nu \right] \right] = \left(\frac{r}{a}\right)^{-(l+1)} \left[(l+1) \left(\frac{2e}{1-e^2} + \frac{\cos \nu - e \frac{\partial \nu}{\partial e} \sin \nu}{1 + e \cos \nu}\right) + \mathrm{i} (l-2p) \frac{\partial \nu}{\partial e} \right] \exp \left[\mathrm{i} (l-2p) \nu\right],
\end{equation} and
\begin{equation}\label{ec:dnude}
\frac{\partial \nu}{\partial e} = \frac{\sin \nu}{1-e^2} (2 + e \cos \nu).
\end{equation}

The expressions corresponding to $(lmpq) = (201q)$ are:
\begin{align}
\left(\frac{r}{a}\right)^{-3} &= \sum_{q=-\infty}^{\infty} G_{21q} (e) \cos \left(q \, M\right), \label{ec:Ge21q} \\
\left(\frac{r}{a}\right)^{-3} 3 \left(\frac{2e}{1-e^2} + \frac{\cos \nu - e \frac{\partial \nu}{\partial e} \sin \nu}{1 + e \cos \nu}\right) &= \sum_{q=-\infty}^{\infty} \frac{\mathrm{d}G_{21q} (e)}{\mathrm{d}e} \cos \left(q \, M\right), \label{ec:dGe21q}
\end{align} while those corresponding to $(lmpq) = (220q)$ are:
\begin{align}
\left(\frac{r}{a}\right)^{-3} \left[
\begin{array}{c}
\cos \\
\sin
\end{array}\right] \left(2 \, \nu\right) & = \sum_{q=-\infty}^{\infty} G_{20q} (e) \left[
\begin{array}{c}
\cos \\
\sin
\end{array}\right] \left[(2+q) M\right], \label{ec:Ge20q} \\
\left(\frac{r}{a}\right)^{-3} \left[3 \left(\frac{2e}{1-e^2} + \frac{\cos \nu - e \frac{\partial \nu}{\partial e} \sin \nu}{1 + e \cos \nu}\right) \left[
\begin{array}{c}
\cos \\
\sin
\end{array}\right] \left(2 \, \nu\right) + 2 \frac{\partial \nu}{\partial e} \left[
\begin{array}{c}
-\sin \\
\cos
\end{array}\right] \left(2 \, \nu\right)\right] &= \sum_{q=-\infty}^{\infty} \frac{\mathrm{d} G_{20q}}{\mathrm{d}e} (e) \left[
\begin{array}{c}
\cos \\
\sin
\end{array}\right] \left[(2+q) M\right]. \label{ec:dGe20q}
\end{align} The question of how many terms are to be kept in order to obtain a reliable representation of the functions on the left-hand side of Eqs.~\eqref{ec:Ge21q}, \eqref{ec:dGe21q}, \eqref{ec:Ge20q}, and \eqref{ec:dGe20q} can be answered by comparing them with their corresponding Fourier series representation. The criterion adopted in this work is the computation of the root mean squared difference (RMSD) between the left and right-hand sides of the latter equations corresponding to different values of $q_{\mathrm{max}}$, which is the maximum number of coefficients. Mathematically expressed, let $g(e,\nu)$ be a function that represents each of those on the left-hand side of Eqs.~\eqref{ec:Ge21q} and \eqref{ec:dGe21q} as well as the real and imaginary parts of Eqs~\eqref{ec:Ge20q} and \eqref{ec:dGe20q}. Also, let $\tilde{g} (e,M)$ be the function representing the corresponding right-hand side of the same equations. The RMSD is given by:
\begin{equation}\label{ec:def_rsmd}
\mathrm{RMSD} = \sqrt{\frac{\sum_{j=1}^N \left[g (e,\nu_i) - \tilde{g} (e,M_i) \right]^2}{N}}
,\end{equation} where $N$ is the number of values of $M$ and $\nu$ within the interval $[-\pi,\pi]$, each of them labeled with $M_i$ and $\nu_i$ being the respective value of the true anomaly.

\begin{table*}
 \caption{RMSD between the left and right-hand sides of Eqs.~\eqref{ec:Ge21q}, \eqref{ec:dGe21q}, \eqref{ec:Ge20q}, and \eqref{ec:dGe20q}. Columns four through seven correspond to the RMSD of the real ($\Re$) and imaginary ($\Im$) parts of the left and right hands of Eqs.~\eqref{ec:Ge20q} and \eqref{ec:dGe20q}, respectively. This results are obtained with $e=0.1$.}
 \label{tab:qmax_rmsd}
 \centering
 \begin{tabular}{c c c c c c c} 
 \hline\hline
 \multirow{2}{*}{$q_{\mathrm{max}}$} & \multicolumn{6}{c}{RMSD} \\
 \cline{2-7}
 & \eqref{ec:Ge21q} & \eqref{ec:dGe21q} & $\Re$ \eqref{ec:Ge20q} & $\Im$ \eqref{ec:Ge20q} & $\Re$ \eqref{ec:dGe20q} & $\Im$ \eqref{ec:dGe20q} \\ 
 \hline
 2 & $4.756939 \times 10^{-3}$ & $1.443418 \times 10^{-1}$ & $1.237304 \times 10^{-2}$ & $1.237193 \times 10^{-2}$ & $3.704831 \times 10^{-1}$ & $3.704534 \times 10^{-1}$ \\
 4 & $9.884915 \times 10^{-5}$ & $4.962669 \times 10^{-3}$ & $4.149271 \times 10^{-4}$ & $4.148743 \times 10^{-4}$ & $2.064961 \times 10^{-2}$ & $2.064789 \times 10^{-2}$ \\
 6 & $1.997451 \times 10^{-6}$ & $1.399011 \times 10^{-4}$ & $1.175513 \times 10^{-5}$ & $1.175534 \times 10^{-5}$ & $8.183466 \times 10^{-4}$ & $8.182644 \times 10^{-4}$ \\
 8 & $4.064722 \times 10^{-8}$ & $3.563261 \times 10^{-6}$ & $3.117166 \times 10^{-7}$ & $3.051717 \times 10^{-7}$ & $2.708817 \times 10^{-5}$ & $2.717803 \times 10^{-5}$ \\
10 & $1.000686 \times 10^{-8}$ & $1.142291 \times 10^{-7}$ & $7.273942 \times 10^{-8}$ & $1.078445 \times 10^{-7}$ & $9.280304 \times 10^{-7}$ & $9.066432 \times 10^{-7}$ \\
12 & $9.977323 \times 10^{-9}$ & $7.538773 \times 10^{-8}$ & $7.213027 \times 10^{-8}$ & $1.076484 \times 10^{-7}$ & $4.663512 \times 10^{-7}$ & $4.368208 \times 10^{-7}$ \\
\hline 
\end{tabular} 
\end{table*}

Of course, the number of terms of each of the Fourier series expansion strongly depend on the eccentricity. As can be observed in Table~\ref{tab:qmax_rmsd}, for $e=0.1$, there is no significative reduction of the RMSD for $q_{\mathrm{max}} > 10$. We have considered the maximum value of the eccentricity from the estimated uncertainty for the Kepler-21 system. Thus, we consider that it is safe to take $q_{\mathrm{max}} = 10$.
It is worth mentioning that the eccentricity functions and its first derivatives with respect to the eccentricity were computed by using the Newcomb operators, as explained in Appendix~\ref{sec:eccfuncs}.

\subsection{Final form of the equations of motion}

The equations governing the orbital evolution, Eqs.~\eqref{ec:evorb_avg} can be rewritten in an even more convenient form that could be suitable for its implementation in a computational code specially for low (or even zero) eccentricities and inclinations. As pointed out by \citet{brouwer_clemence_1961}, the divergences in the aforementioned equations ``are more apparent than real'' and that observation remains valid in this context. First, it can be noted that in Eqs.~\eqref{ec:dw1dt_avg}, \eqref{ec:dw2dt_avg}, \eqref{ec:dOm1dt_avg}, and \eqref{ec:dOm2dt_avg} each derivative of the tidal potentials, generated by the star and by the planet, with respect to the inclination, appear divided by the sine of the corresponding inclination. By observing the expressions of the latter derivatives, given by Eqs.~\eqref{ec:SdUkdik_sec}, then we see that the inclinations functions appear multiplied by its derivative. In consequence, observing the third column of Table~\ref{tab:inc_fns}, we also note that there are no singularities in the respective terms of the aforementioned equations, provided that the dividing sine of the inclination is canceled. 

Something analogous occurs with the eccentricity dividing the derivatives of the tidal potentials with respect to the eccentricity. We can observe in Eqs.~\eqref{ec:SdUkde_sec} that each term of the aforementioned derivatives contains products of the corresponding eccentricity function and its derivative. It can be shown that the these factors, $G_{2pq} (e) \mathrm{d} G_{2pq} (e) /\mathrm{d}e$, do not contain terms of order $e^0$.

We can introduce further modifications to Eqs.~\eqref{ec:dedt_avg}, \eqref{ec:di1dt_avg} and \eqref{ec:di2dt_avg}. In order to avoid divergences that are also present in the equations mentioned previously, we define the following variables: $\xi = \sqrt{1-e^2}$ and $x = \cos i$. 
Therefore, we have:
\begin{equation}\label{ec:dxidt_def}
\frac{\mathrm{d}\xi}{\mathrm{d}t} = - \frac{e}{\sqrt{1-e^2}} \frac{\mathrm{d}e}{\mathrm{d}t}
,\end{equation}
\begin{equation}\label{ec:dxdt_def}
\frac{\mathrm{d}x}{\mathrm{d}t} = - \sin i \frac{\mathrm{d}i}{\mathrm{d}t}
.\end{equation} Thus, the final form of the equations describing the orbital evolution, upon considering the simplifications discussed before, are:
\begin{subequations}\label{ec:ec_mov_orb_sec_mod}
\begin{multline}\label{ec:dadt_sec_mod}
\left\langle \frac{\mathrm{d}a}{\mathrm{d}t} \right\rangle = - \frac{n \, a}{2} \left(\frac{m_2}{m_1} \left(\frac{R_1}{a}\right)^5 \left[\left(\frac{3 \, x_1^2 - 1}{2}\right)^2 \sum_{q=-q_\mathrm{max}}^{q_\mathrm{max}} G_{21q}^2 (e) \, q \, K_{\mathrm{I}}^{(1)} \left(2,\omega_{201q}^{(1)}\right) + 3 \left(\frac{1 + x_1}{2}\right)^4 \sum_{q=-q_\mathrm{max}}^{q_\mathrm{max}} G_{20q}^2 (e) \, (2+q) \, K_{\mathrm{I}}^{(1)} \left(2,\omega_{220q}^{(1)}\right)\right] \right. \\ \left. + \frac{m_1}{m_2} \left(\frac{R_2}{a}\right)^5 \left[\left(\frac{3 \, x_2^2 - 1}{2}\right)^2 \sum_{q=-q_\mathrm{max}}^{q_\mathrm{max}} G_{21q}^2 (e) \, q \, K_{\mathrm{I}}^{(2)} \left(2,\omega_{201q}^{(2)}\right) + 3 \left(\frac{1 + x_2}{2}\right)^4 \sum_{q=-q_\mathrm{max}}^{q_\mathrm{max}} G_{20q}^2 (e) \, (2+q) \, K_{\mathrm{I}}^{(2)}  \left(2,\omega_{220q}^{(2)}\right)\right]\right).
\end{multline}
\begin{equation}\label{ec:dxidt_sec_mod}
\begin{split}
\left\langle \frac{\mathrm{d}\xi}{\mathrm{d}t} \right\rangle = \frac{n}{4} \left(\xi \left(\frac{m_2}{m_1} \left(\frac{R_1}{a}\right)^5 \left[\left(\frac{3 \, x_1^2 - 1}{2}\right)^2 \sum_{q=-q_\mathrm{max}}^{q_\mathrm{max}} G_{21q}^2 (e) \, q \, K_{\mathrm{I}}^{(1)} \left(2,\omega_{201q}^{(1)}\right) + 3 \left(\frac{1 + x_1}{2}\right)^4 \sum_{q=-q_\mathrm{max}}^{q_\mathrm{max}} G_{20q}^2 (e) \, (2+q) \, K_{\mathrm{I}}^{(1)} \left(2,\omega_{220q}^{(1)}\right)\right] \right. \right. \\ \left. \left. + \frac{m_1}{m_2} \left(\frac{R_2}{a}\right)^5 \left[\left(\frac{3 \, x_2^2 - 1}{2}\right)^2 \sum_{q=-q_\mathrm{max}}^{q_\mathrm{max}} G_{21q}^2 (e) \, q \, K_{\mathrm{I}}^{(2)} \left(2,\omega_{201q}^{(2)}\right) + 3 \left(\frac{1 + x_2}{2}\right)^4 \sum_{q=-q_\mathrm{max}}^{q_\mathrm{max}} G_{20q}^2 (e) \, (2+q) \, K_{\mathrm{I}}^{(2)} \left(2,\omega_{220q}^{(2)}\right)\right]\right) \right. \\ \left. - \frac{3}{2} \left[\frac{m_2}{m_1} \left(\frac{R_1}{a}\right)^5 \left(\frac{1 + x_1}{2}\right)^4 \sum_{q=-q_\mathrm{max}}^{q_\mathrm{max}} G_{20q}^2 (e) \, K_{\mathrm{I}}^{(1)} \left(2,\omega_{220q}^{(1)}\right) + \frac{m_1}{m_2} \left(\frac{R_2}{a}\right)^5 \left(\frac{1 + x_2}{2}\right)^4 \sum_{q=-q_\mathrm{max}}^{q_\mathrm{max}} G_{20q}^2 (e) \, K_{\mathrm{I}}^{(2)} \left(2,\omega_{220q}^{(2)}\right)\right]\right).
\end{split}
\end{equation}
\begin{eqnarray}\label{ec:dMdt_sec_mod}
\begin{split}
\left\langle \frac{\mathrm{d}M}{\mathrm{d}t} \right\rangle &= n + \frac{3}{2} \, n \left(\frac{m_2}{m_1} \left(\frac{R_1}{a}\right)^5 \left[\left(\frac{3 \, x_1^2 - 1}{2}\right)^2 \sum_{q=-q_\mathrm{max}}^{q_\mathrm{max}} G_{21q}^2 (e) \, K_{\mathrm{R}}^{(1)} \left(2,\omega_{201q}^{(1)}\right) + 3 \left(\frac{1 + x_1}{2}\right)^4 \sum_{q=-q_\mathrm{max}}^{q_\mathrm{max}} G_{20q}^2 (e) \, K_{\mathrm{R}}^{(1)} \left(2,\omega_{220q}^{(1)}\right)\right]\right. \\ & \left. + \frac{m_1}{m_2} \left(\frac{R_2}{a}\right)^5 \left[\left(\frac{3 \, x_2^2 - 1}{2}\right)^2 \sum_{q=-q_\mathrm{max}}^{q_\mathrm{max}} G_{21q}^2 (e) K_{\mathrm{R}}^{(2)} \left(2,\omega_{201q}^{(2)}\right) + 3 \left(\frac{1 + x_2}{2}\right)^4 \sum_{q=-q_\mathrm{max}}^{q_\mathrm{max}} G_{20q}^2 (e) K_{\mathrm{R}}^{(2)} \left(2,\omega_{220q}^{(2)}\right)\right]\right) \\ & - \frac{n \, \xi}{4} \left(\frac{m_2}{m_1} \left(\frac{R_1}{a}\right)^5 \left[\left(\frac{3 \, x_1^2 - 1}{2}\right)^2 \sum_{q=-q_\mathrm{max}}^{q_\mathrm{max}} \left(\frac{1}{e} G_{21q} (e) \frac{\mathrm{d}G_{21q} (e)}{\mathrm{d}e}\right) K_{\mathrm{R}}^{(1)}\left(2,\omega_{201q}^{(1)}\right) \right. \right. \\ & \left. \left. + 3 \left(\frac{1 + x_1}{2}\right)^4 \sum_{q=-q_\mathrm{max}}^{q_\mathrm{max}} \left(\frac{1}{e} G_{20q} (e) \frac{\mathrm{d} G_{20q} (e)}{\mathrm{d}e}\right) K_{\mathrm{R}}^{(1)} \left(2,\omega_{220q}^{(1)}\right)\right]\right. \\ & \left. + \frac{m_1}{m_2} \left(\frac{R_2}{a}\right)^5 \left[\left(\frac{3 \, x_2^2 - 2}{2}\right)^2 \sum_{q=-q_\mathrm{max}}^{q_\mathrm{max}} \left(\frac{1}{e} G_{21q} (e) \frac{\mathrm{d} G_{21q} (e)}{\mathrm{d}e}\right) K_{\mathrm{R}}^{(2)} \left(2,\omega_{201q}^{(2)}\right) \right. \right. \\ & \left. \left. + 3 \left(\frac{1 + x_2}{2}\right)^4 \sum_{q=-q_\mathrm{max}}^{q_\mathrm{max}} \left(\frac{1}{e} G_{20q} (e) \frac{\mathrm{d}G_{20q} (e)}{\mathrm{d}e}\right) K_{\mathrm{R}}^{(2)} \left(2,\omega_{220q}^{(2)}\right)\right]\right) \\ & - \frac{3}{2} n \, J_2 \left(\frac{R_2}{a}\right)^2 \left(\frac{3 \, x_2^2 - 1}{2}\right) \xi^{-3}.
\end{split}
\end{eqnarray}
\begin{equation}\label{ec:dx1dt_sec_mod}
\left\langle \frac{\mathrm{d}x_1}{\mathrm{d}t} \right\rangle = \frac{3}{2} \frac{m_2}{m_1} \left(\frac{R_1}{a}\right)^5 \left[\frac{G \, m_1 \, m_2}{a \, C_1 \, \dot{\theta}_1} - \frac{n}{\xi}\right] \left(\frac{1 + x_1}{2}\right)^4 \left(1-x_1\right) \sum_{q=-q_\mathrm{max}}^{q_\mathrm{max}} G_{20q}^2 (e) K_{\mathrm{I}}^{(1)} \left(2,\omega_{220q}^{(1)}\right).
\end{equation}
\begin{equation}\label{ec:dx2dt_sec_mod}
 \left\langle \frac{\mathrm{d}x_2}{\mathrm{d}t} \right\rangle = \frac{3}{2} \frac{m_1}{m_2} \left(\frac{R_2}{a}\right)^5 \left[\frac{G \, m_1 \, m_2}{a \, C_2 \, \dot{\theta}_2} - \frac{n}{\xi}\right] \left(\frac{1 + x_2}{2}\right)^4 \left(1-x_2\right) \sum_{q=-q_\mathrm{max}}^{q_\mathrm{max}} G_{20q}^2 (e) K_{\mathrm{I}}^{(2)} \left(2,\omega_{220q}^{(2)}\right).
\end{equation}
\begin{equation}\label{ec:dw1dt_sec_mod}
\begin{split}
\left\langle \frac{\mathrm{d}\omega_1}{\mathrm{d}t} \right\rangle &= \frac{3}{4} \frac{m_2}{m_1} \left(\frac{R_1}{a}\right)^5 \left[\frac{G \, m_1 \, m_2}{a \, C_1 \, \dot{\theta}_1} + \frac{n \, x_1}{\xi}\right] \left[x_1 \left(\frac{3 \, x_1^2 - 1}{2}\right) \sum_{q=-q_\mathrm{max}}^{q_\mathrm{max}} G_{21q}^2 (e) K_{\mathrm{R}}^{(1)} \left(2,\omega_{201q}^{(1)}\right) +  \left(\frac{1 + x_1}{2}\right)^3 \sum_{q=-q_\mathrm{max}}^{q_\mathrm{max}} G_{20q}^2 (e) K_{\mathrm{R}}^{(1)} \left(2,\omega_{220q}^{(1)}\right)\right] \\ & + \frac{n \, \xi}{4} \left(\frac{m_2}{m_1} \left(\frac{R_1}{a}\right)^5 \left[\left(\frac{3 \, x_1^2 - 1}{2}\right)^2 \sum_{q=-q_\mathrm{max}}^{q_\mathrm{max}} \left(\frac{1}{e} G_{21q} (e) \frac{\mathrm{d}G_{21q} (e)}{\mathrm{d}e}\right) K_{\mathrm{R}}^{(1)} \left(2,\omega_{201q}^{(1)}\right) \right. \right. \\ & \left. \left. + 3 \left(\frac{1 + x_1}{2}\right)^4 \sum_{q=-q_\mathrm{max}}^{q_\mathrm{max}} \left(\frac{1}{e} G_{20q} (e) \frac{\mathrm{d}G_{20q} (e)}{\mathrm{d}e}\right) K_{\mathrm{R}}^{(1)} \left(2,\omega_{220q}^{(1)}\right)\right] \right. \\ & \left. + \frac{m_1}{m_2} \left(\frac{R_2}{a}\right)^5 \left[\left(\frac{3 \, x_2^2 - 1}{2}\right)^2 \sum_{q=-q_\mathrm{max}}^{q_\mathrm{max}} \left(\frac{1}{e} G_{21q} (e) \frac{\mathrm{d}G_{21q} (e)}{\mathrm{d}e}\right) K_{\mathrm{R}}^{(2)} \left(2,\omega_{201q}^{(2)}\right) \right. \right. \\ & \left. \left. + 3 \left(\frac{1 + x_2}{2}\right)^4 \sum_{q=-q_\mathrm{max}}^{q_\mathrm{max}} \left(\frac{1}{e} G_{20q} (e) \frac{\mathrm{d}G_{20q} (e)}{\mathrm{d}e}\right) K_{\mathrm{R}}^{(2)} \left(2,\omega_{220q}^{(2)}\right)\right]\right) \\ & + \frac{3}{2} n J_2 \left(\frac{R_2}{a}\right)^2 \frac{3 \, x_2^2 - 1}{2} \xi^{-4}.
\end{split}
\end{equation}
\begin{equation}\label{ec:dw2dt_sec_mod}
\begin{split}
\left\langle \frac{\mathrm{d}\omega_2}{\mathrm{d}t} \right\rangle &= \frac{3}{4} \frac{m_1}{m_2} \left(\frac{R_2}{a}\right)^5 \left[\frac{G \, m_1 \, m_2}{a \, C_2 \, \dot{\theta}_2} + \frac{n \, x_2}{\xi}\right] \left[x_2 \left(\frac{3 \, x_2^2 - 1}{2}\right) \sum_{q=-q_\mathrm{max}}^{q_\mathrm{max}} G_{21q}^2 (e) K_{\mathrm{R}}^{(2)} \left(2,\omega_{201q}^{(2)}\right) + \left(\frac{1 + x_2}{2}\right)^3 \sum_{q=-q_\mathrm{max}}^{q_\mathrm{max}} G_{20q}^2 (e) K_{\mathrm{R}}^{(2)} \left(2,\omega_{220q}^{(2)}\right)\right] \\ & + \frac{n \, \xi}{4} \left(\frac{m_1}{m_2} \left(\frac{R_2}{a}\right)^5 \left[\left(\frac{3 \, x_2^2 - 1}{2}\right)^2 \sum_{q=-q_\mathrm{max}}^{q_\mathrm{max}} \left(\frac{1}{e} G_{21q} (e) \frac{\mathrm{d}G_{21q} (e)}{\mathrm{d}e}\right) K_{\mathrm{R}}^{(2)} \left(2,\omega_{201q}^{(2)}\right) \right. \right. \\ & \left. \left. + 3 \left(\frac{1 + x_2}{2}\right)^4 \sum_{q=-q_\mathrm{max}}^{q_\mathrm{max}} \left(\frac{1}{e} G_{20q} (e) \frac{\mathrm{d}G_{20q} (e)}{\mathrm{d}e}\right) K_{\mathrm{R}}^{(2)} \left(2,\omega_{220q}^{(2)}\right)\right] \right. \\ & \left. + \frac{m_2}{m_1} \left(\frac{R_1}{a}\right)^5 \left[\left(\frac{3 \, x_1^2 - 1}{2}\right)^2 \sum_{q=-q_\mathrm{max}}^{q_\mathrm{max}} \left(\frac{1}{e} G_{21q} (e) \frac{\mathrm{d}G_{21q} (e)}{\mathrm{d}e}\right) K_{\mathrm{R}}^{(1)} \left(2,\omega_{201q}^{(1)}\right) \right. \right. \\ & \left. \left. + 3 \left(\frac{1 + x_1}{2}\right)^4 \sum_{q=-q_\mathrm{max}}^{q_\mathrm{max}} \left(\frac{1}{e} G_{20q} (e) \frac{\mathrm{d}G_{20q} (e)}{\mathrm{d}e}\right) K_{\mathrm{R}}^{(1)} \left(2,\omega_{220q}^{(1)}\right)\right]\right) \\ & - \frac{3}{2} J_2 \left(\frac{R_2}{a}\right)^2  \xi^{-4} \left[\left(\frac{G \, m_1 \, m_2}{a \, C_2 \dot{\theta}_2} + \frac{n \, x_2}{\xi}\right) x_2 \, \xi + n \left(\frac{3 \, x_2^2 - 1}{2}\right)\right].
\end{split}
\end{equation}
\begin{multline}\label{ec:dOm1dt_sec_mod}
\left\langle \frac{\mathrm{d}\Omega_1}{\mathrm{d}t} \right\rangle = - \frac{3}{4} \frac{m_2}{m_1} \left(\frac{R_1}{a}\right)^5 \left[\frac{G \, m_1 \, m_2}{a \, C_1 \dot{\theta}_1} x_1 + \frac{n}{\xi}\right] \left[x_1 \left(\frac{3 \, x_1^2 - 1}{2}\right) \sum_{q=-q_\mathrm{max}}^{q_\mathrm{max}}  G_{21q}^2 (e) K_{\mathrm{R}}^{(1)} \left(2,\omega_{201q}^{(1)}\right) \right. \\ \left. + \left(\frac{1 + x_1}{2}\right)^3 \sum_{q=-q_\mathrm{max}}^{q_\mathrm{max}} G_{20q}^2 (e) K_{\mathrm{R}}^{(1)} \left(2,\omega_{220q}^{(1)}\right)\right].
\end{multline}
\begin{multline}\label{ec:dOm2dt_sec_mod}
\left\langle \frac{\mathrm{d}\Omega_2}{\mathrm{d}t} \right\rangle = \left[\frac{G \, m_1 \, m_2}{a \, C_2 \dot{\theta}_2} x_2 + \frac{n}{\xi}\right] \left( - \frac{3}{4} \frac{m_1}{m_2} \left(\frac{R_2}{a}\right)^5 \left[x_2 \left(\frac{3 \, x_2^2 - 1}{2}\right) \sum_{q=-q_\mathrm{max}}^{q_\mathrm{max}} G_{21q}^2 (e) K_{\mathrm{R}}^{(2)} \left(2,\omega_{201q}^{(2)}\right) \right.\right. \\ \left.\left. + \left(\frac{1 + x_2}{2}\right)^3 \sum_{q=-q_\mathrm{max}}^{q_\mathrm{max}} G_{20q}^2 (e) K_{\mathrm{R}}^{(2)} \left(2,\omega_{220q}^{(2)}\right)\right] + \frac{3}{2} J_2 \left(\frac{R_2}{a}\right)^2 \frac{x_2}{\xi^{3}}\right).
\end{multline}
\end{subequations}

Concerning the time evolution of the spin rate of both the star and the planet, the respective equations of motion take the form:
\begin{subequations}\label{ec:ec_mov_rot_sec_mod}
\begin{equation}\label{ec:d2th1dt2_sec_mod}
\left\langle \frac{\mathrm{d}^2 \theta_1}{\mathrm{d}t^2} \right\rangle = \frac{3}{2} \frac{G \, m_2^2}{a \, C_1} \left(\frac{R_1}{a}\right)^5 \left(\frac{1 + x_1}{2}\right)^4 \sum_{q=-q_\mathrm{max}}^{q_\mathrm{max}} G_{20q}^2 (e) K_{\mathrm{I}}^{(1)} \left(2,\omega_{220q}^{(1)}\right),
\end{equation} and
\begin{multline}\label{ec:d2th2dt2_tt_mod}
\frac{\mathrm{d}^2 \theta_2}{\mathrm{d}t^2} = \frac{3}{2} \frac{G \, m_1^2}{a \, C_2} \left(\frac{R_2}{a}\right)^5 \left(\frac{1 + x_2}{2}\right)^4 \sum_{q=-q_\mathrm{max}}^{q_\mathrm{max}} G_{20q}^2 (e) K_{\mathrm{I}}^{(2)} \left(2,\omega_{220q}^{(2)}\right) \\ + \frac{3}{2} \frac{G \, m_1}{a^3} \left(\frac{B-A}{C_2}\right) \left(\frac{1 + x_2}{2}\right)^2 \sum_{q=-q_\mathrm{max}}^{q_\mathrm{max}} G_{20q} (e) \sin \left[2\left(\omega_2 + \Omega_2 - \theta_2\right) + \left(2+q\right) M\right].
\end{multline}
\end{subequations} Evidently, the lack of angular brackets in the last equation is due to the presence of oscillatory terms originated by the triaxiality-induced torque, which does not average out in the neighborhood of spin-orbit resonances.

As can be observed, the equations of motion thus obtained contain no singularities in the limit of vanishing eccentricities and inclinations. However, we must stress the fact that this particular form of the equations of motion is intended to be translated to a computer code. In this sense, we were interested in the elimination of the singularities present in the equations for the time rate of the eccentricity and of the inclinations -- by defining the variables $\xi$ and $x   $  -- in order to avoid possible instabilities during the numerical integrations. By no means would we pretend to assert that through the aforementioned transformation, the arguments of pericenters and the longitudes of the ascending nodes become non-singular. These parameters still become meaningless when the eccentricities or the inclinations or both are zero. The only non-singular set of variables are the Cartesian-like ones \citep{correiaetal2014,boue_etal2016}.

Regarding the secular evolution of the system, only the values of the major semiaxis, the eccentricity, the inclinations, the argument of the pericenter, and the longitude of the ascending node defined on the planet's equator, as well as the rotation angle of the latter, are needed. The time derivatives of $M$, $\omega_1$, $\omega_2$, $\Omega_1$ and $\Omega_2$ are necessary to compute the tidal forcing frequencies by virtue of Eq.~\eqref{ec:chis}. It is a normal practice to directly use  Eq.~\eqref{ec:tidalmodesapprox}, but as we show later in this work, this approximation is not always valid.

We can summarize the principal characteristic of the dynamical model proposed, whose mathematical expression is given by Eqs.~\eqref{ec:ec_mov_orb_sec_mod} and \eqref{ec:ec_mov_rot_sec_mod}, as follows:
\begin{itemize}
      \item The major semiaxis, the eccentricity and the inclinations evolve in time strictly due to tidal effects as well as the longitude of the ascending node as seen from the star.
      \item The time evolution of the mean anomaly, the arguments of pericenters, and the longitude of the ascending node of the orbit seen from the planet is dominated by both the tidal interaction between both bodies and the $J_2$-related secular terms of the gravitational field of the planet.
      \item The stellar spin rate is dominated only by the secular terms of the corresponding tidal torque.
      \item The spin rate of the planet is driven by the combined action of the secular terms of the tidal and triaxiality-induced torques.
\end{itemize}

\section{\label{sec:prev}Preliminary discussion}

In this section, we briefly present the exoplanetary system studied in this work and we discuss the rheology of each component and the values of physical and rheological parameters we assumed for it, as well as the initial conditions for the numerical simulation we perform in this study, especially for those concerning the rotational evolution of the planet.

\subsection{The Kepler-21 system}

\begin{figure*}
 \sidecaption
  \input{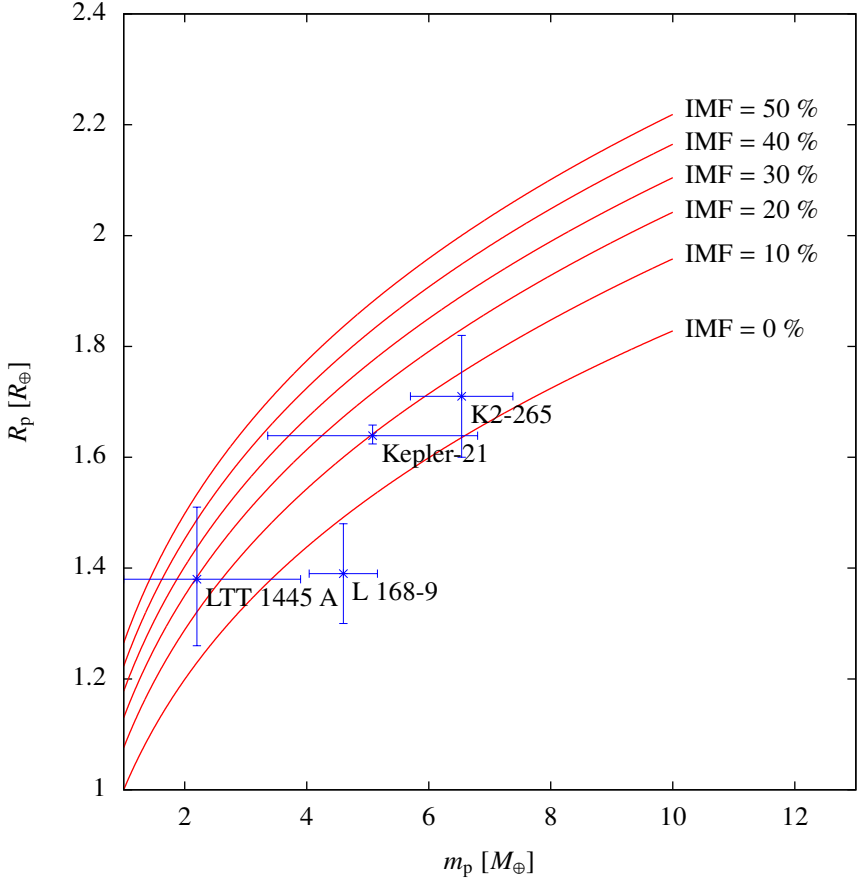}
  \caption{Radius (in Earth radii) vs. mass (in Earth masses) of some super-Earth exoplanets, along with a graph of mass-radius relationships according to several values of the ice-mass fraction (IMF) from \citet{valenciaetal2007c}.}
  \label{fig:relmrst} 
\end{figure*}

The Kepler-21 star, also known as \object{HD 179070}, is a F6IV sub-giant located at a distance of 352 light-years in the Lyra constellation \citep{howelletal2012}. In Table~\ref{tab:physicaldata}, we show the orbital, planetary, and stellar parameters relevant to our work. From the mass and radius of the planet, it can be inferred that it has rocky composition, with a mean density of approximately $6.4 \pm 2.1 \, \mbox{g} \, \mbox{cm}^{-3}$ \citep{LopezMorales_etal_2016}. In Fig.~\ref{fig:relmrst}, we show the mass-radius relationships according to several ice-mass fraction values (which is a measure of the water content of the planet) and considering a planetary composition similar to that of the Earth \citep{valenciaetal2007c}. It also shows the corresponding position of some super-Earths on the graph. The Earth would be placed at the origin. As may be noted, Kepler-21b is expected to have an Earth-like composition with about 10 \% made up of water.

\begin{table}[t]
  \centering
  \caption{\label{tab:physicaldata}Physical and orbital parameters of Kepler-21 system studied in this work \citep{LopezMorales_etal_2016}.}
  \begin{tabular}{ccc}
  \hline\hline
  Parameter & Symbol & Value \\
  \hline
  \multicolumn{3}{c}{Stellar parameters} \\
  \hline
  Mass ($M_{\odot}$)           & $m_1$                & $1.408$ \\
  Radius ($R_{\odot}$)         & $R_1$                & $1.902$ \\
  Rotational period (Days)     & $P_{\mathrm{rot,1}}$ & $12.6$ \\
  \hline
  \multicolumn{3}{c}{Planetary parameters} \\
  \hline
  Mass ($M_{\oplus}$)          & $m_2$                & $5.10$ \\
  Radius ($R_{\oplus}$)        & $R_2$                & $1.639$ \\
  \hline
  \multicolumn{3}{c}{Orbital parameters} \\
  \hline
  Orbital period (Days)        & $P_{\mathrm{orb}}$   & $2.78578$ \\
  Major semiaxis (\textsc{au}) & $a$                  & $0.0427172$ \\
  Eccentricity                 & $e$                  & $0.02$ \\
  \hline
  \end{tabular}
\end{table}

We selected this system based on two main reasons: the proximity of the components suggest that tidal effects are strong at least on the planet and all the relevant physical and orbital parameters are well-determined.

\subsection{The quadrupolar approximation}

We  mentioned in Sect.~\ref{sec:eom} that we want to limit the expansion of the tidal potentials of both the star and the planet up to $l=2$. Despite the fact that the quadrupolar approximation is widely used in the literature of tidal dynamics, such as in studies of the Earth-Moon system \citep{mignard1979,mignard1980}, the Mars-Phobos system \citep{efrolai2007}, and of close-in binary systems \citep{correiaetal2011,correiaetal2014}, we still need to verify if it is secure to use that approximation for the case of the Kepler-21 system as well.

Taking into account Table~\ref{tab:physicaldata}, we have for the Kepler-21 system: $R_1 / a \approx 0.21$ and $R_2 / a \approx 0.0016$. The first of these values seems to be somewhat high. However, we have to keep in mind that in order to neglect the $l=3$ term we have to compare it with the $l=2$ one. Recalling that the $l$-th term of the tidal potential scales as $(R/a)^{2l+1}$, starting with $l=2$ as is shown by Eqs.~\eqref{ec:pottidalplanetlmpqhs} and \eqref{ec:pottidalstarlmpqhs}, for the case of the star we have $(R_1/a)^5 \approx 3.8 \times 10^{-4}$ and $(R_1/a)^7 \approx 1.63 \times 10^{-5}$; while for the case of the planet: $(R_2/a)^5 \approx 1.71 \times 10^{-14}$ and $(R_2/a)^7 \approx 3.13 \times 10^{-20}$. Thus, it seems that it could be necessary to consider terms up to $l=3$ in the expansion of the tidal potential of the star. However, in order to avoid mathematical and numerical complications in our model, we still limit ourselves to the quadrupolar approximation and consider that the order of magnitude of difference between the $l=2$ and the $l=3$ terms of the tidal potential of the star is enough to neglect the latter.

\subsection{Rheology of the components}

Because the tidal theory we adopt in this work is applicable to binary systems, we find it useful to present a discussion of  the rheology of each component. In particular, we consider those that are conformed by a rocky planet and its host star. The reason we limit our study to Earth-like planets is twofold: On one hand, most recent developments in the field of planetary detection methods and their accuracy have led to the discovery of such worlds and, consequently, there is a growing interest in researching whether these celestial bodies offer the conditions that makes life possible. On the other hand, if we can infer that the compositions of the studied planets are similar to that of the Earth, then we can assume that their rheology is also similar and then propose reliable models to describe its internal dissipation.

In Table~\ref{tab:physicaldata}, we show the values of the physical parameters involved in the simulations we performed, that is, the mass and the radius of both components together with those of the major semiaxis and the eccentricity of the system we have selected for applying the model described in Sect.~\ref{sec:eom}, as well as the rotational period of the star and the orbital period of the planet. In what follows, we first discuss the rheology of the host star and then the rheology of the planet.

\subsubsection{Rheology of Sun-like stars}

As discussed by \citet{verasetal2019} stars can be considered to be viscous bodies, despite the fact that they can present some magnetic rigidity. Therefore, the dominant rheological parameter is the shear viscosity $\eta_{1}$. From the work by \citet{efroimsky2012} we know that the complex compliance corresponding to viscous deformation is given by: 
\begin{equation}\label{ec:complexcompvisc} 
\bar{J} (\chi) = - \frac{\mathrm{i}}{\eta_1 \, \chi}. 
\end{equation} Then, we have that its real and imaginary parts evidently are: 
\begin{equation} 
\Re \left[\bar{J} (\chi)\right] = 0 \quad \mathrm{and} \quad \Im \left[\bar{J} (\chi)\right] = - \frac{1}{\eta_{1} \, \chi}. 
\end{equation} As a consequence, the quality functions of the star $K_\mathrm{I}^{(1)} \left(l,\omega^{(1)}_{lmpq}\right)$ and $K_\mathrm{R}^{(1)} \left(l,\omega^{(1)}_{lmpq}\right)$, given in general form by Eq.~\eqref{ec:reklJ} and \eqref{ec:imklJ}, respectively, become \citep{efroimsky2015}: 
\begin{subequations}
\begin{equation}\label{ec:reklJs_visc} 
K_\mathrm{R}^{(1)} \left(l,\omega^{(1)}_{lmpq}\right) = \frac{3}{2} \frac{1}{(l-1)} \frac{1}{1 + \left(B_l \, \eta_1 \, \omega^{(1)}_{lmpq} \right)^2}
\end{equation} and
\begin{equation}\label{ec:imklJs_visc} 
K_\mathrm{I}^{(1)} \left(l,\omega^{(1)}_{lmpq}\right) = \frac{3}{2} \frac{1}{(l-1)} \frac{B_l \, \eta_{1} \, \omega^{(1)}_{lmpq}}{1 + \left(B_l \, \eta_1 \, \omega^{(1)}_{lmpq} \right)^2} \mathrm{sign} \left(\omega_{lmpq}^{(1)}\right).
\end{equation}
\end{subequations}

\subsubsection{Rheology of Earth-like planets}

The rheological response of rocky and icy celestial bodies like planets and satellites -- including the possibility that they can have a subsurface ocean -- is best modeled by the Maxwell-Andrade model, or even better, perhaps, by the \citet{sundcoop2010} model, as asserted by \citet{renaudhenning2018}. Due to the complexity of the latter model and the fact that, in general, the rheological parameters are poorly constrained even for the Earth, we chose the former rheological model to describe the dissipation within the planet. The complex compliance for this rheology is \citep{efroimsky2012,efroimsky2015}: 
\begin{equation}\label{ec:complexcompand} 
\bar{J} (\chi) = J \left[ 1 - \frac{\mathrm{i}}{\chi \, \tau_{\mathrm{M}}} + \left(\mathrm{i} \, \chi \, \tau_{\mathrm{A}}\right)^{-\alpha} \Gamma (1 + \alpha) \right], 
\end{equation} where $\alpha$ is known as the Andrade parameter, $\Gamma$ is the gamma function, $\tau_{\mathrm{M}}$ and $\tau_{\mathrm{A}}$ are the Maxwell and Andrade characteristic time scales, respectively. It worthwhile indicating that $\tau_{\mathrm{M}}$ is defined as the ratio of the shear viscosity over the mean rigidity $\mu$, that is: \begin{equation}\label{ec:def_tauM} \tau_{\mathrm{M}} = \frac{\eta}{\mu}. \end{equation} It should be noted that the first two terms in the first line of Eq.~\eqref{ec:complexcompand} correspond to a viscoeslastic response which is described by the Maxwell rheology, while the last term corresponds to an hereditary reaction caused by dislocation process within the material under stress \citep{efroimsky2012}, as prescribed in Eq.~\eqref{ec:kernJgral}. Andrade's rheology can be applied to a wide range of materials, from rocky planets or satellites partially molten up to icy bodies, even if these bodies have a subsurface ocean. The Andrade's parameter ranges between $[0.15,0.2]$ for the former kind of celestial bodies and between $[0.2,0.4]$ for the latter.

The real and imaginary parts of the complex compliance given by Eq.~\eqref{ec:complexcompand} are: 
\begin{equation} 
\begin{split} 
\Re \left[\bar{J} (\chi) \right] &= \frac{J}{\chi \, \tau_{\mathrm{M}}} \left[ \chi \, \tau_{\mathrm{M}} + \chi^{1 - \alpha} \tau_{\mathrm{M}} \tau_{\mathrm{A}}^{-\alpha} \cos \left(\frac{\pi \alpha}{2}\right) \Gamma (1 + \alpha) \right] \\
                                 &= \frac{J}{\chi \, \tau_{\mathrm{M}}} \mathcal{R} (\chi) 
\end{split} 
\end{equation} and 
\begin{equation} 
\begin{split} 
\Im \left[\bar{J} (\chi) \right] &= - \frac{J}{\chi \, \tau_{\mathrm{M}}} \left[ 1 + \chi^{1 - \alpha} \tau_{\mathrm{M}} \tau_{\mathrm{A}}^{-\alpha} \sin \left(\frac{\pi \alpha}{2}\right) \Gamma (1 + \alpha) \right] \\
                                 &= \frac{J}{\chi \, \tau_{\mathrm{M}}} \mathcal{I} (\chi). 
\end{split} 
\end{equation} Then, again by the virtue of Eqs.~\eqref{ec:reklJ} and \eqref{ec:imklJ}, the quality functions of the planet $K_\mathrm{R}^{(2)} \left(l,\omega^{(2)}_{lmpq}\right)$ and $K_\mathrm{I}^{(2)} \left(l,\omega^{(2)}_{lmpq}\right)$, respectively, are given by: 
\begin{subequations}
\begin{equation}\label{ec:reklJp_and} 
K_\mathrm{R}^{(2)} \left(l,\omega^{(2)}_{lmpq}\right) = \frac{3}{2} \frac{1}{l-1} \frac{\left[\mathcal{R} \left(\omega^{(2)}_{lmpq}\right) + B_l \, \omega^{(2)}_{lmpq} \, \tau_{\mathrm{M}}\right] \, \mathcal{R} \left(\omega^{(2)}_{lmpq}\right) + \mathcal{I}^2 \left(\omega^{(2)}_{lmpq}\right)}{\left[\mathcal{R} \left(\omega^{(2)}_{lmpq}\right) + B_l \, \omega^{(2)}_{lmpq} \, \tau_{\mathrm{M}}\right]^2 + \mathcal{I}^2 \left(\omega^{(2)}_{lmpq}\right)}
\end{equation} and
\begin{equation}\label{ec:imklJp_and} 
K_\mathrm{I}^{(2)} \left(l,\omega^{(2)}_{lmpq}\right) = - \frac{3}{2} \frac{1}{l-1} \frac{B_l \, \omega^{(2)}_{lmpq} \, \tau_{\mathrm{M}} \, \mathcal{I} \left(\omega^{(2)}_{lmpq}\right)}{\left[\mathcal{R} \left(\omega^{(2)}_{lmpq}\right) + B_l \, \omega^{(2)}_{lmpq} \, \tau_{\mathrm{M}}\right]^2 + \mathcal{I}^2 \left(\omega^{(2)}_{lmpq}\right)} \mathrm{sign} \left(\omega_{lmpq}^{(2)}\right). 
\end{equation}
\end{subequations}

\subsection{Initial conditions and physical parameters}

Apart from the values of the parameters given in Table \ref{tab:physicaldata} we must, on one hand, set some other physical and rheological parameters and, on the other hand, choose the values of the initial conditions of the orbital and rotational parameters in order to solve the differential equations of motion.

With regard to the rheological parameters of the host star and the planet, we only have  to set the value of the shear viscosity $\eta_1$ of the former and the shear rigidity $\mu$, the Maxwell time $\tau_{\mathrm{M}}$, and the Andrade's parameter $\alpha$ of the latter. Because we lack a better model, we set $\tau_{\mathrm{A}} = \tau_{\mathrm{M}}$. In the first case, we take the value of the shear viscosity of the host star to be that estimated by \citet{verasetal2019} for the Sun and Sun-like stars, which is $\eta_{1} \approx 10^{12} \, \mathrm{Pa} \, \mathrm{s}$. In the second, based on Fig.~\ref{fig:relmrst}, we can expect that the considered planet has a composition similar to that of the Earth and, therefore, we can take the corresponding values of its rheological parameters. For the Earth, $\mu = 0.8 \times 10^{11}$ Pa, which is the most reliable value among the rheological parameters involved. Keeping in mind both Eq.~\eqref{ec:def_tauM} and the fact that the viscosity depends on the temperature, such that the former decreases when the latter increases, we then have to expect that the Maxwell time  decreases as the planet's internal temperature increases. At the same time, planets that are internally differentiated, such as the Earth, with partially molten cores, are expected to have the corresponding values of Andrade's parameter, that is, $\alpha \in [0.15,0.2]$. Thus, due to this uncertainty, here we perform numerical simulations with different values of the less constrained parameters as follows: $\tau_{\mathrm{M}} = 10, 50, 100$ years and $\alpha = 0.15, 0.2, 0.3$.

Concerning the rest of the physical parameters, the values of the mean densities and surface gravity of both the planet and the host star are computed using the data given in Table~\ref{tab:physicaldata}. The maximum moments of inertia $C_1$ and $C_2$ of the star and the planet, respectively, are computed with: 
\begin{equation}\label{ec:calc_C} 
C_{k} = \xi_k \, m_{k} \, R_{k}^2,
\end{equation} where $\xi_k$ is called the moment of inertia factor. For a homogeneous solid sphere, its value is $2/5$. But for internally differentiated rocky planets, its value is a little lower. In this sense, \citet{zeng_2017} showed that for both the Earth and Earth-like rocky planets, its value is approximately equal to $1/3$. In consequence, we can take $\xi_1 = 2/5$ and $\xi_2 = 1/3$. The last physical parameter to be set is the degree of triaxiality, indicating that we need a value for $(B-A)/C$. Such a parameter can be estimated following the work by \citet{rodriguezetal2012}, namely: 
\begin{equation}\label{ec:est_triax_deg} 
\frac{B-A}{C_2} \approx \frac{15}{4} \frac{m_1}{m_2} \left(\frac{R_2}{a}\right)^3.
\end{equation} For the Kepler-21 system, we obtain $7 \times 10^{-4}$. However, in order to evaluate the effect of this parameter on the rotational evolution, we have also set $(B-A)/C_2 = 10^{-3}$ and zero.

With respect to the values of $J_2$, we have adopted three values, namely: $10^{-3}$, which is on the same order of magnitude of that of the Earth, along with $10^{-4}$, $10^{-5}$ and zero.

Lastly, regarding the orbital parameters we have set the initial values of $a$ and $e$ as those given by Table~\ref{tab:physicaldata}, correspondingly, as we are interested in the study of the fate of the Kepler-21 system and not in its past values. In this sense, we set $i_1 (0) = 0.5 \degr$ and $i_2 (0) = 5 \degr$ in order to estimate planarization and obliquity damping timescales.

Of course, the rotational state of the planet is completely unknown. However, it is believed that rocky planets rotate very fast after their formation \citep{miguelybrunini2010}. For exploratory purposes, we set $\theta_2 (0) = 0$ and $\dot{\theta}_2 / n (0) = 4.1$ for the planet. The reason behind the selection of this particular value for the planet's initial spin rate is soon to become clear. As we shall see, it is very likely that the planet is already captured in synchronous rotation. The initial spin rate of the star is computed from its current estimated rotation period, which is given in Table~\ref{tab:physicaldata}.

\section{\label{sec:charact_dynamical_ev}Semianalytical and numerical characterization of the dynamical evolution}

As we have already pointed out, the methodology developed in this work was applied to the study of the dynamical evolution of the Kepler-21 exoplanetary system, which is known to consist of a planet -- presumably a rocky one -- and its host star. Its orbital parameters and certain physical parameters are shown in Table~\ref{tab:physicaldata}.

We performed 13 numerical experiments varying the rheological and dynamical parameters of the planet because, as we have pointed out, these parameters are the less constrained ones. In these experiments the equations of motion -- Eqs.~\eqref{ec:dadt_sec_mod}, \eqref{ec:dxidt_sec_mod}, \eqref{ec:dMdt_sec_mod}, \eqref{ec:dx1dt_sec_mod}, \eqref{ec:dx2dt_sec_mod}, \eqref{ec:dw1dt_sec_mod}, \eqref{ec:dw2dt_sec_mod}, \eqref{ec:dOm1dt_sec_mod}, \eqref{ec:dOm2dt_sec_mod}, \eqref{ec:d2th1dt2_sec_mod}, and \eqref{ec:d2th2dt2_tt_mod} -- were numerically integrated over a time span of $10^{5}$ years. The adopted values of the rheological and dynamical parameters in each integration (labeled with an ID number) are shown in Table~\ref{tab:int_params}. The aforementioned integrations were divided into three sets: in the first one, only the rheological parameters of the planet ($\alpha$ and $\tau_M$) were varied; in the second one, only the dynamical parameters of the latter were varied ($(B-A)/C_2$ and $J_2$); and in the third one, we only performed a numerical simulation setting $e=0.1$, which is within the uncertainty of the eccentricity value (see Table~\ref{tab:physicaldata}) in order to explore the possibility of entrapment in higher-than-synchronous spin-orbit resonances.
As a general check, we verified that the total angular momentum of the systems is conserved in all the simulations we performed.

\begin{table}[tbp]
\caption{Values of the rheological and dynamical parameters adopted in the numerical integrations performed in this work, together with the corresponding ID number assigned to ease the identification of the result of each simulation.}
\begin{center}
\begin{tabular}{ c c c c c c c }
\hline
\hline
 & ID & $e$ & $\tau_M$ [yr] & $\alpha$ & $(B-A)/C_2$ & $J_2$ \\ 
\hline
      &  1 & $0.02$ &  $10$ &  $0.2$ & $7 \times 10^{-4}$ & $1 \times 10^{-3}$ \\
      &  2 & $0.02$ &  $50$ &  $0.2$ & $7 \times 10^{-4}$ & $1 \times 10^{-3}$ \\
Set 1 &  3 & $0.02$ & $100$ &  $0.2$ & $7 \times 10^{-4}$ & $1 \times 10^{-3}$ \\
      &  4 & $0.02$ &  $10$ & $0.15$ & $7 \times 10^{-4}$ & $1 \times 10^{-3}$ \\
      &  5 & $0.02$ &  $10$ &  $0.3$ & $7 \times 10^{-4}$ & $1 \times 10^{-3}$ \\
\hline
      &  6 & $0.02$ &  $10$ &  $0.2$ & $1 \times 10^{-3}$ & $1 \times 10^{-3}$ \\
      &  7 & $0.02$ &  $10$ &  $0.2$ & $1 \times 10^{-5}$ & $1 \times 10^{-3}$ \\
      &  8 & $0.02$ &  $10$ &  $0.2$ & $7 \times 10^{-4}$ & $1 \times 10^{-4}$ \\
Set 2 &  9 & $0.02$ &  $10$ &  $0.2$ & $7 \times 10^{-4}$ & $1 \times 10^{-5}$ \\
      & 10 & $0.02$ &  $10$ &  $0.2$ & $0$                & $1 \times 10^{-3}$ \\
      & 11 & $0.02$ &  $10$ &  $0.2$ & $7 \times 10^{-4}$ & $0$ \\
      & 12 & $0.02$ &  $10$ &  $0.2$ & $0$                & $0$ \\
\hline
Set 3 & 13 &  $0.1$ &  $10$ &  $0.2$ & $7 \times 10^{-4}$ & $1 \times 10^{-3}$ \\
\hline
\end{tabular}
\end{center}
\label{tab:int_params}
\end{table}

\subsection{Results from the numerical simulations\label{ss:numint}}

\begin{figure}
 \centering
 \begin{subfigure}[b]{0.49\textwidth}
  \centering
  \input{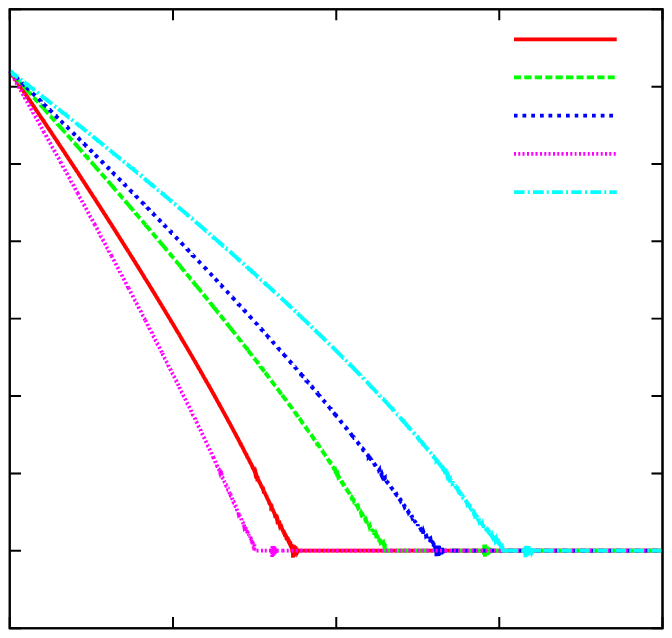}
  \caption{\label{fig:thpn2vst_set1}}
 \end{subfigure}
~
 \begin{subfigure}[b]{0.49\textwidth}
  \centering
  \input{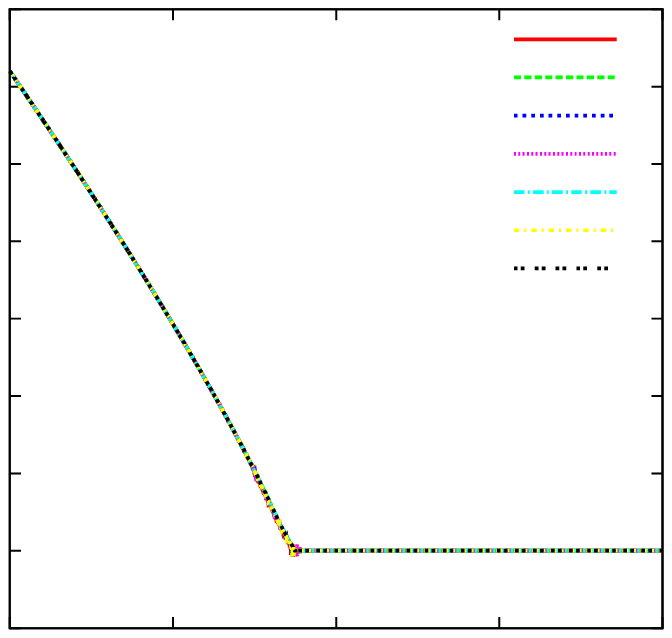}
  \caption{\label{fig:thpn2vst_set2}}
 \end{subfigure}
 \caption{Normalized spin rate of the planet vs. time as resulted from the first set of numerical simulations (\subref{fig:thpn2vst_set1}) and from the second one (\subref{fig:thpn2vst_set2}).}
\label{fig:thpn2vst_set1y2} 
\end{figure}

We focus first on the rotational evolution of the system. In Fig.~\ref{fig:thpn2vst_set1y2} we show the graphs of the normalized spin rate of the planet obtained from the first (Fig.~\ref{fig:thpn2vst_set1}) and second (Fig.~\ref{fig:thpn2vst_set2}) sets of integrations. It can be noted from Fig.~\ref{fig:thpn2vst_set1} that the time necessary to reach the synchronous rotation is greatly affected by the variation of the rheological parameters, while the modification of the dynamical parameter have no consequences in this sense, as can seen in Fig.~\ref{fig:thpn2vst_set2}. However, we found that the synchronism is not always exact. In this sense, we are observing the influence of apsidal and nodal precession rates, as seen from the planet, on the corresponding tidal forcing frequencies, which were computed using the exact expression given by Eq.~\eqref{ec:chis} in contraposition to the widely used approximation presented in Eq.~\eqref{ec:tidalmodesapprox}.

\begin{figure}[t]
  \sidecaption
  \input{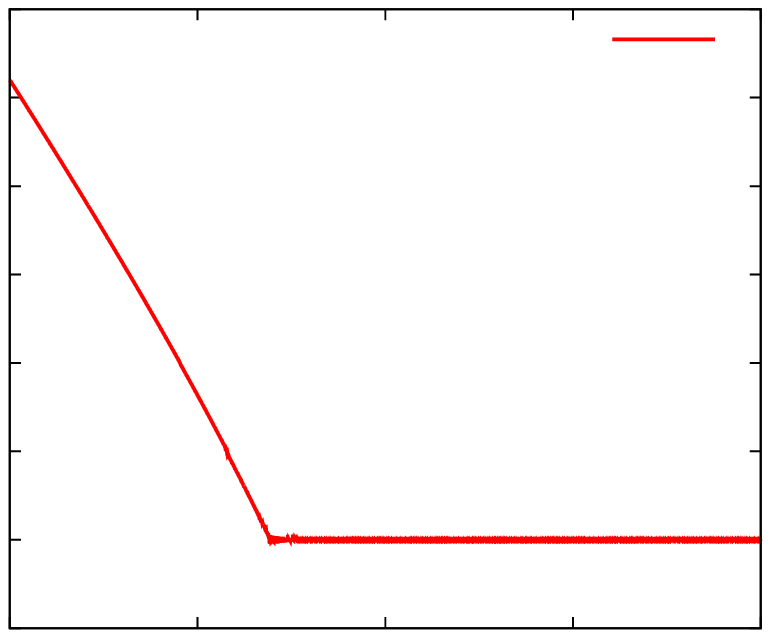}
  \caption{Normalized spin rate of the planet vs. time as result of the third set of numerical simulations (the one with $e=0.1$. We obtain the planet's results captured in the 3:2 spin-orbit resonance.}
  \label{fig:thpn2vst_set3} 
\end{figure}

\begin{figure} 
  \sidecaption
  \input{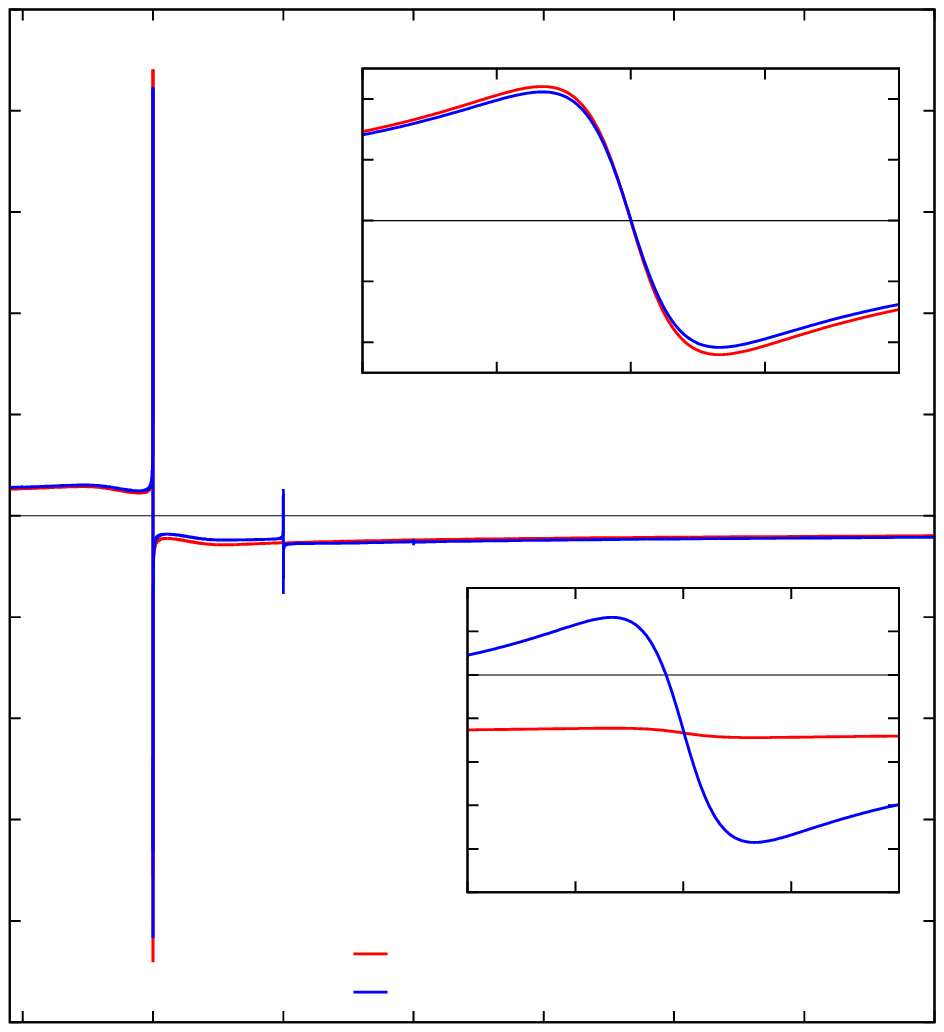}
  \caption{Angular acceleration of the planet due to the secular terms of the tidal torque vs. its normalized spin rate for two values of the eccentricity, the current $e = 0.02$ and $e = 0.1$. We zoomed in on two neighborhoods around the 1:1 and 3:2 spin-orbit resonances for clarity.}
  \label{fig:thppvsthpn} 
\end{figure}

The third set of numerical simulations consist only of an integration setting of $e=0.1$. It can be noted in Fig.~\ref{fig:thpn2vst_set3} that the planet is trapped in the 3:2 spin-orbit resonance as can be expected from Fig.~\ref{fig:thppvsthpn}, where we plot the angular acceleration of the planet due to the secular terms of the tidal torque vs. its normalized spin rate \citep{makefro2013}. In that figure, the solid blue line (which correspond to $e = 0.1$) crosses the horizontal reference axis near $\dot{\theta}_2/n = 1.5$. This is the reason why we pointed out in Sect. \ref{sec:intro} that eccentricity plays a key role in the possibility of capture in higher-than-synchronous resonances. This observation is in agreement with the results obtained by \citet{noyellesetal2014}.

In Fig.~\ref{fig:thppvsthpn} we also may find a justification of this choice in the initial value of the planet's spin rate (i.e., $\dot{\theta}_2 = 4.1 \, n$). As can be noted in the aforementioned figure, the angular acceleration of the planet -- due to the secular terms of the tidal torque -- is practically constant for $\dot{\theta}_2 / n \geq 4$ and thus, if we consider higher initial values of the planet's spin rate, the only difference would be that the latter ends up reaching the 1:1 spin-orbit resonance some time later.

\begin{figure}
 \centering
 \begin{subfigure}[b]{0.49\textwidth}
  \centering
  \input{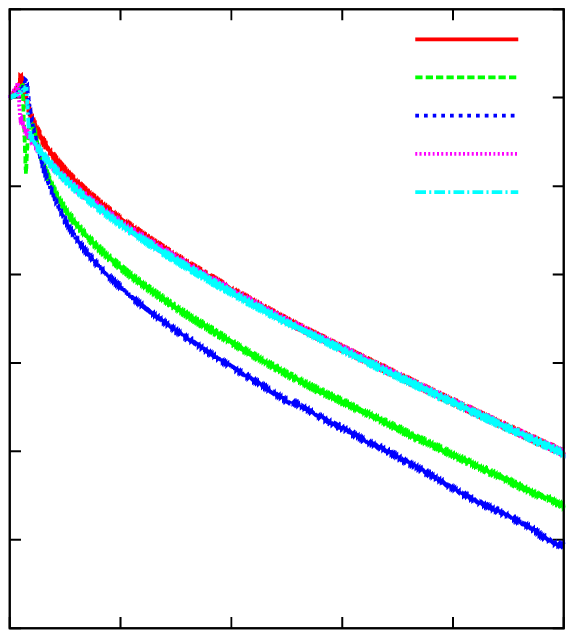}
  \caption{\label{fig:avst_set1}}
 \end{subfigure}
~
 \begin{subfigure}[b]{0.49\textwidth}
  \centering
  \input{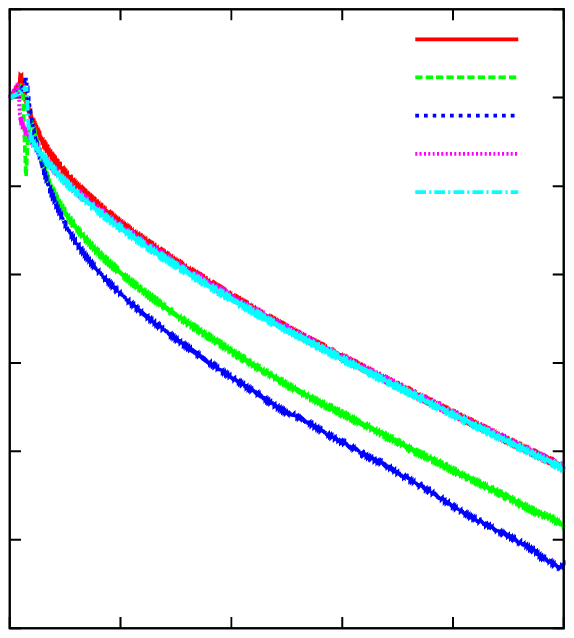}
  \caption{\label{fig:evst_set1}}
 \end{subfigure}
 \caption{Time evolution of $a-a_0$, $e-e_0$ (where $a_0$ and $e_0$ are the initial values of the major semiaxis and the eccentricity, respectively) as the result of the first set of numerical simulations, i.e., those varying the rheological parameters of the planet.} 
 \label{fig:evorb_set1}
\end{figure}

\begin{figure}
 \centering
 \begin{subfigure}[b]{0.49\textwidth}
  \centering
  \input{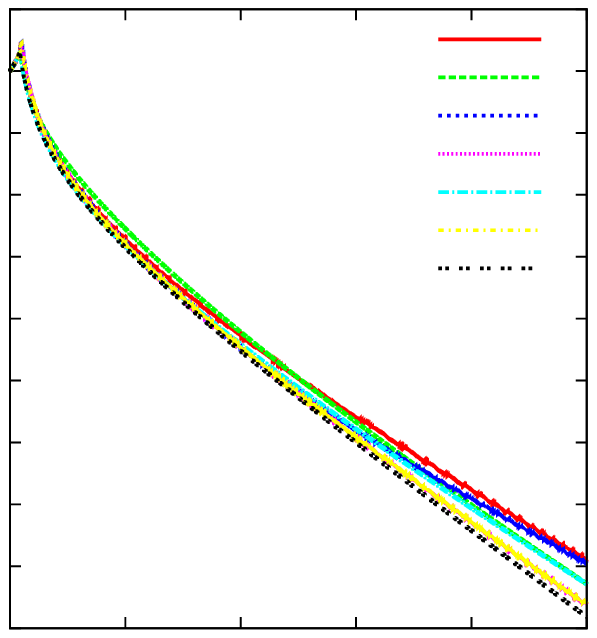}
  \caption{\label{fig:avst_set2}}
 \end{subfigure}
~
 \begin{subfigure}[b]{0.49\textwidth}
  \centering
  \input{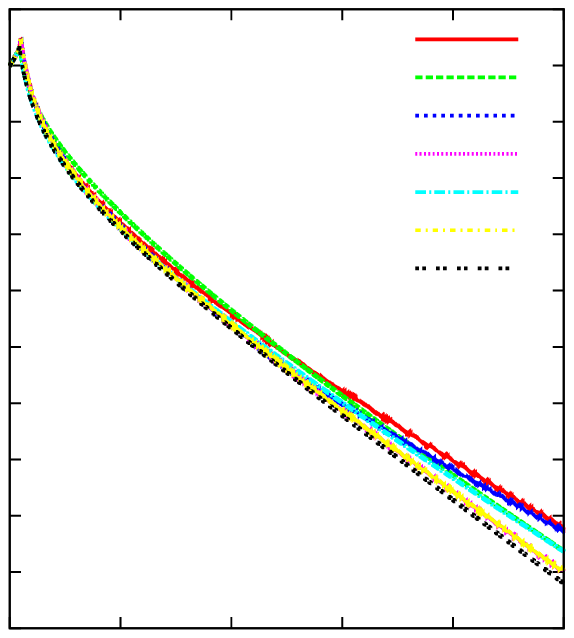}
  \caption{\label{fig:evst_set2}}
 \end{subfigure}
 \caption{Time evolution of $a-a_0$, $e-e_0$ (where $a_0$ and $e_0$ are the initial values of the major semiaxis and the eccentricity, respectively) as result from the second set of numerical simulations, i.e., those varying the dynamical parameters $J_2$ and $(B-A)/C_2$ of the planet.} 
 \label{fig:evorb_set2}
\end{figure}

\begin{figure}
  \sidecaption
  \input{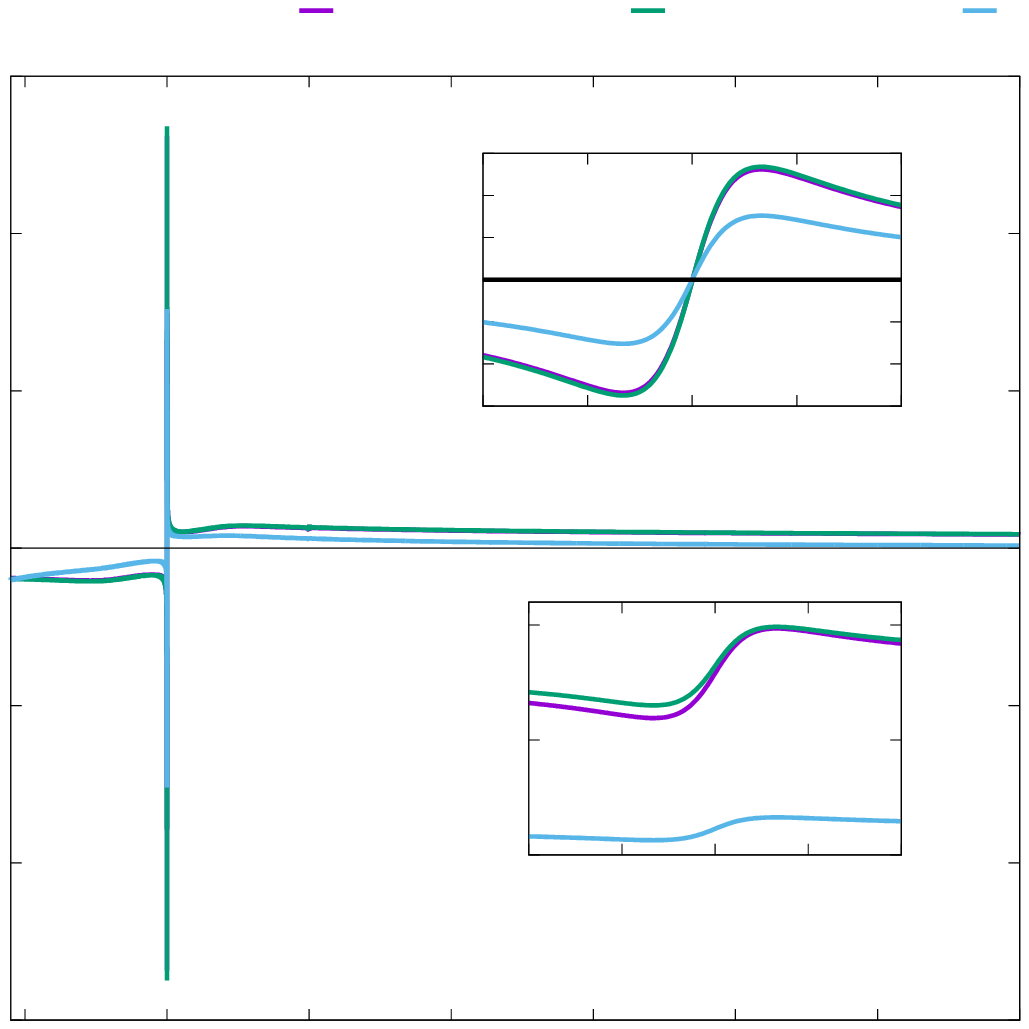}
  \caption{Time derivatives of $a$, $e,$ and $i_2$ due to the secular terms of the disturbing function vs. the normalized spin rate of the planet obtained by direct evaluation of the corresponding equations of motion.}
  \label{fig:apepipvsthpn} 
\end{figure}

Now, we shift the focus to the orbital evolution of the Kepler-21 system. As a general observation, it can be noted that the major semiaxis and the eccentricity evolved in time in a similar way in each simulation when compared to the others, only differing in the final value of the corresponding time rates, as noted in Figs~\ref{fig:evorb_set1} and \ref{fig:evorb_set2}. This similarity can be understood by examining Fig.~\ref{fig:apepipvsthpn}, where we can verify that the system behaves as expected from the same figure, provided that the planet's rotational evolutionary timescale is much shorter than the orbital one. In this sense, we can know in advance how the system is expected to evolve over time before performing any numerical integration. In the case of Kepler-21, we see that the major semiaxis, the eccentricity, and the obliquity grow in time up to the 1:1 spin-orbit resonance and then start decreasing. This result is also consistent with \citet{noyellesetal2014}, who found that Mercury's spin evolved so swiftly that the planet got trapped in its end-state only some 10-20 million of years after its accretion. 

\begin{figure}
 \centering
 \begin{subfigure}[b]{0.49\textwidth}
  \centering
  \input{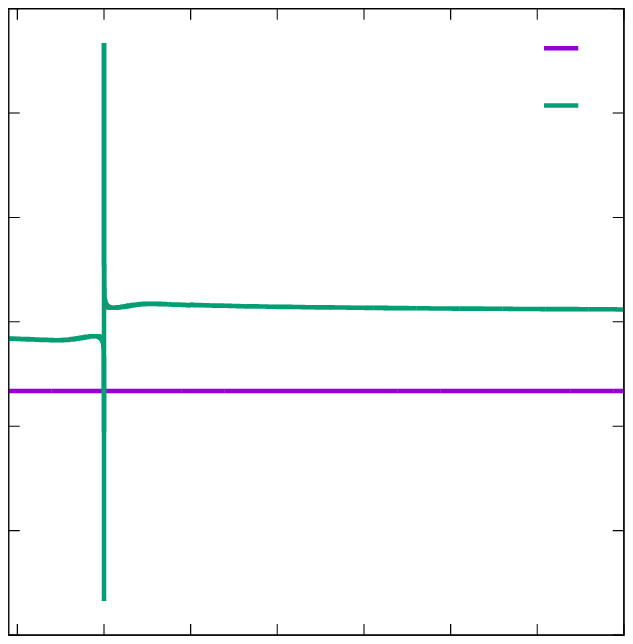}
  \caption{\label{fig:apvsthpn_comp}}
 \end{subfigure}
~
 \begin{subfigure}[b]{0.49\textwidth}
  \centering
  \input{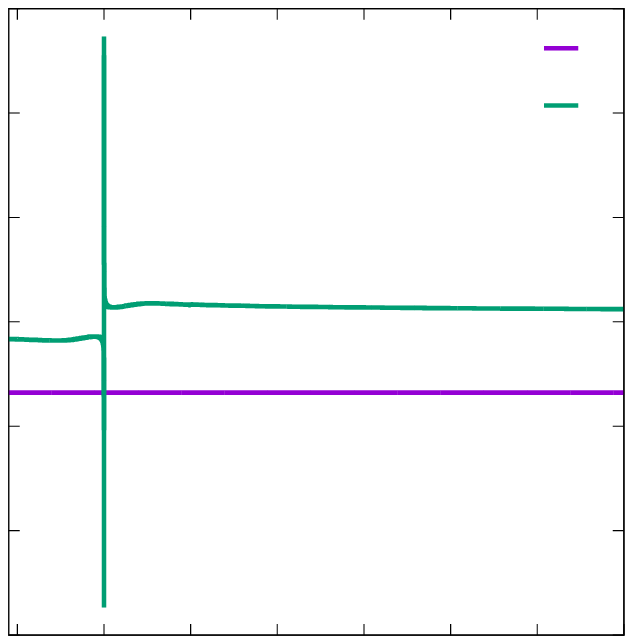}
  \caption{\label{fig:epvsthpn_comp}}
 \end{subfigure}
 \caption{Time derivatives of the major semiaxis (\subref{fig:apvsthpn_comp}) and the eccentricity (\subref{fig:epvsthpn_comp}) as a function of the normalized spin rate of the planet. In this figure, the contributions of the secular terms of the tidal potentials generated by the star and the planet are shown separately.} 
\label{fig:evincs_set2}
\end{figure}

Perhaps it would be interesting to see the individual contributions of the  tidal potentials of the star and of the planet onto the dynamics of the system that they are both a part of. In Figs.~\ref{fig:apvsthpn_comp} and \ref{fig:epvsthpn_comp}, we show the time derivatives of the major semiaxis and of the eccentricity, respectively, due to the secular terms of the star's tidal potential, labeled with superscript (1), and to the planet's tidal potential, labeled with a superscript (2). It can be observed that in both cases, the star's contribution is approximately four orders of magnitude weaker than that of the planet's.

\begin{figure}[t]
  \sidecaption
  \input{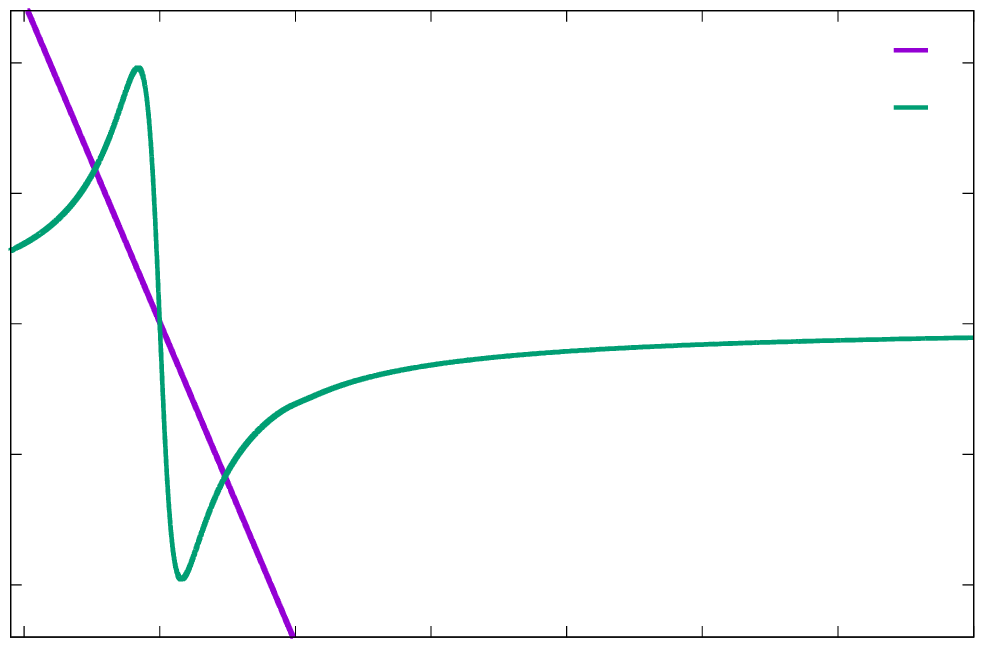}
  \caption{Time derivatives of $i_1$ vs. the normalized spin rate of the planet obtained by direct evaluation of Eq.~\eqref{ec:dx1dt_sec_mod}, considering two different values of the stellar viscosity.}
  \label{fig:i1pvsthpn} 
\end{figure}

Concerning the time evolution of the inclination ($i_1$), we obtained a seemingly counter-intuitive result, provided that its time derivative has a positive value of about $1.35 \times 10^{-10} \, \degr \, \mbox{yr}^{-1}$, which is practically independent of the planet's normalized spin rate. This implies that $$\frac{G \, m_1 \, m_2}{a \, C_1 \, \dot{\theta}_1} > \frac{n}{\xi}$$ for the case of the Kepler-21 system, as can be noted from Eqs.~\eqref{ec:dx1dt_sec_mod} and \eqref{ec:dxdt_def}. As we can also note in Eq.~\eqref{ec:dx1dt_sec_mod}, the time evolution of $i_1$ depends only on the stellar parameters. Thus, the time derivative of the orbital inclination can be plotted against the normalized spin rate of the star, as shown in Fig.~\ref{fig:i1pvsthpn}, where we consider two different values of the stellar viscosity. From Table~\ref{tab:physicaldata} we can verify that $\dot{\theta}_1 / n < 1$ and, as a consequence, the time derivative of $i_1$ is inevitably positive by virtue of the dynamical model developed in this work.  

\begin{figure}
 \centering
 \begin{subfigure}[b]{0.49\textwidth}
  \centering
  \input{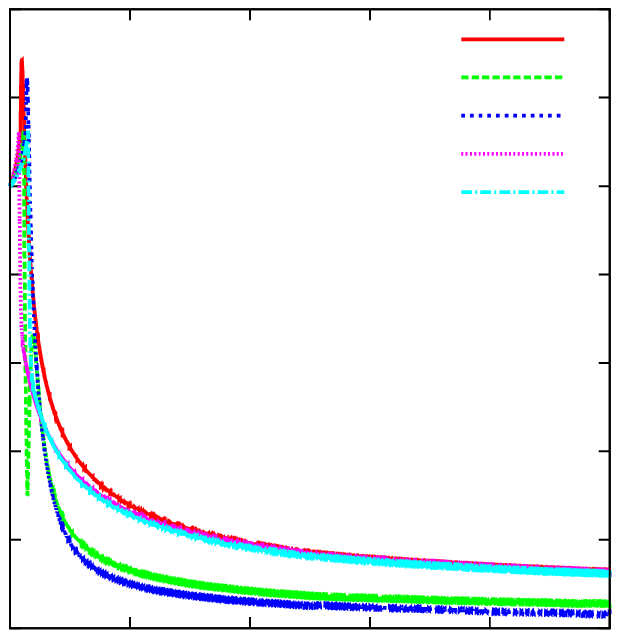}
  \caption{\label{fig:i2vst_set1}}
 \end{subfigure}
~
 \begin{subfigure}[b]{0.49\textwidth}
  \centering
  \input{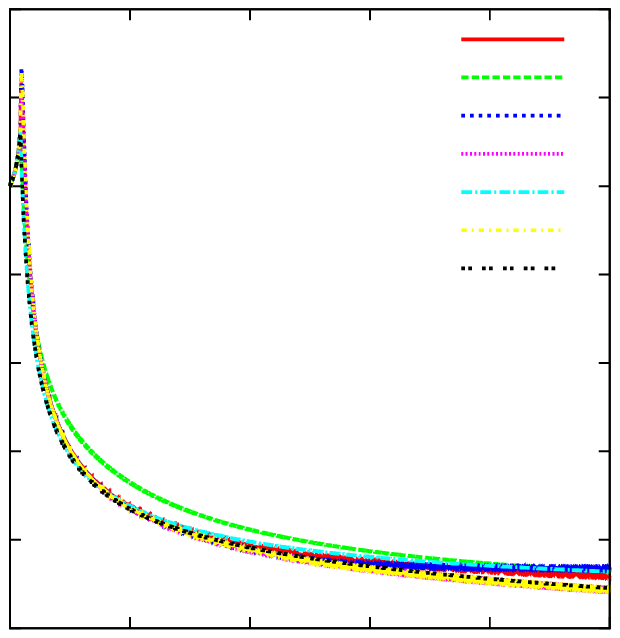}
  \caption{\label{fig:i2vst_set2}}
 \end{subfigure}
 \caption{Time evolution of the planet's obliquity ($i_2$) as result from the first (\subref{fig:i2vst_set1}) -- those varying the rheological parameters -- and second (\subref{fig:i2vst_set2}) -- i.e. those varying the dynamical parameters -- the set of numerical simulations.} 
\label{fig:evincs_set2}
\end{figure}

With respect to the obliquity ($i_2$), it begins to decrease exponentially after the planet gets trapped in the synchronous rotation, as can be observed in Figs.~\ref{fig:i2vst_set1} and \ref{fig:i2vst_set2}. By virtue of Fig.~\ref{fig:apepipvsthpn}, we can also know in advance the general behavior of the time evolution of the obliquity, at least qualitatively as in the case of the major semiaxis and the eccentricity discussed before.

\begin{figure}
  \sidecaption
  \input{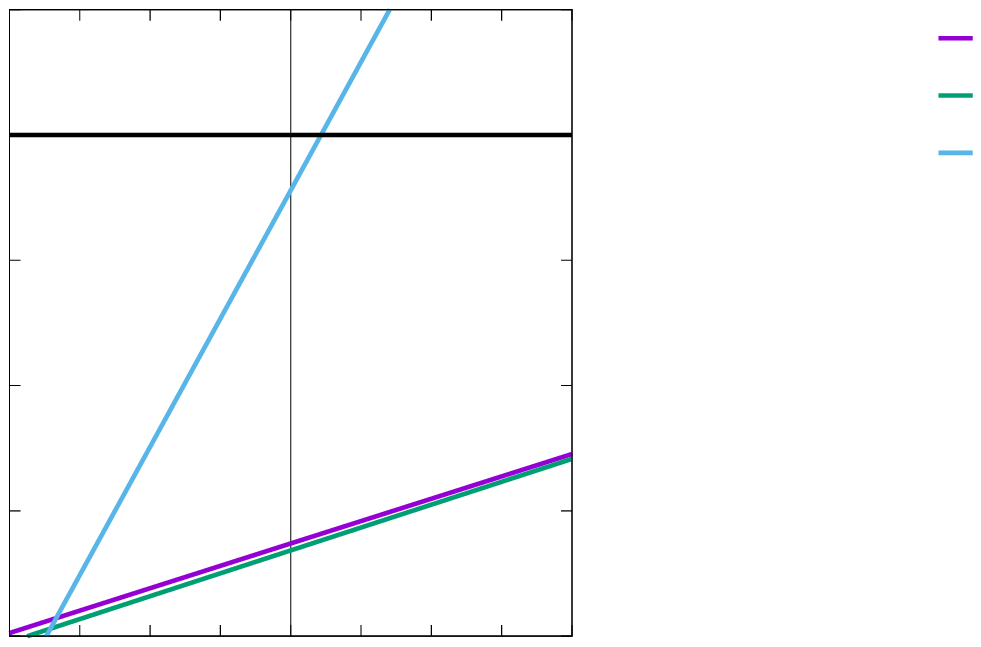}
  \caption{Again, the time derivatives of $a$, $e,$ and $i_2$ due to the secular terms of the disturbing function vs. the normalized spin rate of the planet obtained by direct evaluation of the corresponding equations of motion but in a narrow interval around the 1:1 spin-orbit resonance.}
  \label{fig:apepipvsthpn_zoom} 
\end{figure}

It may seem that in Fig.~\ref{fig:apepipvsthpn} the time derivatives of the major semiaxis, the eccentricity, and the obliquity are equal to zero when the planet is trapped in the 1:1 spin-orbit resonance but, as seen in Figs.~\ref{fig:evorb_set1}, \ref{fig:evorb_set2}, and \ref{fig:evincs_set2}, these orbital elements decrease in time after the planet has reached the synchronous rotation. There are two reasons behind this fact. On one hand, as pointed out before, the synchronism is not exact. On the other hand, if we take a look at a very narrow interval around the 1:1 spin orbit resonance in Fig.~\ref{fig:apepipvsthpn}, as shown in Fig.~\ref{fig:apepipvsthpn_zoom}, we note that the time derivatives of $a$, $e$ and $i_2$ are not exactly zero for $\dot{\theta}_2 = n$.

\begin{figure}
 \sidecaption
 \input{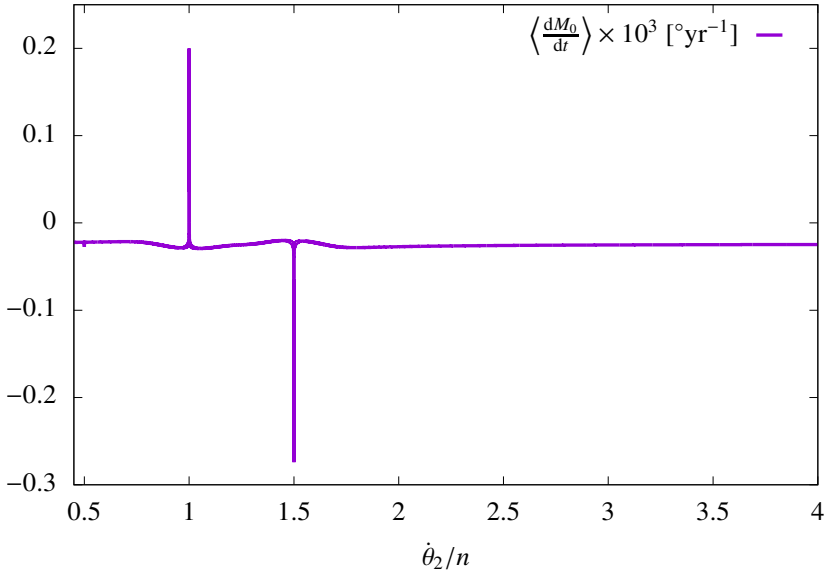}
 \caption{Time derivative of the mean anomaly at the epoch vs. the normalized spin rat of the planet.}
 \label{fig:Mpvsthpn} 
\end{figure}

\begin{figure}
 \centering
 \begin{subfigure}[b]{0.49\textwidth}
  \centering
  \input{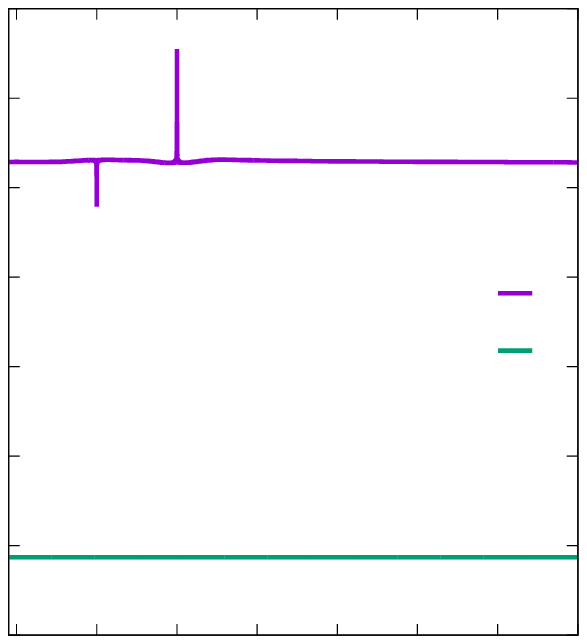}
  \caption{\label{fig:wpomp1vsthpn_comp}}
 \end{subfigure}
~
 \begin{subfigure}[b]{0.49\textwidth}
  \centering
  \input{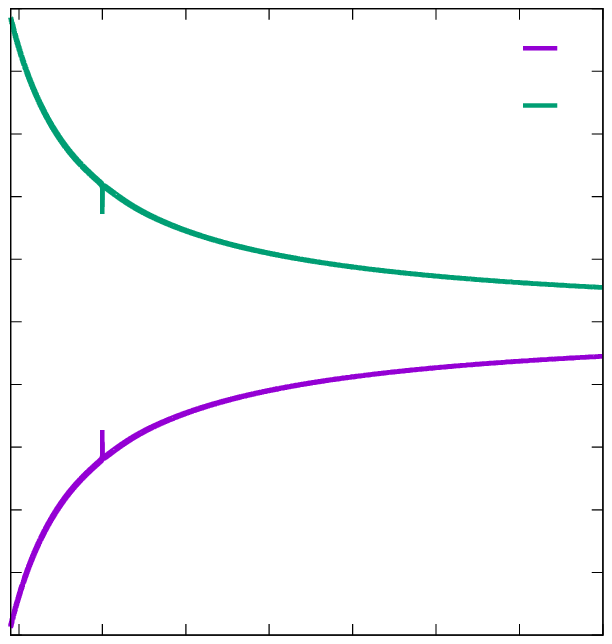}
  \caption{\label{fig:wpomp2vsthpn_comp}}
 \end{subfigure}
 \caption{Time rate of the apsidal and nodal precession, as seen from the star (\subref{fig:wpomp1vsthpn_comp}) and from the planet (\subref{fig:wpomp2vsthpn_comp}), vs. the normalized spin rate of the planet.} 
\label{fig:comp_wpompvsthpn}
\end{figure}

As a final comment on the numerical simulations, it should be noted that this kind of problem involves different timescales of the dynamical evolution. As noted in Figs.~\ref{fig:Mpvsthpn} and \ref{fig:wpomp1vsthpn_comp}, the time rates of the mean anomaly at the epoch, the argument of pericenter, and the longitude of the ascending node, both defined on the stellar equator, are on the order of $10^{-4}$; while the apsidal and nodal precessions as seen from the planet are on the order of $10^1$--$10^{2}$ (Fig.~\ref{fig:wpomp2vsthpn_comp}). Then it is necessary for the numerical integration scheme to treat these timescales separately. 

Taking into account that $n \approx 4.8 \times 10^{4} {}\degr \mbox{yr}^{-1}$, as well, we then have that $\dot{M} \approx n$ and, in consequence, we consider a time step of $10^{-3} P_{\mathrm{orb}}$ ($\Delta t_\mathrm{fast}$) to integrate $\theta_2$, $M$, $\omega_2$, and $\Omega_2$, and a time step of one year ($\Delta t_\mathrm{slow}$) for the rest of the dynamical parameters, that is, $a$, $e$, $i_1$ and $i_2$.

The dynamical variables were integrated together but in different stages within the numerical integration scheme in order to implement the separation of the fast-varying and the slow-varying variables described before. More specifically, the former were integrated using fourth-order Runge-Kutta method ($\Delta t_\mathrm{slow}$), while the latter were integrated using the Bulirsch-Stoer method \citep{press1992} with the corresponding time step $\Delta t_\mathrm{fast}$.

The aforementioned integration scheme starts with an evaluation of the time derivatives of the slow-varying variables at the beginning of the integration interval, that is, the first stage of the Runge-Kutta method, for example, at $t_i$. Then the fast-varying variables are integrated using the Bulirsch-Stoer method up to $t_i + \Delta t_\mathrm{slow}/2. $ At that point, the second stage of the Runge-Kutta method is performed with the updated value of the planet's spin rate in order to evaluate properly the tidal frequencies and the time derivatives of the slow-varying variables. Immediately, the fast-varying variables are integrated again up to the end of the integration interval, that is, $t_i + \Delta t_\mathrm{slow}$ where the new value of $\dot{\theta}_2$ is taken to compute the new values of both the tidal frequencies and the slow-varying variables' time derivatives. Finally, the last stage of the Runge-Kutta method is carried out in order to compute the new values of the slow-varying variables; these are used to start the integration on the next interval together with the last computed values of the planet's spin rate and the rest of the fast-varying variables.

\subsection{\label{ss:timescales}Timescales of orbital decay and circularization}

In general terms, we define the characteristic timescale of the evolution of an orbital element as that which is required for a certain change in that element to take place. Since the orbital inclination and the obliquity of the planet evolve in the way described earlier in this work, we only consider the timescales that correspond to the major semiaxis and the eccentricity. Mathematically expressed, the characteristic timescale of the former, $\tau_{a}$, and the latter, $\tau_{e}$, are defined as: 
\begin{equation}\label{ec:def_tau_a} 
\tau_{a} = \left(r_{\mathrm{Roche}} - a \right) \left(\frac{\mathrm{d}a}{\mathrm{d}t}\right)^{-1}, 
\end{equation} 
\begin{equation}\label{ec:def_tau_e} 
\tau_{e} = - \, e \left(\frac{\mathrm{d}e}{\mathrm{d}t}\right)^{-1},
\end{equation} where $r_{\mathrm{Roche}}$ is the Roche limit and the values of $a$ and $e$ are the current ones. It can be noted that $\tau_{a}$ correspond to the time interval that elapses from the current value of $a$ up to the planet reaches the Roche radius, beyond which it would be completely destroyed by tidal forces. The latter is defined as \citep{murray1999}: 
\begin{equation}\label{ec:def_lim_roche} 
r_{\mathrm{Roche}} = R_{\mathrm{p}} \left( 3 \frac{m_{\mathrm{s}}}{m_{\mathrm{p}}}\right)^{\frac{1}{3}}.
\end{equation} Of course, this definition does not take into account the rigidity of the material that compose the body but only considers self-gravity. Even so, it can be considered as an upper limit. For the Kepler-21 system, $r_{\mathrm{Roche}} \approx 4.54 \times 10^{-3}$ \textsc{au}. 

\begin{table}[tbp]
\caption{Orbital decay and eccentricity damping timescales of the Kepler-21 system obtained through Eqs.~\eqref{ec:def_tau_a} and \eqref{ec:def_tau_e} by evaluating the corresponding equation of motion at $\dot\theta_2 / n = 1$ for different values of the rheological parameters and the eccentricity.}
\begin{center}
\begin{tabular}{ c c c c c c }
\hline
\hline
 $e$ & $\tau_M$ [yr] & $\alpha$ & $\tau_{a}$ [$\times 10^{6}$ yr] & $\tau_{e}$ [$\times 10^{6}$ yr] \\ 
\hline
 $0.01$ &  $10$ &  $0.2$ & $751.32$ & $0.101$ \\
 $0.02$ &  $10$ &  $0.2$ & $629.23$ & $0.345$ \\
 $0.02$ &  $50$ &  $0.2$ & $662.78$ & $0.361$ \\
 $0.02$ & $100$ &  $0.2$ & $675.81$ & $0.367$ \\
 $0.02$ &  $10$ & $0.15$ & $617.57$ & $0.339$ \\
 $0.02$ &  $10$ &  $0.3$ & $679.20$ & $0.368$ \\
 $0.1$  &  $10$ &  $0.2$ & $98.955$ & $1.542$ \\
\hline
\end{tabular}
\end{center}
\label{tab:timescales}
\end{table}

Replacing in Eqs.~\eqref{ec:def_tau_a} and \eqref{ec:def_tau_e} the corresponding values given in Table~\ref{tab:physicaldata} and that of $\left\langle \frac{\mathrm{d}a}{\mathrm{d}t} \right\rangle$ and $\left\langle \frac{\mathrm{d}e}{\mathrm{d}t} \right\rangle$ from Eqs.~\eqref{ec:dadt_sec_mod} and \eqref{ec:dxidt_sec_mod} evaluated at $\dot\theta_2 = n$ for different values of the rheological parameters, we obtain the values of orbital decay and circularization timescales. The results are shown in Table~\ref{tab:timescales} together with the corresponding values of the rheological parameters. As can be observed in that table, since the orbital decay timescale is much larger than the others, when $a \approx r_{\mathrm{Roche}}$, the orbit will be completely circularized.

\subsection{Possibility of detecting tidal effects}

It may be an interesting exercise to consider the possibility of contrasting the results obtained from numerical simulation with those from observations. Within the limits of this work, we can answer the question of whether tidal effects can be detected only in part. In recent years, great advances have been made in observational techniques that allow for highly precise measurements of the physical and orbital parameters of planets orbiting stars other than the Sun. One of these quantities is the orbital period, $P_{\mathrm{orb}}$, of planets, which can be measured by observing several successive transits and is defined by: 
\begin{equation}\label{ec:def_porb} 
P_{\mathrm{orb}} = \frac{2 \, \pi}{n} 
.\end{equation} If we assume that the orbital parameters of a two body system evolve in time only due to the tides each body rises on the other -- as was supposed in the last subsection -- then the orbital period is also expected to evolve as a consequence of the time variation of the major semiaxis. This can be expressed mathematically as: 
\begin{equation}\label{ec:dPorbdt} 
\frac{\mathrm{d} P_{\mathrm{orb}}}{\mathrm{d}t} = - \frac{2 \, \pi}{n^2} \frac{\mathrm{d}n}{\mathrm{d}t}
.\end{equation} From Eq.~\eqref{ec:defn}, we have: \begin{equation*} \frac{\mathrm{d}n}{\mathrm{d}t} = - \frac{3}{2} \frac{n}{a}\frac{\mathrm{d}a}{\mathrm{d}t}. \end{equation*} Then, Eq.~\eqref{ec:dPorbdt} can be rewritten as: 
\begin{equation}\label{ec:dPorbdtdadt} 
\frac{\mathrm{d} P_{\mathrm{orb}}}{\mathrm{d}t} = \frac{3 \, \pi}{n \, a} \frac{\mathrm{d}a}{\mathrm{d}t}. 
\end{equation} If we take the values of the parameters for the Kepler 21 system given in Table~\ref{tab:physicaldata}, compute $n$ with Eq.~\eqref{ec:defn}, and also take the value of $\left\langle \frac{\mathrm{d}a}{\mathrm{d}t} \right\rangle$, obtained from direct evaluation of Eq.~\eqref{ec:dadt_sec_mod}, assuming $\dot{\theta}_2 = n$ and evaluating the tidal frequencies with Eq.~\eqref{ec:tidalmodesapprox}, and then we replace them in Eq.~\eqref{ec:dPorbdtdadt}, we get the following result:
\begin{equation*}
      \left\langle \frac{\mathrm{d} P_{\mathrm{orb}}}{\mathrm{d}t} \right\rangle^{\mathrm{tide}} \approx -0.003879 \, \frac{\mathrm{s}}{\mathrm{yr}},
\end{equation*} which is equivalent to a decrement of nearly $3.88 \, \mbox{ms}$ per year. As a consequence, we would need more than 1000 years to measure a difference of about 4 seconds. The measured orbital period of planet Kepler-21b is $2.78578$ days with an uncertainty of $0.00003$ days \citep{LopezMorales_etal_2016}, which is equivalent to approximately $2.6 \, \mbox{s}$. This result clearly shows the impossibility of detecting tidal effects, if simply by measuring the orbital period of the planet, at least in this case.

In this regard, the question of whether it is possible to detect tidal effects by measuring the orbital period of certain exoplanets has motivated the publication of several recent works. Depending on each particular case, some authors conclude that it is not possible to detect tidal effect, as in \citet{petruccietal2019}, and others not only assert that they measure a decrease of the orbital period, but also that it is a consequence of the orbital decay due to tidal interaction \citep{Yee_2019}.

\section{\label{sec:conclusions}Final discussion and conclusions}

In this work, we develop an accurate method for studying the dynamical evolution of a binary system formed by a rocky planet and its host star. We follow the work by \citet{boue_efroimsky2019} to obtain a set of equations of motion which, on one hand, describe the secular time evolution of the orbital elements attributed to the secular terms of the tidal potentials of both partners and the secular terms related to the $J_2$ coefficient. On the other hand, it gives the time rate of the stellar spin rate due to the secular terms of the corresponding tidal torque and the time evolution of the proper rotation angle of the planet due to the combined action of the secular terms of the tidal torque and the periodic terms of the triaxiality-induced torque. Thus, the main results of this investigation are: Eqs.~\eqref{ec:dadt_sec_mod}, \eqref{ec:dxidt_sec_mod}, \eqref{ec:dMdt_sec_mod}, \eqref{ec:dx1dt_sec_mod}, \eqref{ec:dx2dt_sec_mod}, \eqref{ec:dw1dt_sec_mod}, \eqref{ec:dw2dt_sec_mod}, \eqref{ec:dOm1dt_sec_mod}, \eqref{ec:dOm2dt_sec_mod}, \eqref{ec:d2th1dt2_sec_mod}, and \eqref{ec:d2th2dt2_tt_mod} which contain no singularities in the limit of vanishing eccentricities, inclination, and obliquity. Bearing in mind that the transformation introduced to obtain the equations only moves the latter singularity to $e=1$ and $i = 90 \degr$, we would like to stress the fact that it is a convenient substitution because it allows for the possibility that the eccentricity and the inclinations reach very low values (even zero) without affecting the stability of the numerical integration.

As a case study, we chose the Kepler-21 system. We find that the commonly accepted approximate expression of the tidal modes, given by Eq.~\eqref{ec:tidalmodesapprox}, could no longer be valid in studies such as that carried out in this paper. In other words, at least in the case of the Kepler-21 system, the apsidal and nodal precessions induced by tides and the $J_2$-related terms is appreciable. Also, we  find that the time of tidal synchronization strongly depends on the values of the rheological parameters, as is shown in Fig.~\ref{fig:thpn2vst_set1}.

Regarding the rotational state of planet Kepler-21b, even though it could be or have previously been trapped in the 3:2 spin-orbit resonance, we expect that it has already reached the synchronous rotation, not only because this is its most likely state, but as it orbits very close to a massive star with a radius nearly twice grater than that of our Sun, we can reasonably suppose that the planet is undergoing a strong internal dissipation which can cause an increase of its internal temperature with the consequent decrease of the viscosity and of the Maxwell time. As explained by \citet{makarovetal_2018}, this decrease of the Maxwell time can cause the escape from higher-than-synchronous rotation states in a time interval shorter that the corresponding decrease of the corresponding capture probabilities due to the decrease of the eccentricity in time \citep{makarov_2015}.  

It worth mentioning that from the point of view of tidal theory, the case of Kepler-21 -- after the planet has reached the synchronous rotation -- is analogous to the case of Phobos, which is gradually spiraling inwards towards Mars \citep{efrolai2007}. In this sense, the timescales of orbital decay and circularization given in Table~\ref{tab:timescales} are upper bounds, that is, the rates of orbital decay and circularization become greater as the planet approaches its host star. Even though the results shown in Table~\ref{tab:timescales} suggest that the planet has a short time to exist, we can note that the more rapidly the eccentricity gets damped, the more the time it  takes the planet to reach the Roche limit.

The time evolution of the orbital inclination ($i_1$) seems to be counter-intuitive. However, this is a particularity of the Kepler-21 system, but there could be other systems in which a more ``expectable'' conclusion is reached. As shown in Subsect.~\ref{ss:numint}, tidal interaction can lead the major semiaxis, the eccentricity, and the inclinations to increase or decrease over time, depending on the particular combinations of the values of the rheological and dynamical parameters and variables involved \citep[see][]{verasetal2019}.

We would like to emphasize that even though we have addressed the study of a particular system, the formalism developed in this work is equally applicable to other analogous systems, formed by a star and a terrestrial planet. Moreover, this methodology could be adapted to model the dynamical evolution of binary asteroids. We will carry out the former and the latter types of studies in future works.

The analytical model that we propose in this work takes into account the tidal dissipation of both components of a binary system and the influence of both the $J_2$-related secular terms in the orbital evolution and the permanent-triaxiality-induced torque in the rotational evolution of one of the bodies, while the rotational evolution of the other is considered to be dominated only by the tidal interaction. We are aware that there are other effects that could be equally important to the dynamical evolution and could later be incorporated into the formalism, such as the oblateness of the star, General Relativity \citep{correiaetal2011}, dynamical tides in the star \citep{bolmontetal2016,hodzicetal2018}, atmospheric tides in the planet \citep{correialaskar2003,Leconte2015}, magnetic couplings between the star and the planet \citep{ahuir2019}, and stellar evolution \citep{benbakouraetal2019}.

\begin{acknowledgements} We acknowledge the funding from ANPCyT through PICT 1144-13 and CONICET through PIP 699-15. This investigation was supported by the institutional cluster HOPE at IAFE (\url{http://www.iafe.uba.ar/HOPE/}), where all the numerical simulations were performed, and by the SAO/NASA Astrophysics Data System. Santiago Luna wants to acknowledge the fruitful discussions with Michael Efroimsky, Gwenaël Boué, Dimitri Veras and Valeri Makarov, which helped to reach a better understanding of the dynamical evolution near spin-orbit resonances and to improve the quality of this work. This research has made use of the NASA Exoplanet Archive, which is operated by the California Institute of Technology, under contract with the National Aeronautics and Space Administration under the Exoplanet Exploration Program. 
\end{acknowledgements}

\bibliographystyle{aa} 
\bibliography{referencias}{} 

\begin{appendix}

\section{\label{sec:sum_terms_adim}Expression of the sums of adimensional terms of the derivatives of the disturbing potentials}

In the following, we give the complete expressions of the derivatives of the disturbing potentials considered in this work. Regarding the gradient of the tidal potentials of both bodies, we also give the secular terms of the corresponding derivatives. The latter is identified by the superscript "(sec)". Among the aforementioned expressions, the terms corresponding to $(lmpq) = (201q)$ and $(lmpq) = (220q)$ are shown, in particular. As earlier in this work, such terms are identified with the respective subscripts.

\subsection{Major semiaxis}

\begin{subequations}
\begin{multline}\label{ec:SdUkda}
\frac{\partial U_{lmpqhs}^{(k)}}{\partial a} = \sum_{l=2}^{\infty} \sum_{m=0}^l \sum_{p=0}^l \sum_{q=-\infty}^\infty \sum_{h=0}^l \sum_{s = -\infty}^\infty (l+1) \left(\frac{R_k}{a}\right)^{2l+1} \frac{(l-m)!}{(l+m)!} (2-\delta_{0m}) F_{lmp} (i_k) \, G_{lpq} (e) \, F_{lmh} (i_k) \, G_{lhs} (e) \\ \times \left[\cos \left(v_{lmpq}^{(k)} - v_{lmhs}^{(k)}\right) K_{\mathrm{R}}^{(k)} \left(l,\chi_{lmpq}^{(k)}\right)  + \sin \left(v_{lmpq}^{(k)} - v_{lmhs}^{(k)}\right) K_{\mathrm{I}}^{(k)} \left(l,\chi_{lmpq}^{(k)}\right) \right],
\end{multline}
\begin{equation}\label{ec:SdUkda_sec}
\left\langle\frac{\partial U_{lmpq}^{(k)}}{\partial a}\right\rangle = \sum_{l=2}^{\infty} \sum_{m=0}^l \sum_{p=0}^l \sum_{q=-\infty}^\infty (l+1) \left(\frac{R_k}{a}\right)^{2l+1} \frac{(l-m)!}{(l+m)!} (2-\delta_{0m}) F_{lmp}^2 (i_k) \, G_{lpq}^2 (e) \, K_{\mathrm{R}}^{(k)} \left(l,\chi_{lmpq}^{(k)}\right),
\end{equation}
\begin{equation}\label{ec:SdUkda_201q}
\left\langle\frac{\partial U_{201q}^{(k)}}{\partial a}\right\rangle = 3 \left(\frac{R_k}{a}\right)^5 F_{201}^2 (i_k) \sum_{q=-\infty}^{\infty} G_{21q}^2 (e) \, K_{\mathrm{R}}^{(k)} \left(2,\omega_{201q}^{(k)}\right)
,\end{equation}
\begin{equation}\label{ec:SdUkda_220q}
\left\langle\frac{\partial U_{220q}^{(k)}}{\partial a}\right\rangle = \frac{1}{4} \left(\frac{R_k}{a}\right)^5 F_{220}^2 (i_k) \sum_{q=-\infty}^{\infty} G_{20q}^2 (e) \, K_{\mathrm{R}}^{(k)} \left(2,\omega_{220q}^{(k)}\right)
,\end{equation}

\begin{equation}\label{ec:SdV2da}
\frac{\partial V_{lmpq}^{(2)}}{\partial a} = \sum_{l=2}^{\infty} \sum_{m=0}^l \sum_{p=0}^l \sum_{q=-\infty}^\infty (l+1) \left(\frac{R_2}{a}\right)^{l} F_{lmp} (i_2) \, G_{lpq} (e) \left[C_{lm} \cos \left(v_{lmpq}^{(2)} - m \, \theta_2 + \phi_{lm}\right) + S_{lm} \sin \left(v_{lmpq}^{(2)} - m \, \theta_2 + \phi_{lm}\right)\right],
\end{equation}
\begin{equation}\label{ec:SdV2da_201q}
\frac{\partial V_{201q}^{(2)}}{\partial a} = - 3 J_2 \left(\frac{R_2}{a}\right)^2 F_{201} (i_2) \sum_{q=-\infty}^{\infty} G_{21q} (e) \cos (q \, M),
\end{equation}
\begin{equation}\label{ec:SdV2da_220q}
\frac{\partial V_{220q}^{(2)}}{\partial a} = \frac{3}{4} \frac{(B-A)}{m_2 \, a^2} F_{220} (i_2) \sum_{q=-\infty}^{\infty} G_{20q} (e) \cos \left[2 \left(\omega_2 + \Omega_2 - \theta_2\right) + (2+q) M \right]
.\end{equation}
\end{subequations}

\subsection{Eccentricity}

\begin{subequations}
\begin{multline}\label{ec:SdUkde}
\frac{\partial U_{lmpqhs}^{(k)}}{\partial e} = \sum_{l=2}^{\infty} \sum_{m=0}^l \sum_{p=0}^l \sum_{q=-\infty}^\infty \sum_{h=0}^l \sum_{s = -\infty}^\infty \left(\frac{R_k}{a}\right)^{2l+1} \frac{(l-m)!}{(l+m)!} (2-\delta_{0m}) F_{lmp} (i_k) \, G_{lpq} (e) \, F_{lmh} (i_k) \, \frac{\mathrm{d}G_{lhs} (e)}{\mathrm{d}e} \\ \times \left[\cos \left(v_{lmpq}^{(k)} - v_{lmhs}^{(k)}\right) K_{\mathrm{R}}^{(k)} \left(l,\chi_{lmpq}^{(k)}\right) + \sin \left(v_{lmpq}^{(k)} - v_{lmhs}^{(k)}\right) K_{\mathrm{I}}^{(k)} \left(l,\chi_{lmpq}^{(k)}\right) \right],
\end{multline}
\begin{equation}\label{ec:SdUkde_sec}
\left\langle\frac{\partial U_{lmpq}^{(k)}}{\partial e}\right\rangle = \sum_{l=2}^{\infty} \sum_{m=0}^l \sum_{p=0}^l \sum_{q=-\infty}^\infty \left(\frac{R_k}{a}\right)^{2l+1} \frac{(l-m)!}{(l+m)!} (2-\delta_{0m}) F_{lmp}^2 (i_k) \, G_{lpq} (e) \frac{\mathrm{d}G_{lpq} (e)}{\mathrm{d}e} \, K_{\mathrm{R}}^{(k)} \left(l,\chi_{lmpq}^{(k)}\right),
\end{equation}
\begin{equation}\label{ec:SdUkde_201q}
\left\langle\frac{\partial U_{201q}^{(k)}}{\partial e}\right\rangle = \left(\frac{R_k}{a}\right)^{5} F_{201}^2 (i_k) \sum_{q=-\infty}^\infty G_{21q} (e) \frac{\mathrm{d}G_{21q} (e)}{\mathrm{d}e} \, K_{\mathrm{R}}^{(k)} \left(2,\omega_{201q}^{(k)}\right),
\end{equation}
\begin{equation}\label{ec:SdUkde_220q}
\left\langle\frac{\partial U_{220q}^{(k)}}{\partial e}\right\rangle = \frac{1}{12} \left(\frac{R_k}{a}\right)^{5} F_{220}^2 (i_k) \sum_{q=-\infty}^\infty G_{20q} (e) \frac{\mathrm{d}G_{20q} (e)}{\mathrm{d}e} \, K_{\mathrm{R}}^{(k)} \left(2,\omega_{220q}^{(k)}\right),
\end{equation}

\begin{equation}\label{ec:SdV2de}
\frac{\partial V_{lmpq}^{(2)}}{\partial e} = \sum_{l=2}^{\infty} \sum_{m=0}^l \sum_{p=0}^l \sum_{q=-\infty}^\infty \left(\frac{R_2}{a}\right)^{l} F_{lmp} (i_2) \, \frac{\mathrm{d}G_{lpq} (e)}{\mathrm{d}e} \left[C_{lm} \cos \left(v_{lmpq}^{(2)} - m \, \theta_2 + \phi_{lm}\right) + S_{lm} \sin \left(v_{lmpq}^{(2)} - m \, \theta_2 + \phi_{lm}\right)\right],
\end{equation}
\begin{equation}\label{ec:SdV2de_201q}
\frac{\partial V_{201q}^{(2)}}{\partial e} = - J_2 \left(\frac{R_2}{a}\right)^2 F_{201} (i_2) \sum_{q=-\infty}^{\infty} \frac{\mathrm{d}G_{21q} (e)}{\mathrm{d}e} \cos (q \, M),
\end{equation}
\begin{equation}\label{ec:SdV2de_220q}
\frac{\partial V_{220q}^{(2)}}{\partial e} = \frac{1}{4} \frac{(B-A)}{m_2 \, a^2} F_{220} (i_2) 
\sum_{q=-\infty}^{\infty} \frac{\mathrm{d}G_{20q} (e)}{\mathrm{d}e} \cos \left[2 \left(\omega_2 + \Omega_2 - \theta_2\right) + (2+q) M \right].
\end{equation}
\end{subequations}

\subsection{Mean anomaly}

\begin{subequations}
\begin{multline}\label{ec:SdUkdM}
\frac{\partial U_{lmpqhs}^{(k)}}{\partial M} = \sum_{l=2}^{\infty} \sum_{m=0}^l \sum_{p=0}^l \sum_{q=-\infty}^\infty \sum_{h=0}^l \sum_{s = -\infty}^\infty \left(\frac{R_k}{a}\right)^{2l+1} \frac{(l-m)!}{(l+m)!} (2-\delta_{0m}) F_{lmp} (i_k) \, G_{lpq} (e) \, F_{lmh} (i_k) \, G_{lhs} (e) \\ \times \left[\sin \left(v_{lmpq}^{(k)} - v_{lmhs}^{(k)}\right) K_{\mathrm{R}}^{(k)} \left(l,\chi_{lmpq}^{(k)}\right) - \cos \left(v_{lmpq}^{(k)} - v_{lmhs}^{(k)}\right) K_{\mathrm{I}}^{(k)} \left(l,\chi_{lmpq}^{(k)}\right) \right] (l-2h+s),
\end{multline}
\begin{equation}\label{ec:SdUkdM_sec}
\left\langle\frac{\partial U_{lmpq}^{(k)}}{\partial M}\right\rangle = - \sum_{l=2}^{\infty} \sum_{m=0}^l \sum_{p=0}^l \sum_{q=-\infty}^\infty \left(\frac{R_k}{a}\right)^{2l+1} \frac{(l-m)!}{(l+m)!} (2-\delta_{0m}) F_{lmp}^2 (i_k) \, G_{lpq}^2 (e) \, K_{\mathrm{I}}^{(k)} \left(l,\chi_{lmpq}^{(k)}\right) (l-2p+q).
\end{equation}
\begin{equation}\label{ec:SdUkdM_201q}
\left\langle\frac{\partial U_{201q}^{(k)}}{\partial M}\right\rangle = - \left(\frac{R_k}{a}\right)^{5} F_{201}^2 (i_k) \sum_{q=-\infty}^\infty G_{21q}^2 (e) \, q \, K_{\mathrm{I}}^{(k)} \left(2,\omega_{201q}^{(k)}\right),
\end{equation}
\begin{equation}\label{ec:SdUkdM_220q}
\left\langle\frac{\partial U_{220q}^{(k)}}{\partial M}\right\rangle = - \frac{1}{12} \left(\frac{R_k}{a}\right)^{5} F_{220}^2 (i_k) \sum_{q=-\infty}^\infty G_{20q}^2 (e) \, (2+q) \, K_{\mathrm{I}}^{(k)} \left(2,\omega_{220q}^{(k)}\right),
\end{equation}

\begin{equation}\label{ec:SdV2dM}
\frac{\partial V_{lmpq}^{(2)}}{\partial M} = \sum_{l=2}^{\infty} \sum_{m=0}^l \sum_{p=0}^l \sum_{q=-\infty}^\infty \left(\frac{R_2}{a}\right)^{l} F_{lmp} (i_2) \, G_{lpq} (e) \left[- C_{lm} \sin \left(v_{lmpq}^{(2)} - m \, \theta_2 + \phi_{lm}\right) + S_{lm} \cos \left(v_{lmpq}^{(2)} - m \, \theta_2 + \phi_{lm}\right)\right] (l-2p+q),
\end{equation}
\begin{equation}\label{ec:SdV2dM_201q}
\frac{\partial V_{201q}^{(2)}}{\partial M} = J_2 \left(\frac{R_2}{a}\right)^2 F_{201} (i_2) \sum_{q=-\infty}^\infty G_{21q} (e) \, q \, \sin (q \, M),
\end{equation}
\begin{equation}\label{ec:SdV2dM_220q}
\frac{\partial V_{220q}^{(2)}}{\partial M} = - \frac{1}{4} \frac{(B-A)}{m_2 \, a^2}  F_{220} (i_2) \sum_{q=-\infty}^\infty G_{20q} (e) \, (2+q) \, \sin \left[2(\omega_2 + \Omega_2 - \theta_2) + (2+q) \, M\right].
\end{equation}
\end{subequations}

\subsection{Inclination with respect to the star's and planet's equators}

\begin{subequations}
\begin{multline}\label{ec:SdUkdik}
\frac{\partial U_{lmpqhs}^{(k)}}{\partial i_k} = \sum_{l=2}^{\infty} \sum_{m=0}^l \sum_{p=0}^l \sum_{q=-\infty}^\infty \sum_{h=0}^l \sum_{s = -\infty}^\infty \left(\frac{R_k}{a}\right)^{2l+1} \frac{(l-m)!}{(l+m)!} (2-\delta_{0m}) F_{lmp} (i_k) \, G_{lpq} (e) \, \frac{\mathrm{d}F_{lmh} (i_k)}{\mathrm{d}i_k} \, G_{lhs} (e) \\ \times \left[\cos \left(v_{lmpq}^{(k)} - v_{lmhs}^{(k)}\right) K_{\mathrm{R}}^{(k)} \left(l,\chi_{lmpq}^{(k)}\right) + \sin \left(v_{lmpq}^{(k)} - v_{lmhs}^{(k)}\right) K_{\mathrm{I}}^{(k)} \left(l,\chi_{lmpq}^{(k)}\right) \right],
\end{multline}
\begin{equation}\label{ec:SdUkdik_sec}
\left\langle\frac{\partial U_{lmpq}^{(k)}}{\partial i_k}\right\rangle = \sum_{l=2}^{\infty} \sum_{m=0}^l \sum_{p=0}^l \sum_{q=-\infty}^\infty \left(\frac{R_k}{a}\right)^{2l+1} \frac{(l-m)!}{(l+m)!} (2-\delta_{0m}) F_{lmp} (i_k) \frac{\mathrm{d}F_{lmp} (i_k)}{\mathrm{d}i_k} \, G_{lpq}^2 (e) \, K_{\mathrm{R}}^{(k)} \left(l,\chi_{lmpq}^{(k)}\right),
\end{equation}
\begin{equation}\label{ec:SdUkdik_201q}
\left\langle\frac{\partial U_{201q}^{(k)}}{\partial i_k}\right\rangle = \left(\frac{R_k}{a}\right)^{5} F_{201} (i_k) 
\frac{\mathrm{d}F_{201} (i_k)}{\mathrm{d}i_k} \sum_{q=-\infty}^\infty G_{21q}^2 (e) \, K_{\mathrm{R}}^{(k)} \left(2,\omega_{201q}^{(k)}\right),
\end{equation}
\begin{equation}\label{ec:SdUkdik_220q}
\left\langle\frac{\partial U_{220q}^{(k)}}{\partial i_k}\right\rangle = \frac{1}{12} \left(\frac{R_k}{a}\right)^{5} F_{220} (i_k) \frac{\mathrm{d}F_{220} (i_k)}{\mathrm{d}i_k} \sum_{q=-\infty}^\infty G_{20q}^2 (e) \, K_{\mathrm{R}}^{(k)} \left(2,\omega_{220q}^{(k)}\right),
\end{equation}

\begin{equation}\label{ec:SdV2di2}
\frac{\partial V_{lmpq}^{(2)}}{\partial i_2} = \sum_{l=2}^{\infty} \sum_{m=0}^l \sum_{p=0}^l \sum_{q=-\infty}^\infty \left(\frac{R_2}{a}\right)^{l} \frac{\mathrm{d}F_{lmp} (i_2)}{\mathrm{d}i_2} \, G_{lpq} (e) \left[C_{lm} \cos \left(v_{lmpq}^{(2)} - m \, \theta_2 + \phi_{lm}\right) + S_{lm} \sin \left(v_{lmpq}^{(2)} - m \, \theta_2 + \phi_{lm}\right)\right],
\end{equation}
\begin{equation}\label{ec:SdV2di2_201q}
\frac{\partial V_{201q}^{(2)}}{\partial i_2} = - J_2 \left(\frac{R_2}{a}\right)^2 \frac{\mathrm{d}F_{201} (i_2)}{\mathrm{d}i_2} \sum_{q=-\infty}^\infty \, G_{21q} (e) \cos (q \, M),
\end{equation}
\begin{equation}\label{ec:SdV2di2_220q}
\frac{\partial V_{220q}^{(2)}}{\partial i_2} = \frac{1}{4} \frac{(B-A)}{m_2 \, a^2} \frac{\mathrm{d}F_{220} (i_2)}{\mathrm{d}i_2} \sum_{q=-\infty}^\infty \, G_{20q} (e) \cos \left[2(\omega_2 + \Omega_2 - \theta_2) + (2+q) M\right].
\end{equation}
\end{subequations}

\subsection{Argument of pericenter with respect to the star's and planet's equators}

\begin{subequations}
\begin{multline}\label{ec:SdUkdwk}
\frac{\partial U_{lmpqhs}^{(k)}}{\partial \omega_k} = \sum_{l=2}^{\infty} \sum_{m=0}^l \sum_{p=0}^l \sum_{q=-\infty}^\infty \sum_{h=0}^l \sum_{s = -\infty}^\infty \left(\frac{R_k}{a}\right)^{2l+1} \frac{(l-m)!}{(l+m)!} (2-\delta_{0m}) F_{lmp} (i_k) \, G_{lpq} (e) \, F_{lmh} (i_k) \, G_{lhs} (e) \\ \times \left[\sin \left(v_{lmpq}^{(k)} - v_{lmhs}^{(k)}\right) K_{\mathrm{R}}^{(k)} \left(l,\chi_{lmpq}^{(k)}\right) - \cos \left(v_{lmpq}^{(k)} - v_{lmhs}^{(k)}\right) K_{\mathrm{I}}^{(k)} \left(l,\chi_{lmpq}^{(k)}\right) \right] (l-2h),
\end{multline}
\begin{equation}\label{ec:SdUkdwk_sec}
\left\langle\frac{\partial U_{lmpq}^{(k)}}{\partial \omega_k}\right\rangle = - \sum_{l=2}^{\infty} \sum_{m=0}^l \sum_{p=0}^l \sum_{q=-\infty}^\infty \left(\frac{R_k}{a}\right)^{2l+1} \frac{(l-m)!}{(l+m)!} (2-\delta_{0m}) F_{lmp}^2 (i_k) \, G_{lpq}^2 (e) \, K_{\mathrm{I}}^{(k)} \left(l,\chi_{lmpq}^{(k)}\right) (l-2p),
\end{equation}
\begin{equation}\label{ec:SdUkdwk_201q}
\left\langle\frac{\partial U_{201q}^{(k)}}{\partial \omega_k}\right\rangle = 0,
\end{equation}
\begin{equation}\label{ec:SdUkdwk_220q}
\left\langle\frac{\partial U_{220q}^{(k)}}{\partial \omega_k}\right\rangle = - \frac{1}{6} \left(\frac{R_k}{a}\right)^{5} 
F_{220}^2 (i_k) \sum_{q=-\infty}^\infty G_{20q}^2 (e) \, K_{\mathrm{I}}^{(k)} \left(2,\omega_{220q}^{(k)}\right),
\end{equation}

\begin{equation}\label{ec:SdV2dw2}
\frac{\partial V_{lmpq}^{(2)}}{\partial \omega_2} = \sum_{l=2}^{\infty} \sum_{m=0}^l \sum_{p=0}^l \sum_{q=-\infty}^\infty \left(\frac{R_2}{a}\right)^{l} F_{lmp} (i_2) \, G_{lpq} (e) \left[- C_{lm} \sin \left(v_{lmpq}^{(2)} - m \, \theta_2 + \phi_{lm}\right) + S_{lm} \cos \left(v_{lmpq}^{(2)} - m \, \theta_2 + \phi_{lm}\right)\right] (l-2p),
\end{equation}
\begin{equation}\label{ec:SdV2dw2_201q}
\frac{\partial V_{201q}^{(2)}}{\partial \omega_2} = 0,
\end{equation}
\begin{equation}\label{ec:SdV2dw2_220q}
\frac{\partial V_{220q}^{(2)}}{\partial \omega_2} = - \frac{1}{4} \frac{(B-A)}{m_2 \, a^2} F_{220} (i_2) \sum_{q=-\infty}^\infty G_{20q} (e) \sin \left[2(\omega_2 + \Omega_2 - \theta_2) + (2+q) M\right],
\end{equation}
\end{subequations}

\subsection{Longitude of ascending node with respect to the star's and planet's equators}

\begin{subequations}
\begin{multline}\label{ec:SdUkdOk}
\frac{\partial U_{lmpqhs}^{(k)}}{\partial \Omega_k} = \sum_{l=2}^{\infty} \sum_{m=0}^l \sum_{p=0}^l \sum_{q=-\infty}^\infty \sum_{h=0}^l \sum_{s = -\infty}^\infty \left(\frac{R_k}{a}\right)^{2l+1} m \frac{(l-m)!}{(l+m)!} (2-\delta_{0m}) F_{lmp} (i_k) \, G_{lpq} (e) \, F_{lmh} (i_k) \, G_{lhs} (e) \\ \times \left[\sin \left(v_{lmpq}^{(k)} - v_{lmhs}^{(k)}\right) K_{\mathrm{R}}^{(k)} \left(l,\chi_{lmpq}^{(k)}\right) - \cos \left(v_{lmpq}^{(k)} - v_{lmhs}^{(k)}\right) K_{\mathrm{I}}^{(k)} \left(l,\chi_{lmpq}^{(k)}\right) \right],
\end{multline}
\begin{equation}\label{ec:SdUkdOk_sec}
\left\langle\frac{\partial U_{lmpq}^{(k)}}{\partial \Omega_k}\right\rangle = - \sum_{l=2}^{\infty} \sum_{m=0}^l \sum_{p=0}^l \sum_{q=-\infty}^\infty \left(\frac{R_k}{a}\right)^{2l+1} m \frac{(l-m)!}{(l+m)!} (2-\delta_{0m}) F_{lmp}^2 (i_k) \, G_{lpq}^2 (e) \, K_{\mathrm{I}}^{(k)} \left(l,\chi_{lmpq}^{(k)}\right),
\end{equation}
\begin{equation}\label{ec:SdUkdOk_201q}
\left\langle\frac{\partial U_{201q}^{(k)}}{\partial \Omega_k}\right\rangle = 0,
\end{equation}
\begin{equation}\label{ec:SdUkdOk_220q}
\left\langle\frac{\partial U_{220q}^{(k)}}{\partial \Omega_k}\right\rangle = - \frac{1}{6} \left(\frac{R_k}{a}\right)^{5} 
F_{220}^2 (i_k) \sum_{q=-\infty}^\infty G_{20q}^2 (e) \, K_{\mathrm{I}}^{(k)} \left(2,\omega_{220q}^{(k)}\right),
\end{equation}

\begin{equation}\label{ec:SdV2dO2}
\frac{\partial V_{lmpq}^{(2)}}{\partial \Omega_2} = \sum_{l=2}^{\infty} \sum_{m=0}^l \sum_{p=0}^l \sum_{q=-\infty}^\infty \left(\frac{R_2}{a}\right)^{l} F_{lmp} (i_2) \, G_{lpq} (e) \, m \left[- C_{lm} \sin \left(v_{lmpq}^{(2)} - m \, \theta_2 + \phi_{lm}\right) + S_{lm} \cos \left(v_{lmpq}^{(2)} - m \, \theta_2 + \phi_{lm}\right)\right],
\end{equation}
\begin{equation}\label{ec:SdV2dO2_201q}
\frac{\partial V_{201q}^{(2)}}{\partial \Omega_2} = 0,
\end{equation}
\begin{equation}\label{ec:SdV2dO2_220q}
\frac{\partial V_{220q}^{(2)}}{\partial \Omega_2} = - \frac{1}{2} \frac{(B-A)}{m_2 \, a^2} F_{220} (i_2) \sum_{q=-\infty}^\infty G_{20q} (e) \sin \left[2(\omega_2 + \Omega_2 - \theta_2) + (2+q) M\right].
\end{equation}
\end{subequations}

\section{\label{sec:eccfuncs}Eccentricity functions}

With regard to the eccentricity functions, they are identical to the well-known Hansen's coefficients, $X_{\kappa}^{\nu,\mu} (e)$, which, in general, are defined as: 
\begin{equation}\label{ec:defhansencoefs} 
\left(\frac{r}{a}\right)^{\nu} \exp{\left(\textrm{i} \, \mu \, f\right)} = \sum_\kappa X_{\kappa}^{\nu,\mu} (e) \exp{\left(\textrm{i} \, \kappa \, M\right)}, 
\end{equation} where $\nu$, $\mu$ and $\kappa$ are integers, $r$ is the distance, $f$ is the true anomaly and $\textrm{i} = \sqrt{-1}$. For low to moderate eccentricities they can be computed as a power series of the eccentricity \citep{hughes1981}: 
\begin{equation}\label{ec:hansencoefseries} 
X_{\kappa}^{\nu,\mu} (e) = e^{\left\vert \mu-\kappa \right\vert} \sum_{j=0}^{\infty} N_{\rho,\sigma}^{\nu,\mu} e^{2 j} 
,\end{equation} where $\rho = j + \max (0,\kappa-\mu)$, $\sigma = j + \max (0,\mu-\kappa)$ and $N_{\rho,\sigma}^{\nu,\mu}$ are known as the Newcomb's operators which can be computed recursively by \citep{proulxmacclain1988}: 
\begin{equation}\label{ec:defrecnewcomb} 
\begin{split} 
4 (\rho+\sigma) N_{\rho,\sigma}^{\nu,\mu} = & \, 2(2 \mu - \nu) \, N^{\nu,\mu+1}_{\rho-1,\sigma} - 2(2\mu+\nu) \, N^{\nu,\mu-1}_{\rho,\sigma-1} \\
                        & + (\mu-\nu) \, N^{\nu,\mu+2}_{\rho-2,\sigma} - (\mu+\nu) \, N^{\nu,\mu-2}_{\rho,\sigma-2} \\
                        & + 2(2\rho+2\sigma-4-\nu) \, N^{\nu,\mu}_{\rho-1,\sigma-1} 
\end{split} 
.\end{equation} If $\sigma>\rho$ then $N^{\nu,\mu}_{\rho,\sigma} = N^{\nu,-\mu}_{\sigma,\rho}$. The recursion is initialized by considering that $N^{\nu,\mu}_{\rho,\sigma} = 0$ if $\rho<0$ or $\sigma<0$ and by $N^{\nu,\mu}_{0,0} = 1$.

The relationship between Kaula's eccentricity functions, $G_{lpq} (e)$, and the Hansen coefficients is given by: 
\begin{equation}\label{ec:relGX} 
G_{lpq} (e) = X^{-(l+1),l-2p}_{l-2p+q} (e) 
.\end{equation} Using Eq.~\eqref{ec:relGX} and Ec.~\eqref{ec:hansencoefseries}, the $G_{lpq} (e)$ can be computed by means of: 
\begin{equation}\label{ec:calcGlpq} G_{lpq} (e) = \sum_{j=0}^{\infty} N^{-(l+1),l-2p}_{\rho,\sigma} \, e^{2 j + \vert q \vert} 
,\end{equation} where $\rho$ and $\sigma$ are given now by: $\rho = j+ \max (0,q)$ and $\sigma = j+ \max (0, -q)$, respectively.  Of course, the summation over $j$ must be also truncated at some maximum order in eccentricity, that is, $2 j + \vert q \vert \leq j_{\mathrm{max}}$.

The calculation of the first derivatives of the eccentricity functions are straightforward by virtue of Eq.~\eqref{ec:relGX} is:
\begin{equation}\label{ec:dGelpqde}
\frac{\mathrm{d}G_{lpq} (e)}{\mathrm{d}e} = \sum_{j=1}^{\infty} N^{-(l+1),l-2p}_{\rho,\sigma} \, \left(2 j + \vert q \vert\right) \, 
e^{2 j + \vert q \vert - 1}.
\end{equation}

\end{appendix}

\listofobjects

\end{document}